\documentclass[11pt,a4paper]{article}

\usepackage[T1]{fontenc}
\usepackage[dvips]{graphicx}
\usepackage{color}
\usepackage{times}
\usepackage{amssymb}
\usepackage{rotating}
\usepackage[bf]{subfigure}
\usepackage{setspace}
\usepackage[square, comma, numbers,sort&compress]{natbib}
\usepackage[left=3cm,bottom=3.2cm,right=3.5cm]{geometry}

\title{Analyzing and Modeling Real-World Phenomena with Complex
  Networks: A Survey of Applications}

\author{Luciano da Fontoura Costa$^1$ \and Osvaldo N. Oliveira Jr.$^1$
  \and Gonzalo Travieso$^1$ \and Francisco Aparecido Rodrigues$^1$
  \and Paulino Ribeiro Villas Boas$^1$ \and Lucas Antiqueira$^1$ \and
  Matheus Palhares Viana$^1$ \and Luis Enrique Correa da Rocha$^2$\\
  \small $^1$Instituto de F\'{\i}sica de S\~{a}o Carlos, Universidade
  de S\~{a}o Paulo\\ \small S\~{a}o Paulo, S\~{a}o Carlos, SP, PO Box
  369, 13560-970,\\ \small phone +55 16 3373 9858, FAX +55 16 3371
  3616, Brazil\\ \small $^2$Department of Physics, Ume{\aa}
  University, 90187 Ume{\aa}, Sweden\\ \texttt{luciano@ifsc.usp.br} }

\begin{document}

\maketitle

\begin{abstract}
  The success of new scientific areas can be assessed by their
  potential for contributing to new theoretical approaches and in
  applications to real-world problems. Complex networks have fared
  extremely well in both of these aspects, with their sound theoretical
  basis developed over the years and with a variety of applications.
  In this survey, we analyze the applications of complex networks to
  real-world problems and data, with emphasis in representation,
  analysis and modeling, after an introduction to the main concepts
  and models. A diversity of phenomena are surveyed, which
  may be classified into no less than 22 areas, providing a clear
  indication of the impact of the field of complex networks.
\end{abstract}


\begin{small}
\tableofcontents
\end{small}

\pagebreak

\section{Introduction}

The many achievements of physics over the last few centuries have been based on
reductionist approaches, whereby the system of interest is reduced to a small, isolated
portion of the world, with full control of the parameters involved (\emph{e.g.}, temperature,
pressure, electric field). An interesting instance of reductionism, which is seldom
realized, is the modeling of non-linear phenomena with \emph{linear} models by
restricting the parameters and variables in terms of a linear approximation. In
establishing the structure of matter with the quantum theory in the first few decades of
the 20th century, for example, reductionism was key to reaching quantitative treatment of
the properties of atoms, molecules and then sophisticated structures such as crystalline
solids. Indeed, deciphering the structure of matter was decisive for many developments --
not only in physics but also in chemistry, materials science and more recently in
biology~\cite{Barabasi04:RNG}. Nevertheless, with reductionist approaches only limited
classes of real-world systems may be treated, for the complexity inherent in
naturally-occurring phenomena cannot be embedded in the theoretical analysis.

There is now a trend in science to extend the scientific method to become more
integrationist and deal explicitly with non-linear approaches. The impressive evolution
of the field of complex networks fits perfectly within such a scientific framework. Its
origins can be traced back to Leonhard Euler's solution of the K\"{o}nigsberg bridges
problem (\emph{e.g.}, \cite{Barabasi03:linked}), after which the theory of graphs has been
useful for theoretical physics, economy, sociology and biology.  However, most of such
studies focused on \emph{static} graphs, \emph{i.e.} graphs whose structure remained fixed.
Important developments on dynamic networks were addressed by Erd\H{o}s and
R\'enyi~\cite{ErdosRenyi60:BISI}, among others, particularly for the so-called random
networks, including the model now known as \emph{Erd\H{o}s and R\'enyi} ---
ER~\cite{ErdoRenyi59,Bollobas01,Bollobas90}. This type of network is characterized by the
feature that in a network with $N$ initially isolated nodes, new connections are
progressively established with uniform probability between any pair of nodes. Such
networks are well described in terms of their average degree, implying they have a
relatively simple structure.  Despite the formalism and comprehensiveness of the
theoretical results obtained by Erd\H{o}s and collaborators, random networks ultimately
proved not to be good models for natural structures and phenomena.  Indeed, heterogeneous
structuring, not the relative uniformity and simplicity of ER networks, is the rule in
Nature.  Therefore, it was mainly thanks to the efforts of sociologists along the last
decades (\emph{e.g.}, \cite{Scott00,Wasserman94}) that graph theory started to be systematically
applied to represent and model natural phenomena, more specifically social relations.
These efforts were mainly related to the concept of the \emph{small-world} phenomena in
networks, which are characterized by small average shortest path lengths between pairs of
nodes and relatively high clustering coefficients. Interestingly, the small-world
property turned out to be ubiquitous. The next decisive development in graph applications
took place quite recently, including Faloutsos \emph{et al.}\ characterization of the
Internet power law organization~\cite{Faloutsos1999} and the identification of such a
kind of connectivity in the WWW~\cite{Albert1999}, giving rise to the \emph{scale-free}
paradigm~\cite{Albert:2000}. Subsequent investigations showed that many natural and
human-made networks also exhibited scale-free organization, including protein-protein
interaction networks~\cite{Jeong01:Nature}, domain interaction
networks~\cite{CostaRodrigues06:APL}, metabolic networks~\cite{Guimera05:Nature}, food
webs~\cite{Krause03:Nature}, networks of collaborators~\cite{Newman01_c}, networks of
airports~\cite{Guimera05:PNAS} and roads~\cite{Gastner06:EPJB}.

The success of complex networks is therefore to a large extent a
consequence of their natural suitability to represent virtually any
discrete system. Moreover, the organization and
evolution of such networks, as well as dynamical processes on
them~\cite{Boccaletti06:PR, Fontouracosta2007atc},
involve non-linear models and effects.  The connectivity of networks is ultimately decisive in constraining
and defining many aspects of systems dynamics. The key importance of
this principle has been highlighted in many comprehensive
surveys~\cite{Barabasi:survey, Dorogovtsev02:AP, Newman03:survey,
Boccaletti06:PR, Costa07:AP}. For instance, the behavior of biological
neuronal networks, one of the greatest remaining scientific
challenges, is largely defined by connectivity (\emph{e.g.},
\cite{Strogatz03:book,CostaBarbosa04,sporns2004oda}).  Because of
its virtually unlimited generality for representing connectivity in
the most diverse real systems in an integrative way, complex networks
are promising for integration and unification of several aspects of
modern science, including the inter-relationships between structure
and dynamics~\cite{dafontouracosta2007cbs, Fontouracosta2008scs}.
Such a potential has been confirmed with a diversity of applications
for complex networks, encompassing areas such as ecology, genetics,
epidemiology, physics, the Internet and WWW, computing, etc.  In fact,
applications of complex networks are redefining the
scientific method through incorporation of dynamic and
multidisciplinary aspects of statistical physics and computer science.

This survey is aimed at a comprehensive review of the
myriad of applications of complex networks, discussing how they have
been applied to real data to obtain useful insights. In order to
ensure a coherent, integrated presentation of the related works, the
survey has been organized according to main areas and subareas. The
survey starts with a section describing the basic concepts related to
complex networks theory, such as measurements.

\section{Basic Concepts}\label{Sec:concepts}

\subsection{Network Representation}

A \emph{graph} or \emph{undirected graph} $G$ is an ordered pair $G =
(\mathcal{N}, \mathcal{E})$, formed by a set $\mathcal{N}\equiv
\{n_1,n_2,\ldots,n_N \}$ of \emph{vertices} (or nodes, or points) and
a set $\mathcal{E}\equiv \{e_1,e_2,\ldots,e_E \}$ of \emph{edges} (or
lines, or links) $e_k = \{n_i,n_j\}$ that connect the vertices
\cite{Bollobas85:book,Diestel00,West:2001}. In the Physics literature,
a graph has also been called a
network~\cite{Costa07:AP}.\footnote{Strictly speaking, in graph theory
  a \emph{network} is a directed graph (digraph) with nonnegative
  capacities associated with each edge and distinguished source and
  sink vertices~\cite{West:2001}.}  When the edges between pairs of
vertices have direction, the graph is said to be a \emph{directed
  graph}. In this case, the graph can be represented by $G^\rightarrow$, which is an ordered pair $G^\rightarrow=(\mathcal{N},
\mathcal{E}^\rightarrow)$, where $\mathcal{N}$ is the set of vertices
and $\mathcal{E}^\rightarrow$ is the set of ordered pairs of arcs (or
arrows). If an edge $e_k = (n_i,n_j)$ is a directed edge extending
from the node $n_i$ to $n_j$, $n_j$ is called the \emph{head} and
$n_i$ is referred to as the \emph{tail} of the edge. Also, $n_j$ is a
\emph{direct successor} of $n_i$, and $n_i$ is a \emph{direct
  predecessor} of $n_j$.  A walk (of length $k$) is a non-empty
alternating sequence $n_0e_0 \ldots e_{k-1}n_k$ of vertices and edges
in $G$ such that $e_i = \{n_i, n_{i+1} \}$ for all $i < k$. If $n_0 =
n_k$, the walk is closed. A \emph{path} between two nodes is a walk
through the network nodes in which each node is visited only once. If
a path leads from $n_i$ to $n_j$, then $n_j$ is said to be a
\emph{successor} of $n_i$, and $n_i$ is a \emph{predecessor} of $n_j
$. A \emph{cycle} is a closed walk, in which no edge is repeated. A
graph $G^\star = (\mathcal{N}^\star, \mathcal{E}^\star)$ is a
\emph{subgraph} of $G = (\mathcal{N}, \mathcal{E})$ if
$\mathcal{N}^\star \subseteq \mathcal{N}$, $\mathcal{E}^\star
\subseteq \mathcal{E}$ and the edges in $\mathcal{E}^\star$ connect
nodes in $\mathcal{N}^\star$. If it is possible to find a path between
any pair of nodes, the network is referred to as \emph{connected};
otherwise it is called \emph{disconnected}.

The intensity of connections can also be represented in the graph by
associating weights to edges. Thus, the weighted graph $G^w =
(\mathcal{N},\mathcal{E},\mathcal{W})$ is formed by incorporating, in
addition to the set of $\mathcal{N}$ vertices and $\mathcal{E}$ edges,
the set of $\mathcal{W}\equiv \{w_1, w_2, \ldots, w_E\}$ weights,
\emph{i.e.} real numbers attached to the edges. The weighted graph $G^w$ can
also be directed.  In this case, instead of edges, the nodes are
linked by arcs. Therefore, the most general graph is the direct,
weighted graph $G^{w\rightarrow}$~\cite{Costa07:AP}.

The special category of \emph{geographical networks} is characterized
by having nodes with well-defined coordinates in an embedding
space. Then, the network $G =
(\mathcal{N},\mathcal{E},\mathcal{D})$ incorporates additional
information, given by the set $\mathcal{D}\equiv
\{\vec{p}_1,\vec{p}_2,\ldots,\vec{p}_N \}$, where $\vec{p}_i$ is an
$n-$dimensional vector which gives the position of the node $i$,
generally in the $\mathbb{R}^n$ space.

Graphs can be represented by using \emph{adjacency lists} or
\emph{adjacency matrices}. In the former case, the graph is stored in
a list of edges (represented through head and tail).  This data
structure is frequently used to reduce the required storage space,
allowing the use of sparse matrices. The list may also have a third
element which represents the intensity of the connection. In the
latter case, the graph is represented by an adjacency matrix $A$ whose
elements $a_{ij}$ are equal to $1$ whenever there is an edge
connecting nodes $i$ and $j$, and equal to $0$ otherwise. When the
graph is undirected, the adjacency matrix is symmetric. In order to
represent weighted networks, a generalization of the adjacency matrix
is required.  In this case, weighted networks are represented by the
so-called \emph{weight matrix} $W$, where the matrix element $w_{ij}$
represents the weight of the edge connecting the nodes $i$ and
$j$. From the weight matrix an adjacency matrix can be derived through
a thresholding operation, $A = \delta_T(W)$, that associates for each
element $w_{ij}$ whose value is larger than a threshold $T$ the value
$a_{ij}=1$ ($0$ is otherwise assigned)~\cite{Costa07:AP}.

\subsection{Network Measurements}

In order to characterize and represent complex networks, many measurements have been
developed~\cite{Costa07:AP}. The most traditional ones are the average node degree, the
average clustering coefficient and the average shortest path length. The \emph{degree}
$k_i$ of a node $i$ is given by its number of connections. For undirected networks, using
the adjacency matrix,
\begin{equation}\label{degree}
k_i = \sum_{j=1}^N a_{ij}.
\end{equation}
The \emph{average node degree} is a global measurement
of the connectivity of the network,
\begin{equation}\label{av_degree} \langle k \rangle =
\frac{1}{N}\sum_{i=1}^N k_i.
\end{equation}

If the network is directed, it is possible to define, for each node
$i$, its \emph{in-degree}, $k^{in}_i = \sum_{j=1}^N a_{ji}$, and
\emph{out-degree}, $k^{out}_i = \sum_{j=1}^N a_{ij}$, as well as the
corresponding averages considering the whole network. The total degree
of a vertex $i$ is given by $k_i = k^{in}_i + k^{out}_i$.

Another measurement related to connectivity is the degree distribution
$P(k)$, which gives the probability that a node chosen uniformly at
random has degree $k$.  This has been found to follow a power law for
many real world networks, as discussed later.  For directed networks,
there are two distributions, for incoming links, $P(k^{in})$, and
outgoing links, $P(k^{out})$. The \emph{clustering coefficient} is
related to the presence of triangles (cycles of order three) in the
network~\cite{Watts98:Nature}. The clustering coefficient of a node
$i$ (with degree $k_i>1$) is given by the ratio between the number of
edges among the neighbors of $i$, denoted by $e_i$, and the maximum
possible number of edges among these neighbors, given by
$k_i(k_i-1)/2$. Thus,
\begin{equation} c_i = \frac{2e_i}{k_i(k_i - 1)} = \frac{\sum_{j=1}^N
\sum_{m=1}^N a_{ij}a_{jm}a_{mi} }{k_i(k_i -1)}.
\end{equation} The corresponding global measurement frequently used to
characterize the graph is the \emph{average clustering coefficient},
which is given as
\begin{equation} \langle c \rangle = \frac{1}{N}\sum_{i=1}^N c_i.
\end{equation}

The \emph{length} of a path connecting the vertices $i$ and $j$ is
given by the number of edges along that path. The \emph{shortest path}
(or \emph{geodesic path}) between vertices $i$ and $j$ is any of the
paths connecting these two nodes whose length is
minimal~\cite{Watts:book}. For the whole network, it is possible to
represent the geodesic distances by a distance matrix $D$, in which
the entry $d_{ij}$ represents the length of the shortest paths between
the nodes $i$ and $j$. The \emph{average shortest path length} is
obtained from such a matrix,
\begin{equation}\label{shortest_path} \ell =
\frac{1}{N(N-1)}\sum_{i=1}^N \sum_{j=1}^N d_{ij},
\end{equation} where the sum considers $i \neq j$ and disregards pairs
that are not in the same connected component.

All the measurements discussed above can be extended to weighted
networks, in which case the \emph{node strength} is defined with the weight matrix $W$~\cite{Barrat04:PNAS},
\begin{equation} s_i = \sum_{i=1}^N w_{ij}.
\end{equation} The average strength is defined considering the nodes
in the whole network, \emph{i.e.}
\begin{equation} \langle s \rangle = \frac{1}{N}\sum_{i=1}^N s_i.
\end{equation} The \emph{weighted clustering coefficient} of a vertex
$i$ can be defined as~\cite{Barrat04:PNAS},
\begin{equation} C^w_i = \frac{1}{s_i(k_i-1)} \sum_{j>k}
\frac{w_{ij}+w_{ik}}{2} a_{ij}a_{ik}a_{jk},
 \label{nodeweightedcluster}
\end{equation} where the normalizing factor $s_i(k_i-1)$ ensures that
$0~\leq~C^w_i \leq 1$. The average weighted clustering coefficient is
given by,
\begin{equation} \langle C^w \rangle = \frac{1}{N}\sum_i
C^w_i.  \label{weightedcluster}
\end{equation}
The average shortest path length for weighted networks is determined
similarly as in Equation~(\ref{shortest_path}), considering the weight
of the edges. In this case, the \emph{weighted shortest path length},
$d_{ij}^w$, is defined as the smallest sum of edges lengths throughout
all the possible paths from $i$ to $j$~\cite{Boccaletti06:PR}.

Another important structural aspect of complex network
characterization is the analysis of how vertices with different
degrees are connected. The degree correlation can be determined from
the Pearson correlation coefficient of the degrees at both ends of the
edges \cite{Newman01:PRL}:
\begin{equation}
  r = \frac{
            \frac{1}{M} \sum_{j>i}k_i k_j a_{ij} -
            \left[ \frac{1}{M} \sum_{j>i}
                   \frac{1}{2} (k_i+k_j) a_{ij} \right]^2
           }
           {
            \frac{1}{M}\sum_{j>i}\frac{1}{2}(k_i^2+k_j^2) a_{ij} -
            \left[ \frac{1}{M}\sum_{j>i}
                   \frac{1}{2}(k_i+k_j) a_{ij}  \right]^2
           },
  \label{pearson}
\end{equation}
where $M$ is the total number of edges and $0 \leq r \leq 1$. If $r>0$ the network is
assortative (vertices with similar degrees tend to be connected); if $r<0$, the network
is disassortative (highly connected vertices tend to connect to those with few
connections); for $r=0$ there is no correlation between vertex degrees, and the network
is called non-assortative.

The measurements above can be used for local analysis, in
terms of node measurements, or global analysis, in terms of average
measurements for the whole network. However, an intermediate analysis
is possible by taking into account the modular structures in networks.
Such structures, called \emph{communities}, are common in many real
networks, formed by sets of nodes densely connected among themselves
while being sparsely connected to the remainder of the
network~\cite{Newman04:EPJ,Girvan02:PNAS}.  Communities play an
important role in network structure, evolution and dynamics, defining
modular topologies. Unfortunately, their identification is an
NP-complete problem~\cite{Danon05:JSM}, so that many heuristic
algorithms have been proposed for their
identification~\cite{Newman04:EPJ,Danon05:JSM,Costa07:AP}.

Depending on the application, subgraphs can be essential to
characterize network structures. For instance, modular structures in
networks can be associated with different functions, such as scientific
collaboration areas. In addition to communities, other types of
subgraphs are found in complex networks, such as
motifs~\cite{Milo02:Science}, cycles~\cite{bagrow2006nsr} and
chains~\cite{villasboas2008cmt}. Motifs are subgraphs that appear more
frequently in real networks than could be statistically
expected~\cite{Milo02:Science, Shenorr2002nmt} (see
Figure~\ref{Fig:typemotifs}, page~\pageref{Fig:typemotifs}). Other
types of motifs include chain motifs (handles and
tails)~\cite{villasboas2008cmt} or border
trees~\cite{villasboas2007btc}.

A fundamental issue related to networks measurement is the sampling
bias.  If the data used to generate a particular network present a
large quantity of noise or incompleteness, it becomes critical to
consider measurements that are not much sensitive to
perturbations. Such measurements must reflect the differences in
distinct networks structures~\cite{Villasboas2008scn}.  This analysis
is crucial for the theory of complex networks and constitutes a
promising research field.

Multivariate statistical and pattern recognition methods are useful
for understanding the network structure. For instance, the ability of
a given model to reproduce real-world networks can be evaluated by
canonical variable analysis with Bayesian decision
theory~\cite{Costa07:AP}. In addition, structures of networks can be
classified with these methods, \emph{e.g.} the simplicity of networks can be
determined by searching for nodes with similar
measurements~\cite{Fontouracosta2007ssc}.

\subsection{Network Models}

In addition to the characterization of networks in terms of
informative sets of measurements, it is important to construct models
capable of reproducing the evolution and function of real systems or
some of their main features. Among the many network models,
three important models are those of Erd\H{o}s and R\'{e}nyi, Watts and
Strogatz, and Barab\'{a}si and Albert. The \emph{random graph} of
Erd\H{o}s and R\'{e}nyi (ER) uses what is possibly the simplest way to
construct a complex graph: starting with a set of $N$ disconnected
vertices, edges are added for each pair of vertices with probability
$p$~\cite{ErdoRenyi59,Bollobas85:book}. Consequently, the degree
distribution follows a Poisson distribution for large $N$, with
average degree $\langle k \rangle = p(N-1)$ and average clustering
coefficient $\langle c \rangle = p$. Random graphs are simple but
unsuitable to model real networks because the latter are characterized
by heterogeneous connections and an abundance of cycles of order
three~\cite{Watts98:Nature}.

The model developed by Watts and Strogatz, referred to as
\emph{small-world networks}, overcomes the lack of cycles of order
three in random graphs, but does not provide non-uniform distribution
of connectivity~\cite{Watts98:Nature}. To construct a small-word
network, one starts with a regular lattice of $N$ vertices in which
each vertex is connected to $\kappa$ nearest neighbors. Next, each
edge is randomly rewired with probability $p$.  When $p = 0$ there is
an ordered lattice with high number of cycles of order three but large
average shortest path length, and when $p\rightarrow 1$ the network
becomes a random network.

In order to explain the uneven distribution of connectivity present in
several real networks, Barab\'{a}si and Albert developed the so-called
scale-free network model, henceforth abbreviated as BA model, which is
based on two rules: \emph{growth} and \emph{preferential
  attachment}~\cite{Barabasi:survey}. The process starts with a set of
$m_0$ vertices and, at each subsequent step, the network grows with
the addition of new vertices.  For each new vertex, $m$ new edges
connecting it to previous vertices are inserted. The vertices
receiving the new edges are chosen according to a linear preferential
attachment rule, \emph{i.e.} the probability of the new vertex $i$ to
connect to an existing vertex $j$ is proportional to the degree of
$j$, \emph{i.e.}
\begin{equation}
\mathcal{P}(i\rightarrow j) = \frac{k_j}{\sum_u k_u}.
\end{equation}
This evolution is related to the ``the rich get richer'' paradigm,
\emph{i.e.} the most connected vertices have greater probability to receive
new vertices. Several networks are believed to be well-modeled by the
Barab\'{a}si and Albert approach, which means there is preferential
attachment in these networks, as we shall discuss in the next
sections.

Network models have been increasingly considered to investigate different types of
dynamics.  Indeed, the relation between the function and structure of networks may help
understand many real-world phenomena, such as the association between biological networks
and the products of such interaction or between society and epidemic spreading. Many
dynamical processes have been studied by complex networks researchers, including
synchronization~\cite{barahona2002ssw, hong2002ssw},
spreading~\cite{VOLCHENKOV_EPIDEMIC_PRE_02, BOGUNA_EPIDEMIC_PRE_02}, random walks
dynamics~\cite{costa2007ecn, Fontouracosta2008snl, dafontouracosta2007ral},
resilience~\cite{Albert:2000, Chi04:IJMPB, Holme2002}, transportation~\cite{Tadic2007}
and avalanches~\cite{Fontouracosta2008aaa}. A good review on dynamical processes in
complex networks appeared in~\cite{Boccaletti06:PR}.


\section{Social Networks}

Since ancient times, the way individuals establish relations among
themselves has been crucial to guide the cultural and economical
evolution of society. Hidden and clear relationships have always
defined different social, diplomatic, commercial and even
cultural networks. In various of those ancient networks, it was
possible to qualitatively identify relevant structural
properties~\cite{Ormerod04:physicaA}, such as the importance of
strategic individuals to intermediate or decide negotiations, or
experience the power of ideological/religious thoughts within groups
of people.

Though the quantitative study of social systems dates back to the
seventeenth century~\cite{Ball02:nature}, the systematic study of
social relations using mathematical methods possibly began in the
first decades of the last century with the study of children
friendship in a school in $1926$~\cite{Wellman26}, later followed by
Mayo~\cite{Scott00} with an investigation into interactions among
workers in a factory. The motivation for understanding ``social
networks'' increased in the following years, especially with analysis
of empirical data. We shall not review those early works, but they are important because the methods developed by social network researchers are now adopted by the so-called complex network community. In addition, some of the problems now treated with complex networks had already been addressed by social researchers. These included the phenomena described by Simon in his
seminal paper on a model to generate highly skewed distribution
functions~\cite{Simon55:biometrika}, the study of citation networks by
Price~\cite{Price65}, which converged in a model of network growth
able to generate power law degree distributions~\cite{Price76},
Freeman's measurement of centrality (betweenness), which quantified
the amount of geodesic paths passing through a
node~\cite{Freeman77,Freeman79}, and the \textit{small-world} effect,
which emerged in the famous Milgram social
experiment~\cite{Milgram67,Travers69}, to name but a few. Further
information and results from the sociological point of view can be
found in specialized books~\cite{Wasserman94, Scott00, Watts03,
Degenne99, Freeman04} and papers in journals such as ``Social
Networks''~\cite{SocialNetwork_journal}.

Although sociometric research has contributed to the understanding of
society, the collected data are still subjected to criticism because
of the difficulty to define and associate intensities to some types of
relation between two persons. Personal relations based on feelings,
thoughts, trust or friendship deeply depend on the cultural
environment, the sex and/or age of the actors involved and even on the
current political and economical context. As an example, if the
Milgram classical experiment~\cite{Milgram67,Travers69} had been done
in Brazil, it would be needed to be redesigned since in this country
most people are called by the first name and relationships are
less formal. In other words, the concept of acquaintance seems to vary
according to the country. In addition, in sexual networks, Liljeros \emph{et
al.}~\cite{Liljeros01} suggested that men may overtell their sexual
partners because of social expectations. In music, the level of
similarity between two bands may be divergent if assigned by
musical experts or enthusiastic fans~\cite{Cano06}.
Complex networks theory has also been considered to model regional interaction patterns in archaeological contexts. This approach allows to treat the interactions between sites in geographical space in terms of a network which minimizes an associated Hamiltonian~\cite{Evans06, Knappett08}.

Trying to overcome this bias, sociologists prepared extensive
questionnaires and cross-compared the responses obtained to achieve
reliable data. However, the interview process is time-consuming and
expensive if one wants to get a significant sample size. On the other
hand, researchers have also investigated systems, such as the
collaboration and citation social networks, where the rules specifying
the relation between the actors are quite clear, which guarantees some
common ground for defining the network.  Both types of approaches have
benefited from the increase of World Wide Web
popularity~\cite{Watts07:nature,Wellman01:science}. The pleasure to be
world visible, to contribute to global knowledge, to share thoughts,
or only to make unusual friends or find partners have contributed to
an increasing number of members in all types of virtual social
environments. Tools like blogs, photoblogs, messengers, emails, social
network services and even a complete social environment such as
``Second Life''~\cite{SecondLife} are now widespread for all ages and
genders~\cite{Wellman01:science,Gonzalez07:nature}. Although these
virtual networks reflect only a piece of the world population and
somehow specific types of relations, they usually provide a
significant statistical sample and possibly unbiased features of the
social relations they represent. In addition, there are extensive
electronic databases about music, theater, sports, scientific papers
and other fields which have contributed to construct reliable social
networks faster and more accurately than
ever~\cite{Watts07:nature,Gonzalez07:nature}. One needs nevertheless
to be cautious when inferring behavioral and social conclusions from
those specific networks, especially when dealing with dynamical
variables.

In the following sections, we shall focus on describing the results
on social networks by using mainly the methods adopted or
developed by the complex network community in the last $10$~years
\cite{Barabasi:survey,Newman03:survey,Boccaletti06:PR,Costa07:AP},
considering real-world data. We present the most
important and common structural properties in several types
of social networks such as degree distributions, community structure
and the evolution of topological measurements, reporting their
relation with social features when available.

\subsection{Personal Relations}

Personal relations are possibly the most important and oldest network
type from the sociological point of view.  Since people can establish
contact with other individuals in several ways, networks of this kind
ultimately provide information about the structure of
society. Personal relations can be divided into several classes
ranging from friendship to professional relations, including sexual
\cite{Liljeros01} and trust networks \cite{Guardiola02,Boguna04}, or
email \cite{Guimera03PRE} and blogs \cite{Bachnik05,Zakharov06}. In a
search for a universal behavior, the concept of personal relations was
extended to other species such as wasps~\cite{Valverde06} and
dolphins~\cite{Lusseau03}. Since this topic includes the majority of
social networks ever studied, we separate the various subjects into
sub-sections for the sake of better organization.

\subsubsection{Movie Actors}

An important class of social acquaintances is related to professional
actors participating in a movie. In terms of the artistic scene, such
a network can provide a glimpse of the popularity of one actor along
his career as well as individual fame. Despite the vast selection of
movies in the Internet Movie Database (\texttt{http://us.imdb.com}),
the film actors network is structurally small-world with high
clustering coefficient. This means a well-defined tendency of actors
to play with common partners~\cite{Watts98:Nature}. Possibly because
of the number of actors in a single movie and the number of movies an
actor takes part along the career, the average degree is significantly
large ($\langle k\rangle \sim 29$). The popularity of some actors
and the short life time of the majority could explain the power-law
regime ($\gamma \sim 2.3$) observed for large $k$ in the distribution of
actors partners~\cite{Barabasi1999}. Amaral and
collaborators~\cite{Amaral2000:pnas} pointed out that this
distribution indeed is truncated by an exponential tail which could be
an effect of aging, \emph{i.e.}, all actors naturally end their careers at
some point. This conclusion is emphasized by the results of Zhang
\emph{et al.}~\cite{Zhang06:physicaA} who argue that when considering
multiple edges between actors, a stretched exponential distribution
fits the data (the same functional form was used to obtain a
good fit on a network of Chinese recipes~\cite{Zhang06:physicaA}).

\subsubsection{Acquaintances}

Popular folklore is fascinated by the so-called ``six degrees of separation'' concept,
resulting from the Milgram's famous social experiment, which suggests that any two
randomly chosen people are separated by six intermediate individuals on average. Since
the experiment was done completely inside the USA, a question remained of whether the six
degrees also applied to the whole world. Taking advantage of the email system, Dodds
\emph{et al.}\ organized a similar worldwide experiment involving any interested
person~\cite{Dodds03} and found, in accordance with Milgram's results, that the diameter
of social acquaintances varied between $4.05$ and $7$, whether only completed chains were
considered or not (in the latter case, they estimated the value). The carefully organized
experiment detected that successful chains depend on the type of relationship between
senders and receivers, and not on the connectivity of the individuals. The most useful
category of social tie was medium-strength friendships that originated in the workplace.
Geography clearly dominated in the early stages of the chain, while occupation tended to
dominate the final stages.

Although interesting, this kind of social experiment depends largely on peoples'
motivation to participate (being limited by lack of interest, time or incentive)
\cite{Dodds03}. To overcome this experimental bias, already-established ties of
acquaintances are usually investigated. The ``blog'' services were used by Bachnik
\emph{et al.}, who found power-law in out- and in-degree distributions (exponents ranging
from nearly $2$ to $3$) and small-world property for providers with three different
numbers of members \cite{Bachnik05}. The ``blogs'' databases are large, but are also
considerably sparse, with most members isolated. On the other hand, Zakharov investigated
the LiveJournal service and found a small-world network with a large giant component
containing nearly $4\cdot 10^6$ users~\cite{Zakharov06}.  Possibly the biggest social
network ever investigated, this structure took 14~days and 2~computers to be crawled. He
found a power-law ($\gamma=3.45$) for $k > 100$ which corresponds to actively
participating users, while not so active users with few connections led to a plateau in
the distribution. The effect of opinion formation in a network like that would be
particularly interesting to investigate since it has a high level of clustering ($\sim
0.3$) and connection reciprocity (nearly $80\%$ of the edges are bi-directional).
However, this network has a limitation in the total number of friends per user ($750$)
and only $150$ of them are listed on the users' info page, which clearly affects the flow
of information. Upon using a new diffusion method, Zakharov identified some communities
considerably fast \cite{Zakharov06}.

Using phone calls records within a period of 18 months, involving approximately 20 per
cent of a country's population, Onnela \emph{et
  al.}~\cite{Onnela07:njp,Onnela07:pnas} constructed a network with
$4.6\cdot 10^6$ nodes and $7\cdot 10^6$ edges. The nodes represent users
and a connection is established when two users called and received
calls from each other. Moreover, the weights represented the duration
of calls between two users. The degree and strength distributions
could be approximated by power-laws (respectively, $\gamma=8.4$ and
$\gamma=1.9$). A careful analysis of degree and weight correlations
was carried out \cite{Onnela07:njp}, in which strong ties appeared
between members of circles of friends, while most connections between
different communities were weak. As a consequence, the removal of
strong ties has basically local effects (within communities). In
contrast, when weak connections are removed, communication between
different communities may be disrupted, causing collapse of the whole
network. It is worth noting that information diffusion was tested in
such weighted network using an equivalent SI (Susceptible
Infected) epidemiologic model. They found that the majority of the
nodes were first infected through ties of intermediate strength with a
peak at 100 seconds~\cite{Onnela07:pnas}, which means that most of the nodes were infected after 100 seconds.

A student affiliation network was built by Holme and collaborators,
from officially registered classmates during a period of 9 years in
Ajou University~\cite{Holme07:physicaA}. Considering the weight of the
network as inversely proportional to the number of students attending
the course (which is related to the social proximity between the
students) and assuming that old courses contribute less than new ones
to this social strength, they found that students become more
peripheral with time in the network, whose core comprises mainly freshman
students. Interestingly, fellow students become strongly tied over
time owing to a decrease in the number of classmates. Using
questionnaires data from middle and high school students in the USA,
Gonzalez \emph{et al.}\ analyzed the resulting friendship
network~\cite{Gonzalez07:physicaA}. They found that when considering
only mutual connections (bi-directional edges), the network results in
a set of various connected components. The friendship networks were
identified as assortative with a degree distribution with a sharp
cutoff. However, the c-networks built with connections between
communities identified with the k-clique percolation method were
disassortative, with scale-free degree distribution. From a comparison
of node pairs with given ethnicities in relation to the random
network, they found that the common behavior for each ethnic group is
to nominate friends of the same ethnicity than from any of the other
ethnicities. Interestingly, an asymmetry with respect to the
composition was observed in case of inter-ethnic nominations. For
instance, both blacks and whites show increasing homophily as they get
into minorities; however, minor black groups get more integrated than
the white ones when involved in other ethnic
majorities~\cite{Gonzalez07:physicaA}.

\subsubsection{Email}

The establishment of networks encouraged communication between
individuals. Guimer\`{a}  \emph{et al.}\ obtained an email network for
the University at Rovira i Virgili~\cite{Guimera03PRE}. This
intrinsically directional network was converted to an undirected
version by assigning connections between two individuals whenever they
sent and received messages from the same partner. As in the
LiveJournal case, the network displayed small-world and high
clustering ($0.25$) in the giant component, but with an exponential
cumulative degree distribution ($P(k)\propto
\exp(-k/k^*)$ with $k^* = 9.2$ for $k>2$).  On the other hand, the
community structure was self-similar with heavily skewed
community-size distribution $\gamma \sim 0.48$ in the range $2$--$100$
and sharp subsequent decay~\cite{Guimera03PRE,Arenas04}. The
self-similarity means that the organization is similar at different
levels, \emph{i.e.} individuals form teams, teams join to form departments,
departments join to form colleges, and so on. Valverde \emph{et
al.}~\cite{Valverde06} constructed a network using data from the email
traffic between members of Open Source Software Communities (OSS) and
found that the weight distribution follows, on average, a power-law
with two regimes ($\gamma \sim 1.5$ and $\gamma \sim 2.4$) in the case
of small communities and with one regime $\gamma \sim 2.27$ for a
large community~\cite{Valverde06}. The latter community also presented
a power-law in the strength distribution ($\gamma \sim 1.73$).

\subsubsection{Trust}

From the hierarchical point of view, the network of trust can be
viewed as a special sub-network of the acquaintances network, where
the strongest connections in the latter could be related to
connections of the former network. The lack of reliable data about
trustful partnerships motivated the studies to concentrate on
electronic trust ties. The PGP (Pretty Good Privacy) encryption
algorithm lets one user certify another one by signing his public
encryption key; in other words, the first user creates a directional
connection to the second one if he trusts that the second is really
what he says he is. By looking the in- and out-degree,
Guardiola \emph{et al.}\ found a power-law degree distribution with
exponents, respectively, $\gamma = 1.8$ and
$1.7$~\cite{Guardiola02}. Interestingly, the network is composed of
many strongly connected components in which the clustering
coefficients are independent of the component size. The network was
found extremely fragile against attack while the strongly connected
structure remained essentially unaffected.  Bogu\~n\'a \emph{et al.}\
considered only bidirectional signatures in the PGP web of
trust~\cite{Boguna04} which resulted in a stronger sense of trust
since both sides must sign each other's keys.  The giant component
presented a two regime power-law degree distribution with exponents
$2.6$ (for $k<40$) and $4$. The clustering is large and nearly
constant as a function of the degree, but the network is
assortative. Finally, using the Girvan and Newman community detection
algorithm, a scale-free community distribution ($\gamma \sim 1.8$) was
identified. The evolutionary Prisoner's Dilemma has been investigated
in a substrate of this network and in an e-mail network as well. It
was found that the connectivity between communities and their internal
structure affect the evolution of cooperation~\cite{Lozano08:plos1}.

\subsubsection{Sexual Relations}

Even though sexual relations between individuals do not necessarily
correspond to acquaintance relations, sexual partners can become
social partners and \textit{vice-versa}. Analyzing a random sample of
individuals aged between $18$ and $74$ years old in Sweden, Liljeros
\textit{et al.}~\cite{Liljeros01} found, in a study within a period of
$12$ months, a cumulative power-law distribution of partners with
exponents $2.54$ and $2.31$, respectively, for women and men. These
exponents are quite close, but the average number of partners is
larger for men than for women, which is explained by the fact that men
may inflate their number of partners because of social
expectations. Interestingly, for the whole life-time another
cumulative power-law distribution was found, but with smaller
exponents, $2.1$ (women) and $1.6$ (men). The scale-free structure of
such network was suggested to emerge from the increased skills in
acquiring new partners as the number of previous partners also grows,
in the same way of ``the rich get richer'' paradigm.  The main result
is that the core-group (\emph{i.e.} groups of risk) concept may be arbitrary
in such networks since there is no well-defined boundary between core
groups and other individuals and therefore, safe-sex campaigns should
be focused on highly connected individuals to prevent propagation of
sexually transmitted diseases.  However, by using statistical methods
in the same and different data, Jones \emph{et al.}~\cite{Jones03}
argued that the method used by Liljeros and collaborators to find the
exponent of the number of partners distribution had statistical
problems. They concluded that the preferential attachment process may
not be the only cause generating real sexual networks as
proposed. They explained that targeting at-risk core groups has a
proven efficiency in reducing disease incidence although degree-based
interventions have already been proposed in the past and could be also
adopted to lower the reproductive rate of sexually transmitted
diseases. Finally, they suggested that other structural properties
such as concurrency and local clustering have significant impact on
epidemic processes since infinite-variance networks have different
internal structures, affecting the spreading processes.

In a review paper~\cite{Liljeros04:PhysicaA}, Liljeros explores the complexity of the
mechanisms contributing to spreading of sexually transmitted diseases, and concludes that
one single solution for the problem is far from being found, even though the underlying
structure of the sexual web contributes to this dynamics. Indeed, broad and targeted
interventions have both been proven to be effective.

Other networks were constructed using smaller
datasets~\cite{Latora06:jmv,Gonzalez06:epj,Blasio07:pnas,Britton07:tpb} which described
populations of different countries and specific regions. The uncertain scale-free
behavior of such a distribution of partners has motivated interesting discussions on the
internal structure~\cite{Handcock04:tpb} and dynamics of sexual interactions
~\cite{Gonzalez06:epj}. For instance, Gonz\'alez and collaborators suggested a model to
capture structural features of a homo- and a heterosexual network. They found that the
clustering coefficient, the number of triangles and squares are relatively small, which
can be better explained with their model than with the BA model~\cite{Gonzalez06:epj}.
Analyzing different periods of time, Blasio \emph{et al.} found evidences of nonrandom,
sublinear preferential attachment for the growth in the number of partners for
heterosexual Norwegians~\cite{Blasio07:pnas}. It is worth noting for epidemic spreading
that sexual behavior is observed to be slightly different for men and women. For
instance, men tend to have more sexual partners on
average~\cite{Liljeros01,Britton07:tpb} and, in a specific region~\cite{Britton07:tpb},
the number of sex contacts decreases with the number of partners for women while being
kept constant for men. At the same time that individuals with many partners can
potentially spread a disease to more people, the chance to transmit to a specific partner
is lower because of less intense sexual activity.

Also worth analyzing is the network of romantic communication, though
it is not necessarily related to sexual contacts. Holme and
collaborators~\cite{Holme04:socialnetworks} observed that during the
period considered in their study, both the number of members and the
average degree grew with time but with decreasing growth rate. This is
partly explained by the fact that old members log on for the first
time during the sampling period. Reciprocity depends on
the type of relationship established in the virtual community and is
rather low. Furthermore, the number of triangles is smaller than the
number of 4-circuits and the network apparently presents
dissasortative mixing. The cumulative degree distributions were
highly skewed, but not with pure power-law
form~\cite{Holme04:socialnetworks}.

\subsubsection{Sports}

The sports field is particularly interesting because of its dynamical
nature, as teams constantly change players, there are dozens of
collective sports involving interaction between people, championships
range from local to world size scale, and so on. However, little
attention has been given to such dynamical systems in terms of
networks. As one of the tests of their community detection algorithm,
Girvan and Newman investigated the community structure of the United
States college football in the 2000 season~\cite{Girvan02:PNAS}. The
nodes represented teams connected by edges expressing regular-season
games between those teams. Although teams are divided into conferences
which imply more games between those members, interconference games
are usual, which generates a complex structure involving all
participants.  With their algorithm, they were able to identify the
conferences with high precision. The North American college football
league was also a motivation for a ranking system based on a directed
complex network where the direction represented wins or losses of a
specific team (the nodes)~\cite{Park05:jsm}. The method was based on
the idea that if a team A can beat a team B and B can beat a team C,
then A most probably will beat C. The resulting ranking for a
season was compared with the official method with good
agreement.

A complex network study of Brazilian soccer players was carried out by
Onody and Castro~\cite{Onody04}.  Initially, a bi-partite network
was built where one type of node represented the teams while the other
represented the players. They found an exponential law $P(N)\sim
10^{-0.38N}$ in the probability that a player has worked in a given
number of clubs $N$. To investigate the topological properties, they
merged those nodes in such a way that two players became connected if
they were in the same team at the same time. The final network
exhibited an exponential degree distribution such that $P(k)\sim
10^{-0.011k}$. The time evolution of that network showed that although
its size became more than five times larger in the period from $1975$
to $2002$, the network maintained the small-world characteristic. They
suggested that the clustering coefficient had a small decrease because
of the exodus of a considerable number of players in the last decades
while the average degree became 20\% higher, possibly because of the
increase in the transfer rate and/or in the professional
life-time. The network became more assortative with time, suggesting a
growing segregationist pattern where teams with similar importance
preferentially transfer players between themselves.

\subsubsection{Comics}

Differently from the actors network where the social ties are
professional, the comics network are composed by characters whose
connections can be constrained by other factors such as moral rules,
as pointed out by Gleiser~\cite{Gleiser07:jsm}. Therefore, though both
networks are related in some level, they have fundamental differences
which can be captured by structural measurements.

The social relation between Marvel characters was investigated in
terms of a bi-partite network (characters and books) where two
characters were connected through the appearance into the same comic
book, with connections representing both ``professional'' and family
ties. Although the Marvel universe tries to reproduce human relations,
Alberich and collaborators concluded that the clustering coefficient
and average degree showed a considerable smaller value than usual real
social networks~\cite{Alberich02}. The resulting network has a small
diameter. Perhaps as a result of personal fame, Captain America and
Spider Man are examples of hubs which contribute to the power-law with
cutoff observed in the distribution of the number of partners in this
collection of stories~\cite{Alberich02}. Contrasts between this
specific artificial network with real ones were suggested by the
results of Gleiser~\cite{Gleiser07:jsm}. The analysis of degree
correlations indicated no correlation up to $k=200$ and
disassortativity after this value. The clustering coefficient as a
function of the degree clearly showed the existence of a hierarchy of
nodes. In order to extract the intensity of relations between the
characters, a weighted network counting the number of times two
characters appeared together was adopted, leading to a power-law strength
distribution with ($\gamma\sim 2.26$). He also found that the hubs are
basically heroes that connect different communities. The fact that
villains appear only around hubs can be a result of the Comics
Association rules, which state, for instance, that the good has always
to succeed against the evil. These constraints imply that this
artificial social network differs from a real social net.

A different network related to comics stories was built by considering
the cross-talk or comic dialogue (Xiangsheng) folk art of
China~\cite{Xu07:physicaA}. The network is small, but exhibits
interesting features. Most of the comic dialogues
involve two players and the number of dialogues an actor has played
follows a power law ($\gamma\sim 2$). The projection onto a
non-bipartite network showed that the collaboration is highly
clustered and could contain a hierarchical, modular
structure. Moreover, the network has some disassortativity and is
small-world. These results suggested that the main actors tend to
repeat dialogues with the same collaborators.

\subsubsection{Non-human relations}

While investigating networks of wasps in which the weights were given
by the number of dominances (hierarchy) of one wasp over the other,
Valverde and collaborators found networks with weight distribution
similar to the small OSS communities~\cite{Valverde06}. Another
non-human social experiment was carried out by
Lusseau~\cite{Lusseau03}, where social acquaintance of dolphins was
defined as preferred companionship, \emph{i.e.} individuals that were seen
together more often than expected by chance. The data took $7$~years
of observations and showed that the emerging network is small-world
with high clustering coefficient. The data set is relatively small, but it
is argued that the tail of the number of acquaintances distribution
follows a power-law ($\gamma \sim3.45$). The dolphin network is
resilient to random attacks, but targeted attacks increase the
diameter of the network, although the random attacks are not enough to
fragment the network into small connected components. This effect is
possibly a consequence of the non-power-law interval for a small
number of acquaintances. The hubs were identified to be mainly adult,
old females~\cite{Lusseau03}.

\subsection{Music}

Professional music relations could be included in the section of
\textit{personal relations}, but we have singled music out because
different types of networks have been studied. In terms of
professional relations, Gleiser and Danon~\cite{Gleiser03} proposed a
network where two musicians were connected if they played in the same
jazz band. This connection mechanism resulted in an assortative,
small-world network with high clustering and degree distribution
following $P(k)\sim (1+0.022k)^{-1.38}$. For the information collected
dating back to the 1930's and 1940's, interesting social aspects were
extracted from the network. Studying the community structure with the
Girvan and Newman algorithm, they found a clear segregation pattern of
black and white musicians in this collaboration network and that the
cumulative distribution of community sizes presents a power-law with
$\gamma=0.48$ \cite{Gleiser03}. The network of collaboration between
rappers was constructed by Smith~\cite{Smith06} considering two
rappers connected whenever they recorded together. The resulting
network is also a small-world with high clustering and a partnership
distribution following a power-law ($\gamma=3.5$).

No correlations were found between topological measurements such as
betweenness and node degree, or between an index of record sales and
starting release year, \emph{i.e.} high level of collaboration is not
related to commercial popularity. Differently from the jazz music
network, the community analysis used a weighted network and then a
``clearing algorithm'' to convert the weighted network to a non-weighted
version such that the most important edges were
extracted~\cite{Smith06}. Considering only the connections, the fast
modularity community structure algorithm~\cite{Clauset04} was applied
and could only identify small and peripheral rap groups. When the
clique percolation method was applied~\cite{Palla05:Nature}, groups and
geographical regions were identified correctly but the same did not
apply to music labels. The level of communities (groups, music labels,
regional/community affiliation) identified could be controlled by a
parameter in the clearing algorithm and then applying the clique
percolation method. Both jazz and hip-hop networks are relatively
small, with considerable high average degree.

Another collaborative network was built by Park \emph{et
al.}~\cite{Park06} who used data extracted from the
\emph{allmusic.com} database. They constructed a network of
similarity between artists using the same database and investigated
the topological properties of both networks and the resulting networks
obtained when intersecting them, \emph{i.e.} the networks obtained from only
common nodes in both networks. The method is interesting for
comparison of the same dataset for different social ties. However, the
overall properties did not change considerably, and the collaboration
network before and after the intersection process presented power-law
$\gamma\sim 3$, while the similarity network had exponential degree
distribution with $P(k)\sim \exp(-0.12k)$ (before) and $P(k)\sim
\exp(-0.15k)$ (after intersection). The small-world and high
clustering features were maintained after the intersection
process. Interestingly, only 464 common edges (about $4\%$ of the
total) were identified in the intersection of both networks which
indicates that having worked together does not necessarily translate
into being classified as musically similar. The similarity network was
assortative while the collaboration was partly assortative. The
cumulative fraction of betweenness displayed a power-law in all
networks studied.

The similarity between artists was extensively investigated by Cano
\emph{et al.}~\cite{Cano06} by using four online databases which
differ in the way similarity is assigned (by user habits or musical
knowledge and by musical experts). Those networks are larger than the
musicians networks above, though preserving the small-world property
with high clustering. They found that user rating networks resulted in
power-law in-degree distributions $\gamma\sim 2.4$, while experts
classification mechanisms resulted in networks with exponential
decay. The out-degree distributions follow the same shape with
cut-offs due to clear limitations in the web-pages usability
constraint (the recommendations should fit in one web-page). The
power-law behavior for the degree distribution was confirmed in an
experiment where users sent playlists such that two artists were
connected if they appeared together. An exponential decay emerged in
the case of a user selecting the most similar artist to a given one in
a list of 10 possibilities \cite{Cano06}.

Lambiotte and Ausloos analyzed a website dedicated to sharing musical
habits~\cite{Lambiotte05:pre,Lambiotte06:epjb} using a percolation-based method to
identify social groups and music genres according to personal habits. Non-trivial
connections between the listeners and music groups were identified, with some empirical
subdivisions obeying standard genre classification while others did not. By analysing the
original bi-partite network, a power-law ($\alpha\sim1.8$) was observed for the
distribution in the number of listeners of a specific music group while an exponential
($\exp^{-x/150}$) was fitted in the case of the number of music groups per
user~\cite{Lambiotte05:pre}. They further improved the methodology mapping music groups
into genres using online listeners records. Analyzing different tags given by listeners
to classify music groups, they observed that similar groups tend to be listened by the
same people~\cite{Lambiotte06:epjb}.

The Brazilian popular music network was constructed by connecting two
song writers with a common singer \cite{Silva04}. The network presents
small-world feature, high clustering and high average
connectivity. The cumulative degree distribution is fitted by an
exponentially truncated power-law with $\gamma=2.57$.

At the level of personal ties, a network of jazz bands was built with two bands being
connected if they had a musician in common, from which another segregation was found
between black and white musicians~\cite{Gleiser03}. That network presented small-world,
high clustering, with the degree distribution following an exponential function $P(k)\sim
\exp(-(k/32.8)^{1.78})$. Using the Girvan and Newman community detection algorithm, they
found a polarization in two big communities representing the major records locations,
namely, Chicago and New York. The New York community split up into two other communities
corresponding to another segregation between black and white musicians~\cite{Gleiser03}.
In these networks, an example of errors in the acquired data is the inclusion of the same
musician appearing with different names, as pointed out in the jazz
network~\cite{Gleiser03}. Similarly, errors occur in the Brazilian musician network,
which includes anonymous musicians concentrating a large number of
connections~\cite{Silva04}.

\subsection{Collaborations}

The topic of scientific collaboration can be understood as a sub-topic of personal
relations, being linked to professional (academic) ties between scientists. It deserves
special section for its relevance in terms of knowledge dissemination.  From the several
databases investigated, common features were identified, even though the sizes and some
structural properties depend on the field of interest. Newman built networks from
databases of sizes varying from $10^4$ to $10^7$ papers and compared the topological
properties in a given time interval. The distribution of number of papers is well fitted
by a power-law for two databases ($\gamma\sim2.9$ for Medline and $\gamma\sim3.4$ for
NCSTRL) and by an exponentially truncated power-law for the Los Alamos
Archive~\cite{Newman01_a,Newman01_c}. The evolutionary computation (EC) co-authorship
network had a power-law ($\gamma\sim2$)~\cite{Cotta05} as well. The distributions of
number of authors displayed a power-law shape in all four databases investigated by
Newman ($2 \leq \gamma \leq6$) and in the EC data ($\gamma\sim 5.27$), with the
largest collaboration project containing $1681$ people. In case of arXiv (cond-mat) data
restricted to network papers, this distribution is fitted by an exponential
($\exp^{-n/1.5}$)~\cite{Lambiotte05:pre2}. Purely theoretical papers appear to be the
work of two scientists on average, according to results for mathematics and sociology.
The latter were also found to contain a small number of collaborations although
suggesting an increase in the collaboration rate in the last
decades~\cite{Barabasi02,Moody04}. On the other hand, experimental and interdisciplinary
subjects have higher average collaboration rate, reaching about $9$
authors per paper in high-energy physics. An intermediate collaboration rate ($\sim 4$)
was identified in topic-specific fields (``networks'' and
``granular-media'')~\cite{Ausloos07:epjb}.

The histogram of the number of collaborators per author fits a
power-law in many databases with exponents varying from
$\gamma\sim2.1$ to $\gamma\sim2.58$
\cite{Cotta05,Moody04,Barabasi02}. The only difference appears in some
networks investigated by Newman, which exhibited two power law slopes
($\gamma\sim2$ and $\gamma\sim3$). It is also not clear to fit a
topic-specific database~\cite{Ausloos07:epjb}. Interestingly, Cotta
and collaborators pointed out that the most prolific authors are not
necessarily the most connected, but in fact they have diverse
interests which motivate collaborations. They also found that the
number of collaborators strongly correlates with the number of
papers~\cite{Cotta05}.

Only the network of sociologists has been found not to be small-world,
possibly because of the large number of research areas with little
collaboration between them. The conclusion is reinforced by the small
clustering observed~\cite{Moody04}. The Medline database presents
small clustering, as a result of the tree-like structure (``principal
investigator'' with several post-docs) akin to biomedical
research~\cite{Newman01_a,Newman01_c}. The other databases are
characterized by a high degree of clustering~\cite{Cotta05}. The
clustering coefficient decays and the degree of an author increases
with time in the case of neuroscience and mathematics \cite{Barabasi02}.
The growth of the giant component of these two latter subjects is
consistent with the increasing collaboration rate over the last
years. Newman verified that for most authors, the paths between them
and other scientists go through just one or two of their collaborators
(funneling effect). The average distance to all authors decreased with
increasing number of collaborations
~\cite{Newman01_b,Newman01_c,Cotta05}. In terms of knowledge
diffusion, the average distance is important because it measures the
centrality of an author in terms of its access to information. On the
other hand, betweenness is a measure of author's control over
information flowing between other
scientists~\cite{Newman01_b,Newman01_c}.

Girvan and Newman tested their algorithm to detect communities in a database of
co-authorship of papers, books and technical reports of the Santa Fe Institute between
1999 and 2000~\cite{Girvan02:PNAS}. They found that scientists are grouped into two types
of similarity, either of research topic or methodology. The same method was used by
Arenas and collaborators to identify communities in two networks of
co-authors~\cite{Arenas04}. Initially, they considered the panels contributions of
Statistical Physics conferences in Spain over a period of $16$~years and took the final
co-authorship network. They found that collaboration is more common within the same
institution and that the emergent community size distribution follows a power-law
($\gamma\sim1$). The arXiv network has a two-slope power law in the same distribution
with exponents $\gamma\sim1$ (with strength $s < 60$) and $\gamma\sim0.5$
($60<s<1000$)~\cite{Arenas04}. The community in collaboration network was also addressed
by considering hierarchical measurements in a scientific institutional collaboration
network~\cite{silva2007cca}. In this case, the authors identified different patterns of
authorship emerging from different research areas. Using arXiv data (cond-mat only)
restricted to papers whose abstract contain the word ``network'', Lambiotte and
Ausloos~\cite{Lambiotte05:pre2} found that the distribution of connected components
roughly follows a power-law ($\alpha\sim 2$). Analyzing the collaboration network of
scientists restricted to two fields, they observed a transition in the largest connected
component from one of the fields to the other along a period of time, \emph{i.e.} most of the
collaboration was dominated by one of the fields in each phase~\cite{Ausloos07:epjb}.
Moreover, other smaller drastic events were observed during the period of analysis.

Li \emph{et al.}\ considered a different weight function while
investigating the collaboration network between econophysics
authors from 1992 to 2003/2004~\cite{Li05}. The nodes represented
scientists and the weights between two scientists were obtained by
summing three functions, where each function represents one type of
relation between them. The first counts the number of co-authorships,
the second counts the number of citations and the last, the
acknowledgements. Each function expresses a different contribution to
the final weight value according to their relevance in the final
report. They found that the degree and weight Zipf plots have an
undefined shape and did not change when considering two snapshots
(with one year of difference between them) of the collaboration
database, though individual nodes changed considerably their
importance in the network (measured by their
betweenness)~\cite{Fan04,Li05}.  The network has a small giant
component (with about $25\%$ of the total nodes), small clustering and
large average shortest paths when compared to other collaboration
networks. They found that the weight affects slightly the edge and
node betweenness distributions. By using only the amount of
collaboration between co-authors, they investigated the community
structure in the giant component of such a network. They used the
Girvan and Newman algorithm and hierarchical clustering. The first
algorithm provided the best results, classifying scientists belonging
to the same university, institute or interested in similar research
topics.  The second one classified modules, but the result was not
consistent with real data. They found that the members of each
community changed considerably depending on whether the weights are
taken into account~\cite{Zhang05}.

Barab\'{a}si and collaborators suggested that for the evolution of the
collaboration network, the small and finite time interval might affect
the results leading to incorrect conclusions because of incomplete
data.  Such a trend was identified by observing that the average
shortest path decreased with time (and size of the network) while it
was expected to increase~\cite{Barabasi02}. Although important,
proceedings editorship cannot be seen in the same level of
peer-reviewed papers. Cotta and collaborators argued that proceedings
collaborations should be removed since they bias considerably the
shortest path distribution, as they include authors from different
thematic subjects and create long-distance edges. They also suggested
to analyze separately the distinct types of scientific reports since
different types of collaboration are reflected in such
reports~\cite{Cotta05}. Some care should be taken when considering the
original data because two authors might appear with the same name or
one author might identify himself in different ways on different
papers. To overcome this problem, Newman ran the measurements twice,
first considering the network obtained by the author's surname and
first initial only, and a second version considering the surname and
all initials~\cite{Newman01_a}.

\subsection{Religion}

The importance of religion in the ancient and contemporary society is unquestionable;
despite some fluctuations along the years, religion always played a central role in the
history of humanity.  Choi and Kim used a Greek and Roman mythology dictionary to analyze
the relation between the mythological and fascinating characters of that
time~\cite{Choi07:physicaA}. Using a directed network, they associated outgoing links
with a specific entry of a character in the dictionary and the incoming links were
related to characters appearing in the corresponding explanation field of that specific
character. It means that Heracles, for instance, has the most number of outgoing links
which reflects that he gave origin to many great events by himself. On the other hand,
Zeus concentrates a high number of incoming links since he appeared as a supporting
character in different myth tales. The degree distributions resulted in power-laws with
exponents between $2.5$ and $3.0$. The close relation between different characters is
visible by looking the small-world feature and the high clustering of common neighbors.
Furthermore, a hierarchical structure was observed, which had two main origins:
genealogical tree and the native class to which a myth character belongs, \emph{e.g.}, gods,
titans, heroes. The distribution of local cyclic coefficients displayed two peaks that
correspond to tree-like and triangular patterns~\cite{Choi07:physicaA}.

In the medieval church, there was a big concern about heresy, which is
fundamentally an opinion at variance with established religious
beliefs professed by a baptized church member. Since heresies were
against the official beliefs of the church, their propagation was
punished and avoided. The inquisitors recognized that they should
target the most connected people, \emph{i.e.} the people responsible of
propagating those ideas, which resulted on much more effective
results~\cite{Ormerod04:physicaA}. A qualitative analogy of historical
facts and epidemic spreading on networks led Ormerod and Roach to
suggest that the spreading of medieval heresy resembles a disease
diffusion in a scale-free network~\cite{Ormerod04:physicaA}.

At an individual level, Amaral \emph{et
al.} analyzed a group of 43 Mormons and suggested that a Gaussian
distribution could explain their acquaintance
network~\cite{Amaral2000:pnas}. This result agrees with another
friendship network of high school students, also mentioned in the same
paper~\cite{Amaral2000:pnas}.

\subsection{Organizational Management}

The growth of interdisciplinary research can be identified by data related to national or
continental research funding proposals. Based on data from the National Natural Science
Foundation of China (NSFC), Liu and collaborators constructed a network of related
research areas~\cite{Liu06}. Since each proposal to the foundation provides two fields to
specify different areas, they took each research area as a node, with two nodes being
connected if they appeared in the same proposal. The cumulative degree distribution
followed the function $P(k)= \exp(-0.13k)$ for the period between $1999$ and $2004$. The
growth of interdisciplinarity could be identified by the increase in the average degree
and clustering, and by the decrease in the average distance. The network tends to be
disassortative with time.

At the level of organizations, Barber \emph{et al.}~\cite{Barber06}
and Lozano \emph{et al.}~\cite{Lozano06} built networks based on
the European-Union Framework Programs (FP) for Research and
Technological Development. They got a bi-partite
network from the first four FPs, with one node-type representing the
research projects and the other type representing the organizations.
The connections were assigned by considering organizations which
collaborate in common projects. Network properties were extracted from
the intersection graph. The distributions of project (number of
organizations taking part in one project) and organization (number of
projects in which an organization takes part) sizes exhibited
power-laws, respectively, with $\gamma\sim4$ and $\gamma\sim2$ for
larger values, which remain stable over the three last periods
considered. This indicates that the organizations participating in a
particular number of programs in each FP were not altered in spite of
changes in the research activities. The network is highly clustered,
small-world, and the number of triangles increased linearly with the
degree of the node~\cite{Barber06}. Using data from the 6th FP, Lozano
\emph{et al.}\ obtained a different network where two organizations
were connected whenever they had collaborated at least in two
projects. The degree distribution had an undefined shape and the
analysis focused on detecting communities by a new algorithm, which
was able to classify communities according to their type
(industries or research centers), nationality or services
providers~\cite{Lozano06}.

The network of the United States House of
Representatives from the 101st to 108th Congresses was investigated by
Porter and collaborators~\cite{Porter06}. Initially, they built a
network with two types of nodes which corresponded to Representatives
and committees, with edges connecting each Representative to the
committees. Then, another network was created whose nodes represent
the committees while the edges indicate common membership between
committees. They found evidence for several levels of hierarchy within
the network of committees and identified some close connections
between committees without incorporating any knowledge of political
events or positions. They identified correlations between committee
assignments and political positions of the Representatives. Finally,
they verified that the network structure across different Congresses
has changed, especially after the shift in the majority party from
Democrats to Republicans in the 104th House.

The networks of product development for different industries were
characterized by Braha and
Bar-Yam~\cite{Braha04:PRE,Braha04:book,Braha07:management}. In these
networks, two tasks are connected by arcs if the first task feeds
information to the second task. The resulting network was highly
clustered with small average distance, which reflects the
optimization of the networks and unavoidable iterative nature of the
design process. They found that in-degree distributions always
presented a scale-free behavior with cutoff, while out-degree
distributions were scale-free with and without cutoff, which can be
related to the fact that transmission of information is often less
constrained than
reception~\cite{Braha04:PRE,Braha04:book,Braha07:management}.


\section{Communication}

The study of the structure and function of social networks has
always been constrained by practical difficulties of mapping the
interactions of a large number of individuals. The construction of
these networks is based on questionnaire data, which reaches only a
few number of individuals and depends on the personal opinion about
their ties. With the advent of the Internet and with the use of
phone and mobiles, a large amount of data can be recorded for
further analysis. These communication networks are useful not only
for providing better data for the study of social networks but also
for economical reasons. Here we discuss the email, call graph, and
wireless networks.

\subsection{Communication by email}

Email has become one of the most important means of communication, being largely used in
business, social, and technical relationships. Email exchanges therefore provide
plentiful of data on personal communication in an electronic form, amenable to build
social networks automatically. Email networks also provide one of the major means of
computer virus spreading.  Another important related aspect concerns email topic
classification~\cite{keila2005see}.

There are two ways to construct email networks: (i) vertices are email
addresses and there is a directed arc from vertex $i$ to vertex $j$ if
$i$ sends at least one email to $j$ (this network is obtained by the
log files of email servers)~\cite{ebel2002sft, tyler2003esa,
wang2004pem}; (ii) vertices are also email addresses but there is a
directed arc from vertex $i$ to vertex $j$ if $j$ is in the address
book of vertex $i$ (this network is obtained from the email book of several
users of a specific institution)~\cite{newman2002ena,
wang2004pem}.  Ebel \emph{et al.}~\cite{ebel2002sft}, however, studied
these networks by considering them as with undirected arcs where email
addresses are connected if at least an email was exchanged between
them.

All collected email networks studied were obtained from university email
servers~\cite{ebel2002sft, newman2002ena, Guimera03PRE, wang2004pem}
except the email network from the HP Labs mail
server~\cite{tyler2003esa}. These networks can be very large, with the
largest being the email network of the Kiel
University~\cite{ebel2002sft} with 59,812 vertices, and the smallest,
the email network of University at Rovira i Virgili in Tarragona,
Spain~\cite{Guimera03PRE}, which contains only 1,667 vertices. All of these
email networks have the small-world and the scale-free
properties~\cite{ebel2002sft, tyler2003esa, wang2004pem,
newman2002ena, Guimera03PRE}. Tyler \emph{et al.}~\cite{tyler2003esa}
and Gimer\`a \emph{et al.}~\cite{Guimera03PRE} also found that email
networks are composed of communities and proposed methodologies to
find them.  Braha and Bar-Yam \cite{braha2006ctf} showed that vertex
degree and betweenness of such kind of networks change dramatically
from day to day, suggesting a reinterpretation of ``hubs'' in
dynamical networks.

A model of an evolving email network was proposed by Wang and
Wilde~\cite{wang2004pem} and is based on addition and deletion of
links between users, and on the user email checking time. Zou et
al.~\cite{zou2004ewm} proposed a model for simulating the email
spreading of virus and immunization, which is also based on the user
email checking time and his/her probability of opening email
attachments. They showed that viruses spread more quickly on
scale-free networks than on small-world and random networks, and
that the immunization defense is more effective on the first kind of
network than on the other two.

The free software package EmailNet for email traffic mining
generates the email network for further analysis, and provides an interface for
visualization of the networks (available at
\texttt{http://ipresearch.net/emailnet}~\cite{vanalstyne2003esa}).

\subsection{Telephone}

Telephone call graphs are obtained from telephone calls completed
during a specified time period.  The vertices are the telephone
numbers, and a connection from vertex $i$ to vertex $j$ is present if
during the specified period there was a call from $i$ to $j$; the
established arc has naturally a direction, with $i$ at the tail and
$j$ at the head. Such graphs can be very large; for example, a one-day
call graph used by Abello \emph{et al.}\ has more than 50~million
vertices and 170~million edges~\cite{Hayes00}.  Aiello, Chung and
Lu~\cite{Aiello00} found power laws for the in-degree and out-degree
distributions, as well as for the distribution of components with a
given size (excluding the giant component) in a telephone call
graph. A similar study was done for mobile phones by Nanavati \emph{et
  al.}~\cite{Nanavati06}.  They analyzed local calls in four different
regions for two different periods (one week and one month); despite
social differences in the regions and the different periods used, the
results are similar. The average degree varies from $3.6$ to $8.1$, with
a clustering coefficient from $0.1$ to $0.17$.  The distribution of
in-degrees and out-degrees is close to a power-law, with exponents $2.9$
and $1.7$, respectively.  There is a significant correlation between
the in-degree and the out-degree of vertices, which means that people that
receive many calls also generate many calls.  With respect to nodes
connected by a call, assortativity is present for the in-degree of the
caller with in- and out-degree of the receiver; for the out-degree of the
caller, weak disassortativity with in- and out-degree of the called was
detected.  That is to say, vertices that receive many calls
tend to call vertices that also receive and generate many calls, and
vertices that receive few calls tend to be connected with vertices that
receive and generate few calls; on the other hand, the number of calls
generated by the caller is not a good predictor for the connectivity
of the receiver.  A giant strongly connected component and a power law
distribution of the remaining components were also detected.

Onnela \emph{et al.}~\cite{Onnela07:pnas,Onnela07:njp} considered the
call graph of a mobile operator, but included only reciprocal calls in
the graph, \emph{i.e.} an undirected edge is present between nodes $i$ and
$j$ only if $i$ called $j$ and $j$ called $i$ during the
considered time interval; calls for numbers of other operators were
not included. The largest connected component of this network has a
degree distribution with a tail of the form $P(k) =
a(k+k_0)^{-\gamma},$ but with significantly larger exponent
($\gamma=8.4$) than the networks discussed above.  The number of calls
made between two telephone number and the total duration of the calls
were used as two different weighting criteria. The distribution of
strengths shows that most users make few, brief calls, but some make a
large number of calls, with some pairs of users chatting for
hours. The network is assortative, confirming previous results for a
social network.  The authors also found that the overlap of the
neighbors of two vertices is proportional to the weight of their
connection, therefore corroborating experimentally the ``weak ties
hypothesis'' of Granovetter~\cite{Granovetter1973}.

Lambiotte \emph{et al.}~\cite{lambiotte2008gmg} analysed the call
graph of a Belgian mobile phone company with 2.5 million customers and
810 million calls or text messages over a period of 6 months. In order
to eliminate accidental calls, a link between customers $i$ and $j$ is
considered only if there were at least six reciprocal calls between
them during the time considered. The authors showed that this network
has power-law degree distribution with $\gamma=5$ and that the
probability of two customers being connected is proportional to
$1/d^{-2}$, where $d$ is the geographic distance between them. Blondel
\emph{et al.}~\cite{blondel2008fuc} concluded that the same network
also exhibits hierarchical communities.

In order to understand the telephone traffic, a model which represents
the customer network behavior in real world has to be taken into
account. Unlike classic traffic analysis where a fully connected
customer network is considered, Xia \emph{et al.}~\cite{xia2005sfu}
employed a model for the customer network based on the scale-free
property and showed that the structure of the customer network is
more likely to cause call blockings than the limited capacity of
the telephone network.

The data available of mobile phone companies can also be used to
analyze the pattern behavior of the customers. Gonz\'alez \emph{et
  al.}~\cite{gonz2008uih} used the trajectory of 100,000 mobile phone
users over a period of six months to understand their mobility pattern
behavior and found that, despite the prediction of random
trajectories in~\cite{brockmann2006mn}, the human trajectories have
some regularities, like being characterized by small traveling length
and a high probability of returning to few locations. These patterns
are important for disease prevention, urban planning, and social
modeling.

\subsection{Wireless}

In wireless networks, mobile stations (mobile nodes) connect through
access points to the Internet.  Each mobile node can connect at
different access points at different times.  Hsu and
Helmy~\cite{Hsu05,Hsu06} studied the patterns of connectivity of
mobile nodes to access points, and constructed a network based on
encounter between nodes.  Two nodes are said to encounter each other
if there is an overlap of their connections to the same access
point. Each mobile node is a node in the network, and a
link is created between two nodes if there is an encounter between
them in a specified interval of time.  The authors found that the
network was sparse, with nodes having encounter with just
a few neighbors in average (with a node connected to about 1.88\% to
5.94\% of all the nodes, depending on the data set considered),
following a bi-Pareto distribution.  The graph has short average
distances and high clustering coefficient.  The high clustering is
attributed to the fact that nodes have a ``home'' access point, and
all nodes with the same access point have a high probability of
being connected, giving rise to cliques.  The average distance is
small due to the presence of some nodes in these cliques that connect
to nodes in other access points, generating a ``small-world''
effect.

In conclusion, communication networks, besides being important means to represent social
networks, also deserve special attention for economical reasons, since such kind of
networks include everybody. Communication networks are generally huge, with millions of
vertices, have the small-world property, degree distribution with power law tail, and
community structure. The main results obtained for these networks are: (i) a definition
of a hub has to reinterpreted in email networks, because the vertex degree and
betweenness change dramatically from day to day; (ii) in the case of call graphs, the
overlap of two neighbors is proportional of the weight of their connection; (iii) the
probability of two vertices being connected in call graphs is inversely proportional to
the square of their geographic distance; (iv) the mobility behavior of customers in
mobile call graphs is not random; instead, it has some regularities. Further works could
extend the analysis of such data and develop models to describe the features observed.


\section{Economy}

Trade, currency, industrial production, wealth distribution and
tourism are important target studies of economy. These systems can be
considered as formed by discrete parts that interact in a defined
way~\cite{Goyal07}. For instance, trade/commerce is the voluntary exchange of goods,
services, or both.  In this way, a network can be generated by this
activity associating nodes to countries and considering the in- and
out-degree as representing, respectively, the imports and exports
between these countries.  In this section, we show how complex
networks theory can be used to model economical relationship.

\subsection{Trade networks}

Economic relations are currently enlarged owing to the even
increasing number of commercial partners due to globalization. Trades
at a personal level have substantially increased with e-commerce, but
the gross market between countries is still a business of
companies. Trying to unveil this fascinating economical complex system
of trades between countries, Serrano and Bogu\~n\'a studied the
so-called world trade web~\cite{SERRANO_TRADE_PRE_2001}. A network was
built by assigning a node to each country and considering the in- and
out-degree as representing, respectively, the imports and exports
between countries.  For the non-weighted directed network, they found
the in- and out-degree correlation to be very high ($r=0.91$), with
$0.61$ of reciprocity. A power-law function emerged with exponent
$\gamma \sim 2.6$ for $k>20$ in the number of commercial partners,
regardless of whether the in-, out- or undirected degree distribution
were considered. The intense trade activity between countries exhibits
the small-world property, a high clustering coefficient, and large
average degrees. A positive correlation between the number of trade
channels and the country's economical wealth was identified, with some
exceptions for developing countries. The system seems
to present a hierarchical architecture of highly interconnected
countries belonging to influential areas, which in turn connect to
other influential areas through hubs. This last property is suggested
by the strong clustering coefficient dependence with the degree
(power-law behavior with $\gamma\sim0.7$) and the disassortative
degree structure~\cite{SERRANO_TRADE_PRE_2001}.  In another paper,
Serrano and Bogu\~n\'a \cite{SERRANO_TRADE_PRE_2001} showed that
the topology of the "World Trade Web" exhibited the scale-free,
small-world effect and modular structure, similar to the Internet and
WWW networks.

Ausloos and Lambiotte \cite{AUSLOOS_GDP_PHYSA_2006} studied the correlation between the
\emph{G7 country Gross Domestic Product} (GDP) evolution of the 23 most developed
countries. The authors built a network from a correlation matrix and found subgraphs with
high cluster coefficient as indicative of globalization effects. Mi\'{s}kiewicz and
Ausloos \cite{MISKIEWICZ_GDP_SPRINGER_2006} studied the globalization process and Gligor
and Ausloos \cite{GLIGOR_EPJB_GDP_2008} studied the structure of macroeconomics of the
European Union (EU) countries using network concepts, with the correlation matrix of the
GDP/capita annual rates of growth between 1990 and 2005 being used to establish the
weight of the connections among the 25 countries (nodes) in EU. They introduced the
\textit{overlapping index}, $O_{ij}$, to find a hierarchy of countries in the network
analyzed. In non-weighted networks, this quantity is proportional to the number of
neighbors shared by $i$ and $j$. The extension of the overlapping index for a weighted
network was used to identify clusters and the hierarchy of EU countries.

\subsection{Currency}

In 2004 Li \emph{et al}.\ \cite{LI_WEB_PHYSA_2004} extended the studies of
Serrano and Bogu\~n\'a to ``World Exchange Arrangements Web,''
with a method to build a bipartite network where the nodes can
represent countries or currencies. In the network, one country is
linked to one currency if the currency circulates in the country.  The
authors showed that the network possesses a scale-free behavior with
exponent $\gamma=1$. A non-bipartite network was also analyzed where
the nodes were the currencies only. Two currencies were linked if
there was one or more countries where these currencies circulate. This
network also presented scale-free behavior with coefficient $1.3$,
small-world effect, disassortative and modular structure. The power
law behavior of the degree distribution had also been observed by
Gorski \emph{et al}.\ \cite{GORSKI_CURRENCY_APPB_2006} in the complex network
built from FOREX database (the largest financial market in the
world). The robustness of currency complex networks was investigated
recently by Naylor \emph{et al.}, who used hierarchical methods to show that
the currency network topology minimizes the effects caused by economic
crises, such as the Asian crisis from 1 August 1997 to 31 October
1998.

\subsection{Industry}

Andrade \emph{et al.}\ \cite{ANDRADE_PLANTS_PHYSA_2006} showed that the
topology of the oil refineries network is scale-free. In order to
build the network, the authors took devices and unitary processes
(\emph{e.g.}, valves, pumps, tanks) as nodes, which are linked by pipes.
They observed a small-world effect and hierarchical organization on
two networks, and argued that the network topology may provide
a useful tool to design, characterize and evaluate refinery plants.

\subsection{Wealth}

The effects of network topology on the evolution of a dynamic process
have been widely studied.  For example, Souma \emph{et al.}\
\cite{SOUMA_WEALTH_CONDMAT_2001} built a model to investigate the
wealth distribution with networks displaying the small-world effect. The authors
considered a multiplicative stochastic process and studied the effect
of deletion and rewiring of edges in the wealth distribution. Their
results indicated two phases for the wealth distribution, depending on
two parameters, viz. the probability of rewiring edges and fraction of
rewiring edges. The distribution was log-normal for the first phase
and power law for the second phase. It had an intermediate state
characterized by log-normal distribution with a power law tail, which
is also observed in real world economy. The break of the wealth
clusters also occurred if the edges were rewired. A similar study was
performed by Matteo \emph{et al.}\ \cite{MATTEO_WEALTH_CONDMAT_2003} using an
additive stochastic process. The authors showed that the shape of the
wealth distribution was defined by the degree distribution of the
network used. The results were consistent with real data of income
distribution in Australia.

\subsection{Tourism}

Tourism comprises a multitude of activities that form one of the world's fastest growing
economic industries. The sector is so strongly characterized by intense linkages through
which informational exchanges occur, that it seems natural to apply network theoretical
methods to its study \cite{baggio2006csi,Baggio2008}. Nonetheless only quite recently
quantitative techniques have been applied, mainly in analyzing tourism destinations,
complex localized clusters of public and private companies and
organizations~\cite{scott07}. The major characteristics of tourism destination networks
have been measured both from a static and a dynamic point of view. The results showed
that topological measurements can highlight their main structural features such as
product clusters, structural divides and central
organizations~\cite{baggio2007wgt,Scott2008dnf}. For instance, Scott \emph{et
al.}~\cite{Scott2008dnf} analyzed the tourism in Australia and showed that complex
network theory can help define the weaknesses in destination structures, which can be
improved by policy and management approaches. The underlying social and economic system
evolutionary history has been also assessed by using network dynamical growth
models~\cite{Baggio:ATMC}. In general the topology of the networks is comparable to the
one exhibited by similar systems: a marked scale-free structure.  However, some
differences are found, mainly due to the relatively poor connectivity and cauterization.
These results are interpreted by considering the formation mechanisms and the
peculiarities of the economic sector and of the real cases examined. Clustering and
assortativity coefficients were also proposed as quantitative estimations of the degree
of collaboration and cooperation among the destination stakeholders~\cite{baggio2007wgt}.
Moreover, networks were used to perform numerical simulations to investigate possible
scenarios of information and knowledge diffusion among the components of a tourism
destination. It has been shown that the highest improvement in efficiency for this
process is obtained by optimizing the network to increase the degree of local
clustering~\cite{Scott08:book}.

Tourism networks have also been addressed by analyzing the relation
between their structure and dynamics. Costa and
Baggio~\cite{Costa:Baggio:08} studied the Elba (Italy) tourism
destination network using complex networks topological measurements and
considered dynamical process over the network, \emph{i.e.} the
inward/outward activations and accessibilities, according to the
superedges framework~\cite{Fontouracosta2008scs}.  They showed that
the type and size of companies has strong influence over their
activations and accessibilities. On the other hand,
the geographical position of companies tends not to affect
dynamical features. With the characterization of the tourism networks
structure, the authors concluded that Elba tourism network is
fragmented and heterogeneous.


\section{Financial market}
\label{sec:fin_mark}

A corporation, by definition, has a group of owners, called shareholders, that share its
stocks. A stock representing part of the assets and profits of a company can be bought
and sold in a stock exchange, and any person can in principle own part of a publicly
traded company. Stock prices are constantly changing, following unstable market and
political conditions all around the globe. Thus, the financial market is a highly
complex, evolving system, difficult to grasp and predict, being sensitive to economical
instabilities such as that of Black Monday (October 19, 1987), when the Dow Jones
Industrial Average (DJIA) decreased sharply (this index measures the performance of
companies in the USA stock market). The organization of the financial market may be well
understood by representing it as a network. Given the high accuracy of financial data
available, it is possible to build networks that reliably reflect the real market, in
contrast to other fields that lack high quality data, such as the Internet and social
networks. Financial networks can be constructed, for example, from stock prices or stock
ownerships. In the first case, each pair of stocks is connected to each other by a
weighted edge that encodes the distance between them. This distance can be computed as a
function of the correlation coefficient, taken between the time series of stock prices.
From this complete network (\emph{i.e.} having all possible edges) a hierarchical structure is
frequently obtained, which is usually a minimal spanning tree, also called in this
context ``asset tree''. A minimal spanning tree is a connected subnetwork with no cycles,
which includes every node of the original network with minimal cost (\emph{i.e.} minimal sum of
edge weights). On the other hand, a network of stock ownership encodes the stockholders
of a group of companies. It can be a directed network indicating who is the owner of a
stock and who is the company that shares the stock, where the owner can be an individual
or even another company. Examples of the characterization of these two types of networks
are discussed in the following paragraphs.

Mantegna \cite{Mantegna1999} studied the stocks traded in the New
York Stock Exchange (NYSE) from 1989 to 1995, and obtained a
taxonomy that organizes stocks according to economical activity. A
network of stocks was computed, where each pair of nodes was
connected by a weighted edge that encodes the distance between two
stocks. This distance was calculated as a function of the
correlation coefficient, taken between two daily time series of
stock prices. The hierarchical structure was identified in terms of
the minimal spanning tree of the network of stocks. Bonanno \emph{et
al.}~\cite{Bonanno2003} employed the same methodology to study stock
prices in the NYSE, recorded from 1987 to 1998. The authors compared
the topology of the minimal spanning tree resulting from real data
with an analogous one obtained from simulated data using market
models. A power-law for the real data was observed, with exponent
$2.6$, a feature that was not captured well in simulations, since high and
low degrees differ significantly. This means that the hierarchical
structure obtained from real data is not captured by
stock price models.

Onnela \emph{et al.}~\cite{Onnela2003} obtained a similar exponent
($\gamma = 2.1$) to the one computed by Bonanno \emph{et al.},
although during crash periods the exponent changes, such as in the
Black Monday, where $\gamma = 1.8$. The authors used daily stock
prices recorded from 1980 to 1999 in the NYSE. The asset tree
underwent other particular changes in periods of crisis, such as the
decreasing in its length. Onnela \emph{et al.}~\cite{Onnela2003a}
investigated another network obtained from correlations of stock
prices. In this case, the network is not necessarily a tree, being
obtained by selecting only the $N-1$ edges with lower weights of the
complete network (note that a minimal spanning tree also has $N-1$
edges). This network, generally called ``asset graph'', does not
clearly exhibit a power-law degree distribution. Moreover, results
indicate that the asset graph is more robust and stable than the
asset tree, since the former changes less than the latter in
consecutive periods and under extreme conditions. However, the asset
graph does not necessarily have a hierarchical organization, thus
preventing a taxonomic analysis of the financial market.
Furthermore, Onnela \emph{et al.}~\cite{Onnela2004} focused on the
details of the construction of the asset graph, and measured how the
clustering coefficient changes in this process. The authors were
interested in finding what portion of the edges in the asset graph
represent information, and not noise. They found a different
behavior of the clustering coefficient in the empirical graph, in
comparison to a random graph, and this observation was applied to
estimate that only 10\% of the edges in the asset graph convey real
information. The Indian financial market, referred to as National Stock
Exchange (NSE), was analyzed by Pan and Sinha \cite{Pan2007} using
correlations between time series of stock prices (from 1996 to
2006). The network employed connects stocks correlated above a given
threshold, while the threshold was chosen to generate the highest
number of connected components, which were named clusters by the authors. It
was observed that the majority of Indian stocks is not grouped
according to business sectors, \emph{i.e.} correlations inside sectors are
in general much weaker than market-wide correlations. Nevertheless,
when the whole data of stock prices were divided into two time
intervals, it was found that the Indian market is evolving into
clusters corresponding to business sectors.

Battiston \emph{et al.}~\cite{Battiston2005} considered a network of
European firms where a directed edge $(i,j)$ exists if firm $i$ is a
stockholder of firm $j$, and its weight is the amount of investment
from $i$ to $j$. Another version of this network was taken by
joining firms whose headquarters are in the same region. The firm
network shows a power-law distribution of investments, close to the
distribution of firm activities, thus showing a correlation between
investment and activity. On the other hand, the regional network has
a log-normal distribution of activity, investments and degree,
though again investment and activity are correlated. Garlaschelli
\emph{et al.}~\cite{Garlaschelli2005} studied shareholders of stocks
traded in the NYSE, in the Italian stock market and in NASDAQ
(National Association of Securities Dealers). A network description
of investors and assets was used, where companies were linked
through a directed edge to investors that held some of their shares.
Notice that this is not a bipartite network since a company can be a
shareholder of another company. The three datasets showed power-law
distributions for both node in-degree (called portfolio
diversification) and node in-strength (sum of the weights of ingoing
links, called portfolio volume). Souma \emph{et
al.}~\cite{Souma2005} characterized a similar directed network, this
time considering only Japanese companies along with shareholding
links. Six datasets were used, ranging from 1985 to 2003, and for
all of them a scale-free distribution of out-degrees was observed,
with varying exponents along the years considered.

Kumar and Sinha~\cite{PAN_PRE_STOCK_2007} investigated the stock
price fluctuations in the National Sotck Exchange (NSE) of India.
Here, the authors use the network concept to represent the
intra-market structure, i.e the interactions  within the market. To
do that, they used the filtered correlation matrix to build the
network, where the nodes are the stock types (Automobile \&
transport, Financial, Technology, Energy, Basic materials, Consumer
goods, Consumer discretionary, Industrial, IT \& Telecom, Services,
Healthcare \& Pharmaceutical and Miscellaneous). Two nodes were
linked if the filtered correlation between they exceeded a threshold
$c$. The filtering of the correlation matrix removed the common
market influence and random noise. As main conclusion, the authors
noted that the stocks in emerging markets are more correlated than
in developed markets.

Financial markets have a distinctive feature when considering the broad range of fields
covered by this review: available financial data is accurate, being collected along
several years. This means that researchers in this field can perform highly reliable
analysis of the evolution of financial markets, thus helping comprehending its underlying
mechanisms and also allowing the creation of new market models. As far as we know, the
main approach to create stock networks is the correlation between time series of stock
prices. This approach helps identifying the most correlated stocks, although a special
care is needed to avoid spurious correlations. Efforts have been made to retrieve
interesting structures from correlation matrices, such as the taxonomic organization in
the asset tree. Other types of financial networks involving investors and companies have
shown power-law degree distributions, a common feature of complex networks. Nevertheless,
research on financial markets employing a networked representation is still in its
infancy, if we compare with more traditional subjects in complex networks, such as the
Internet and biological networks. In the context of networks research, a great challenge
for financial studies is to create models of market evolution, thus helping the
prediction of important market changes that could affect all the economy of a country or
even of many countries.


\section{Computer Science}

Computer science is a relatively recent area, as the establishment of
the first departments and award of the first degrees occurred in the
1960s~\cite{Aho1992,Denning2000}. While its fundamental topics
involve computer architecture, data storage and processing, and system
control (programming and algorithm developing)~\cite{Bishop1991}, it
has become essential for developments in a variety of areas. For
computer science provides tools to help solving problems of physics,
mathematics, chemistry, biology, medicine, economy, etc. For instance,
the major problems in biology, \emph{e.g.}, protein folding, function
prediction, phylogeny, and modeling of biological systems (see
Section~\ref{Sec:biological}) can only be solved with computers and
optimized algorithms.

\subsection{Software Architecture}

The importance of computers and software to our lives is
unquestionable. Almost everything we do needs a computer and specific
software. The cost of software development increases with its
complexity, and may exceed a million dollars (\emph{e.g.}, computer games,
compilers, and operating systems). The sub-area of software
engineering is aimed at providing methodologies and tools for
designing and building software efficiently, which can be achieved by
decomposing a problem into many small, distinct but interlocking
pieces, named software components~\cite{VALVERDE_SOFT_EL_02,
  MOURA_SOFT_PRE_03}. Deciding the size of the software components is
difficult and must be planned carefully. Levels of granularity are
defined for these components, \emph{e.g.}, subroutines, classes, source
files, libraries, packages, etc. There are several ways to represent
software as a complex network depending on the size of the software
components. The topology of networks and the hierarchical
relationships among the software components were studied by Valverde,
Sol\'{e} and Cancho in
\cite{VALVERDE_SOFT_EL_02,VALVERDE_SOFT_CONDMAT_03}, where the nodes
were program classes and the edges were the inheritance and
composition among the classes. Networks for various pieces of software
were found to exhibit the same organization pattern, including eMule,
Openvrml, GTK, VTK and the Linux Kernel. The networks studied
displayed scale-free topology and small-world features. Furthermore,
software architecture networks can be generated by a local
optimization process, instead of preferential attachment or
duplication-rewiring. Myers~\cite{MEYERS_SOFT_PRE_03} found similar
results for six software architecture networks VTK, CVS, abiWord, the
Linux Kernel, MySQL and XMMS, where the distribution of incoming and
outgoing links is scale-free.

The same applied to the Linux Kernel, Mozilla,
XFree86 and the Gimp networks
\cite{MOURA_SOFT_PRE_03}, in which the networks were built in a
different way. These software packages were written in the C/C++
computer languages, which have two main types of files, ``.c'' (or
``.cpp'') that are the source files and ``.h'' the headers files. The
former contains the source code of the program and the header files
contain the definition of variables and constants, and description of
data structures. Each source file includes (through the "$\#$include"
clause) a certain number of header files.  The authors considered the
header files as nodes of a network, with two nodes being connected if
they were included in the same source file. The topology of these
networks is scale-free and displays the small-world effect even
when the hubs are deleted.

Challet and Lombardoni studied networks of software components
\cite{CHALLET_SOFT_PRE_04} for Linux packages and found clear evidence
of the asymmetry between the distribution of incoming and outgoing links
. While the first presents a scale-free behavior, with
exponent $\gamma=2.0$, the latter does not. The authors also studied
bug propagation in these networks with a similar model to
\emph{Susceptible, Infected and Resistant} of epidemic spreading. In
the case of one faulty node with propagation only to the next
neighbors, it is easy to fix the problem with a debug process. In
contrast, fixing the problem is much harder when non-local fault
propagation occurs, as in an illegal memory access. In other papers on
software networks, Shannon's information entropy was used to measure
structural complexity \cite{MA_SOFT_SEC_05}, and scale-free distribution
and small-world effects were found for networks built from Internet-
based repositories of open software \cite{LABELLE_SOFT_06}

\subsection{Data Sharing}

Iamnitchi \emph{et al.}~\cite{Iamnitchi2002} considered a scientific
file sharing network, where nodes are scientists and two scientists
are connected if one of them is interested in the data of the
other. Based on the small-world topology of scientific collaboration
networks, where edges are created between researchers that are
co-authors in a paper, the authors claimed that scientific file
sharing networks may be also small-worlds, and thus they proposed a
mechanism of data location that takes advantage of the presence of
local clusters. Indeed, three data sharing networks (a
physics data sharing community, WWW data sharing between Internet
hosts and Kazaa traffic between users) have small-world properties
\cite{Iamnitchi2003,Iamnitchi2004,Leibowitz2003}. Leibowitz \emph{et
al.}~\cite{Leibowitz2003} studied Kazaa traffic in 2003, a period
where Internet traffic was already dominated by peer-to-peer
applications. Data of an Israeli Internet Service Provider (ISP) were
employed, which showed that Kazaa traffic is much concentrated in the
transport of a few popular files, suggesting that caching can be a
solution to decrease redundant traffic. Moreover, some of these
popular files lose their popularity in a few days, whereas other files
remain constantly popular (at least in the time period considered by
the authors).

Peer-to-peer (P2P) overlay networks (\emph{i.e.} virtual networks at
application level that use the Internet infrastructure) were
characterized recently. A power-law degree distribution was
identified in a snapshot of the connected hosts of Gnutella taken in
late 2000, although a more recent snapshot of early 2001 deviates
from a pure power-law \cite{Ripeanu2002}. The P2P network eDonkey
also displayed a power-law for in- and out-degree distributions
\cite{Guillaume2004}. In this case, a directed edge $(i,j)$ exists
if a host $i$ makes a query related to a file provided by $j$. As a
guide for the improvement of Gnutella, it was found that it does not
use efficiently the underlying Internet: although 40\% of its nodes
are inside the top ten ASs, less than 5\% of Gnutella's connections
join nodes inside the same AS \cite{Ripeanu2002}.

\subsection{Spam Filtering}

A network of email contacts was used by Kong \emph{et
al.}~\cite{Kong2005} to create a spam filtering technique. It is a
distributed collaborative system that relies on queries made between
neighboring users (\emph{i.e.} users that exchange emails) to decide
whether a suspicious message is a spam. Thus, the system is based on a
``trust'' algorithm considering spams already identified by other
users. Simulations showed a spam detection rate near 100\% with almost
no false positives.

\subsection{Circuits}

The small-world effect was observed by Ferrer i Cancho
\emph{et al.}~\cite{CANCHO_CIRCUITS_PRE_2001} in networks of electronic
circuits, where each node represented an electronic component (\emph{e.g.},
integrated circuits, resistors, capacitors) and edges were wires in a
broad sense. The authors showed that the degree distribution of this
network follows a power law and the most common degree is $k\approx
4$, corresponding to four neighbors in 2D plane.

Another example of circuit network was used in 2004 by Barab\'{a}si
\emph{et al.} \cite{BARABASI_HOTSPOT_EPJB_04}, with the network formed
by a circuit of a microprocessor Simple12. They showed that the
intensity of the electric current fluxes on the edges of the network
varies strongly, up to four orders of magnitude. Furthermore, the
standard deviation and the average value of the flux are related by
the power law $\sigma=\langle f \rangle^{0.5}$, a relation that
appears in other real transportation networks, with different
exponents.

In \cite{TEUSCHER_NANO_CONDMAT_06}, a study was presented of nano
self-assembled devices that showed random topology and small-world
effects. These circuits have better performance in the
synchronization process, small latency and density of classification
task, in comparison to purely local circuits. The authors also
discussed the marketing and technological viability of building
self-assembled circuits with these features.

\subsection{Image Processing and Analysis}

Image processing, analysis, characterization and classification are
intrinsically related~\cite{Costa:book}. Some of these steps are
particularly difficult to be implemented. For instance, segmentation
techniques require identification of objects in images and may involve
complex algorithms, to the point that sometimes such an identification
may become impossible. Since images are composed by adjacent pixels,
complex networks theory can be applied for their processing and
analysis. Such networks can be obtained by associating pixels to nodes
and connecting them according to some criteria, such as image
neighborhood and similarity of the local properties of the pixels
(\emph{e.g.}, gray-level values).  Costa~\cite{dafontouracosta2004cns}
suggested an image segmentation approach where an image was mapped
into a network, and a community identification method was applied to
obtain the segmented objects. The rationale behind this approach is
the fact that pixels belonging to the same object tend to be densely
connected whereas pixels in different objects are sparsely connected,
which then leads to a criterion for establishing
communities~\cite{rodrigues2007fci}.

Complex network theory has also been considered for boundary shape
analysis~\cite{Backes08}.  The shape boundary characterization
approach models a shape into a small-world network and use degree and
joint degree measurements to extract the shape signature or descriptor
which is able to characterize the shape boundary~\cite{Backes08}.  The
methodology proved to be robust, noise tolerant, and invariant to
scaling and rotations. Other image processing methods derive from
complex networks methodologies and concepts, such as color
segmentation, texture analysis~\cite{Chalumeau06} and image
classification. In the latter case, an image would be represented by a
network and the corresponding measurements are extracted. Then, images
can be classified according to the network measurements. In addition
to images, videos can be addressed in a similar fashion, where each
video frame is characterized independently as an image, while
interconnections are considered between successive frames.


\section{Internet}\label{sec:Internet-dyn}

The origins of the Internet can be traced back to 1969, when the
military networking system ARPAnet was created
\cite{Moschovitis1999}. It has since evolved into a worldwide
computer network no longer restricted to academy, through which many
services are provided, with email and the World Wide Web (WWW) being
the most popular. Thus, since the Internet is such a multi-purpose
integrated tool, it is important to understand how it globally works
to keep it safe in case of failure or attack, or to improve its
performance. A straightforward approach to study the Internet is to
represent it as a graph, with hosts, routers and servers being the
nodes and the physical links connecting them (optical fibers or
copper cables, for example) being the edges.

Complex networks researchers have indeed employed a graph-based
approach to model the Internet structure and simulate its traffic.
Nevertheless, the complete mapping of the Internet is difficult to
achieve, since it is not centrally administrated and changes
constantly. Thus, researchers usually employ coarse-grained maps
that contain only the links between Autonomous Systems (AS), which
are subnetworks separately administrated, or use maps including only
the connections at the router level. Nevertheless, such incomplete
graphs have allowed important findings about the structure and
dynamics of the Internet, as discussed in the following paragraphs.

Faloutsos \emph{et al.}\ were possibly the first to study power-laws in the Internet
\cite{Faloutsos1999}. Three snapshots of the Internet at the AS level, collected between
1997 and 1998, and one instance at the router level, collected in 1995, were examined.
Power-laws were found for the degree distribution, the degree rank, the number of pairs
of nodes within $h$ edges (called ``hops'') and the graph eigenvalues. Faloutsos \emph{et
al.}\ employed AS maps from BGP (Border Gateway Protocol, an inter-domain router
protocol) routing tables, an approach criticized by Chen \emph{et al.}~\cite{Chen2002},
which used instead BGP and IRR (Internet Routing Registry, a routing database). Siganos
\emph{et al.}~\cite{Siganos2003} extended the study of \cite{Faloutsos1999} and found
that those power-laws hold for more snapshots of the Internet, ranging from 1997 to 2002,
as well as for the more complete data of \cite{Chen2002}. BGP-based AS maps from a three
year interval (1997-1999) were also verified to be scale-free
\cite{Pastor-Satorras2001,Vazquez2002}, and the same applied to a dataset with the routes
traced from only one host to any other address in the network \cite{Caldarelli2000}.

The correlation profile, \emph{i.e.} the comparison of the degree
correlations of a network with its null (randomized) version, was
used by Maslov \emph{et al.}~\cite{Maslov2002} to identify
differences in networks with the same degree distribution, such as
the Internet at the AS level (in 2000) and molecular networks. A
power-law degree correlation function, which gives the average degree
of the neighbors of a node with a given degree $k$, was found in
another map of the Internet, which is probably a consequence of a
hierarchical structure \cite{Pastor-Satorras2001}.
Bianconi~\cite{Bianconi2008} also used null versions of the Internet
to gain more insights about its structure. The author considered
approximations of randomized network ensembles that preserve the
degree distribution, the degree correlation and the community
structure of five AS maps (among other real-world networks). The
entropy of an ensemble, which is proportional to the logarithm of
its number of networks, was used to indicate how important to the
Internet is a given feature, like its degree distribution or
community structure. For the Internet maps employed, the lowest
entropies were found in ensembles that preserved degree
distributions. Therefore, the feature that carries more information
about Internet, in the context of this analysis, is the degree
distribution.

Furthermore, Bianconi~\cite{Bianconi2004} assessed the number of
cycles of length 3,~4 and~5 in~13 snapshots of the Internet at the AS
level, collected between 1997 and 2001, and found a scaling behavior
of these quantities over time. Moreover, cycles of size~5, called
pentagons, are more frequent in these AS maps than in random networks
with the same degree distribution. Therefore, pentagons could be
regarded as characteristic motifs of the Internet. The rich-club
phenomenon was also identified in an AS network, which means that
nodes with high degree are well connected with each
other~\cite{Zhou2004}. This property has been addressed not only
considering the degree, but across hierarchical
degrees~\cite{mcauley2007rcp}. Such a tendency to have a strong core
in the Internet was reported in \cite{Carmi2006,Carmi2007}, where the
nucleus of the Internet, \emph{i.e.} its innermost $k$-core (a $k$-core is
the largest subgraph whose nodes have at least degree $k$), is a well
connected cluster with approximately 100 nodes and diameter of 2. The
authors used an extended data set, combining BGP-based AS data with
the one obtained in the DIMES project \cite{DIMES}, and divided the
network into three parts: (i) the aforementioned nucleus, (ii) a
fractal connected component that is not in the nucleus but encompasses
the largest part of the Internet and (iii) dendrite-like components,
which are connected only to the nucleus. The decomposition of the
Internet was done with the $k$-shell decomposition method. The
$k$-shell is a set of nodes that contains all the nodes with degree
larger or equal to $k$, which are recursively removed from the network, where
decomposition is made by creating $k$-shells starting with $k=1$ until
$k=k_{max}$.  The modularity of the Internet was also investigated
using spectral analysis of AS level networks. Eriksen \emph{et
  al.}~\cite{Eriksen2003} found modules that approximately match
single countries, and Rosato \emph{et al.}~\cite{Rosato2008} observed
that the Internet has large and highly clustered regions joined by a
few links.

One common analysis of the Internet involves the study of its
vulnerability to failure or attacks. The vulnerability of complex
networks, including the Internet at AS level, was analyzed in
\cite{Albert:2000} with regard to random node removal (node failure)
or hub removal (network under attack). In this case, the diameter and
size of connected clusters were monitored while nodes were
removed. Results showed that the scale-free networks (BA model,
Internet and WWW) are extremely efficient against random failures,
differently from the ER model. However, the hub-based attack rapidly
breaks scale-free networks into small isolated groups. In
\cite{Holme2002} attacks were performed in complex networks (including
the Internet at the AS level) removing nodes with high degree or high
betweenness centrality, both calculated only at the beginning of the
attack (\emph{i.e.} in the original network) or at each step of
attack. Similar strategies were employed for edge removal.  At each
node or edge removal, the average inverse geodesic length and the size
of the largest connected subgraph were monitored. It was shown that
attacks using the recalculated betweenness are the most harmful to the
Internet considering edge-based attacks. For node-based attacks, the
different strategies are equally harmful to the Internet. The average
inverse geodesic length was also used in another study
\cite{Latora01}, where it was called ``efficiency''. The drop in
efficiency when a node is removed was calculated for every node of
Internet backbone maps of USA and Europe. The authors identified the
nodes with the highest efficiency reduction, which could explain the
problems Internet users faced on September 11, 2001. Another
interesting observation made by the authors is that these important
nodes do not necessarily have high degrees.

A scaling behavior of the form $\sigma \sim \left\langle f
\right\rangle^\xi$ between the mean flux per node $\left\langle f
\right\rangle$ and the flux dispersion per node $\sigma$ in the
Internet router topology was reported by Menezes and Barab\'{a}si
\cite{Menezes2004}, where the flux is the transmission rate of data
packets. To explain the observed scaling law the authors used
diffusion of random walkers and transfer of packets along shortest
paths. The results pointed to a scaling exponent of the Internet
($\xi \simeq 0.5$) reflecting an endogenous behavior. That is to
say, Internet is rather more affected by internal decisions, like
the origin and destination of a packet, than by external agents such
as the variation of the number of Internet active hosts. In
\cite{Menezes2004a} a method was presented to separate internal from
external dynamics. The processing capacity of nodes was also
considered by Duch and Arenas \cite{Duch2006} to study Internet
traffic, using data similar to the one employed in
\cite{Menezes2004}. The authors assumed a queue model with Poisson
packet arrival and exponential processing time per node. Varying the
parameter of this model and the time window for measuring data, they
showed that $\xi$ can vary significantly for the Internet.

Centers of activity were identified in the French-side of the
Internet, in a scientific network called Renater
\cite{Barthelemy2002}. In this case, the exchange of data between
all routers was considered, even if they were not physically
connected, and the Random Matrix Theory (RMT) was applied in a
matrix of correlations between edge flows (measured in bytes per 5
minutes) to identify the most correlated connections. The Internet
has a high heterogeneity of connection quality between hosts
separated by the same distance \cite{Percacci2003}. This result was
measured using the Round-Trip-Time (RTT) and the geographical
distance between pairs of hosts over the Internet. As for the
network growth, it was discovered that in 2002 the overall Internet
performance, measured in RTT per unit distance, was improved in
comparison with the year 2000 \cite{Percacci2003}. In addition, new
links were preferably placed among old nodes than between new nodes
in the Internet, which indicates that redundancy plays an important
role on its growth \cite{Pastor-Satorras2001}.

Rosato \emph{et al.}~\cite{Rosato2008} performed simulations of
traffic flow in an Internet map, and noticed that congestion is
reached for lower traffics in the Internet than in a random
network. The authors argued that the hubs can be responsible for this
behavior, since these nodes participate in a large number of
transmissions between other nodes. Using a model of self-regulated
packet generation with different levels of routing depth, Valverde and
Sol\'{e}~\cite{Valverde2004} simulated Internet's traffic and detected
a critical path horizon that separates a low-efficiency from a
highly-efficiency traffic. This transition happens when the size of
the routing tables approaches the network diameter. Furthermore, the
authors hypothesized that the Internet is working close to this
critical path horizon. Nevertheless, Kim and Motter
\cite{Kim2008a,Kim2008} found that there is a large amount of capacity
available (more than 94\%) in a router-level Internet map, mainly a
consequence of large fluctuations in traffic. They employed the router
map of the ABILENE backbone, MIT and Princeton University networks, as
well as the corresponding average traffic measured in June 2006.

When delivering data packets, routers in the Internet choose
neighboring routers to obtain an estimation of the shortest path
between origin and destination. However, this procedure does not
take into account the load present in each router, which can lead to
unexpected processing delays in routers or even congestion in
extreme cases. Echenique \emph{et al.}~\cite{Echenique2004} showed
in simulations with an AS map of 2001 that if the queue load of
routers is taken into account, along with the cost of paths, the
routing protocol of the Internet can be improved. Moreover, the
authors noticed that an increased clustering coefficient can also
lead to improved efficiency as congested nodes can be avoided when
alternative routers in the neighborhood are available. In a
related research, Krioukov \emph{et al.}~\cite{Krioukov2007} studied
the scaling properties of Internet routing. A good scaling happens
when the size of routing tables grows slowly (\emph{e.g.} logarithmically).
However, the routing in the Internet is far from ideal, since for
optimal routing it is necessary to maintain a global view of the
evolving network in each router. In this case, good scaling of table
sizes is, in principle, impossible. Although the Internet needs
substantial routing modifications, simple solutions may arise, as
Bogu\~n\'a \emph{et al.}~\cite{Boguna2007} analytically proved:
using only local information nodes can work efficiently, especially
in networks with power-law degree distribution and high clustering
coefficient, such as the Internet. Tadi\'{c} \emph{et
al.}~\cite{Tadic2007} also argue that it is not possible to
substantially improve the efficiency of Internet traffic by
considering global information. More specifically, the authors
stated that neighborhoods at a maximum distance of two are
sufficient. Local information was also highlighted by Thadakamalla
\emph{et al.}~\cite{Thadakamalla2007} as important parameters for
transport strategies in geographical scale-free networks such as the
Internet.

An accurate network model should contain the same characteristics of
its real counterpart. Each Internet model usually reflects a few of
these features, which leads to a variety of complementary
models~\cite{rodrigues2007sbi}. Yook \emph{et al.}~\cite{Yook2002},
for example, introduced a model based on empirical observations of
the router and AS levels of the Internet. Their model includes
incremental growth, where the probability of connecting a new node
$i$ to an older node $j$ is linearly proportional to the degree
$k_j$ and to the inverse of the distance $d_{ij}$ between them.
Moreover, positions of nodes form a scale-invariant fractal set,
positively correlated with the population density. Park and Newman
\cite{Park2003} analytically studied ensembles of networks with the
restriction proposed in \cite{Maslov2002}, \emph{i.e.} that the maximum
number of edges between a pair of nodes is one. This mechanism leads
to the negative degree correlation (disassortative mixing) reported
by Pastor-Satorras \emph{et al.}~\cite{Pastor-Satorras2001}, which
inversely associates the average degree of the neighbors of a node
$i$ with its degree $k$. Nevertheless, Park and Newman argue that
this approach is not responsible for all the degree correlations
verified in the Internet.

Rosato and Tiriticco simulated the growth of the Internet at AS
level \cite{Rosato2004} using a mechanism of triad formation
introduced in \cite{Holme2002a}. This mechanism uses preferential
attachment to create the first edge that connects a new node $i$ to
an existing node $j$. Moreover, node $i$ is sequentially connected
to the neighbors of its first assigned neighbor $j$, thus building a
``triangle'', or to any other node following preferential
attachment. This model leads to a scale-free network with adjustable
clustering coefficient. The authors found that simulations of this
model fits Internet data from 1998 to 2000, for quantities such as
diameter, clustering coefficient, degree distribution and average
geodesic length.

Zhou and Mondrag\'on modeled the Internet at the AS level using an
interactive growth mechanism and nonlinear preferential attachment
\cite{Zhou2004a,Zhou2006}. Their approach, called positive-feedback
preference (PFP) model, is also based on data features such as the
initially slow growth of node degree and the establishment of links
between old nodes. Simulations of this model showed that it can
reproduce many characteristics of the Internet, including the degree
distribution, the rich-club phenomenon and the disassortative
mixing.

Serrano \emph{et al.}~\cite{Serrano2006} considered in their model
the number of users (hosts) in each AS and the capability of
adaptation of ASs according to their size. The authors assumed
exponential growth for the number of users, the number of ASs (the
nodes) and the number of links between ASs (the edges of the
network), all inspired by empirical observations of Internet
data~\cite{Barrat04:PNAS}. Another important feature of this model
is that ASs can have connections strengthened by increasing their
edge weights. Furthermore, the model can optionally include a
distance constraint to avoid connections between distant ASs. The
parameters used for the model led to a power-law degree distribution
with exponent $\gamma = 2.2 \pm 0.1$, which is in agreement with
previous works \cite{Faloutsos1999,Pastor-Satorras2001}. The model
explains the average shortest path length, the clustering
coefficient and the $k$-core decomposition of the Internet.

Internet models have also been compared by considering multivariate
statistical methods, as proposed by Rodrigues \emph{et
al}.~\cite{rodrigues2007sbi}. The authors suggested the
consideration of canonical variable analysis and Bayesian decision
theory to determine which model is most suitable to reproduce the
Internet structure. Such an approach allows one to consider a large set
of measurements to obtain an accurate description of the
network structure.  Rodrigues \emph{et al}. analyzed Internet models
at the autonomous systems level and concluded that none of the
models considered were able to reproduce accurately the Internet structure. The
proposed approach is aimed at determining which model is more
suitable to any type of complex network.

The Internet was one of the first and most studied subjects in the field of complex
networks, for which various topological and dynamical features have been already
observed. These include a power-law degree distribution, the rich club phenomenon,
disassortative mixing, large unused bandwidth and far from optimal routing. Many models
were proposed for the Internet, which at most reproduce some of its experimentally
observed features. Despite the incompleteness of Internet maps, the results from analysis
made with different data compiled at distinct time periods and using various
methodologies for collection are consistent. Recurrent features in these analysis are the
power-law degree distribution and the existence of a nucleus set of nodes. Nonetheless, a
more accurate methodology for data collection is still necessary. Researchers and
professionals interested in improving Internet's performance and security may benefit a
lot from these relatively recent discoveries made in the complex networks field. Although
new insights about Internet's structure and dynamics frequently appear, its distributed,
centrally uncontrolled nature, and its constant evolution, still pose challenges for a
complete understanding of its main governing rules.


\section{World Wide Web}\label{sec:www-charac}

The World Wide Web (WWW, or just Web) is perhaps where Internet users spend most of their
time, despite the existence of other highly popular Internet applications such as
electronic mail. The WWW relies on the HyperText Transfer Protocol (HTTP), which allows a
Web browser running on a client computer to communicate with Web servers distributed over
the Internet. Web servers usually contain hundreds (sometimes many more) files that can
be accessed by anyone with a Web browser connected to the Internet. These can be
virtually any computer file, such as those containing text, image or video;
the files, however, that give the WWW a networked structure are the pages, which may contain
pointers (hyperlinks) to other pages available in the WWW. Thus, one can think of the WWW
as a giant network of pages interconnected by hyperlinks. These files are the most
important of the WWW because they are multimedia files that group rich text (\emph{i.e.} with
formatting options), images, videos and even computer programs (\emph{e.g.} scripts and
applets).

The WWW viewed as a network is an interesting subject for complex networks researchers
and for those whose main concern is the Web itself \cite{Berners-Lee2006}. It is huge,
with billions of pages, despite its relatively recent birth (released in 1993 to public
access), and it has an almost uncontrolled growth mechanism, where individuals or
organizations create their interconnected pages depending solely on their will. Grasping
the WWW organization and how users jump from page to page while navigating/surfing the
Web is crucial to optimize search engines and facilitate access to information. Mainly
because of its uncontrolled nature, the WWW is difficult to map completely, even for
sophisticated search engines~\cite{Lawrence1999}. Usually a WWW map is constructed by a
computer program called ``crawler'' that navigates through pages storing their
hyperlinks, \emph{i.e.} identifying source and target pages. Unfortunately, many pages are not
considered by the crawler for reasons such as authorization requirements to obtain a
page, dynamic pages that require filling a form, crawler limitations, broken links or
unreachable Web servers. Usually, Web maps are viewed as directed networks, where each
edge points from one node to another, \emph{i.e.} the page that contains the hyperlink points
to the referenced page. Sometimes the Web is not considered at the page level, but at the
site level, where each site (roughly a group of pages under the same domain address) is
considered as a node, and pairs of sites are connected whenever a hyperlink is found
between some of its pages.

In a number of papers the WWW was reported as a scale-free network
\cite{Albert1999,Barabasi1999,Barabasi2000,Broder2000,Donato2004}. With a network of
325,729 pages taken from the \mbox{nd.edu} domain, Albert, Jeong and Barab\'asi
\cite{Albert1999,Barabasi1999,Barabasi2000} verified that its out- and in-degree
distributions follow a power-law of the form $P(k) \sim k^{-\gamma}$, with exponents
$\gamma_{out} = 2.45$ and $\gamma_{in} = 2.1$, respectively. Broder \emph{et
al.}~\cite{Broder2000} and Donato \emph{et al.}~\cite{Donato2004} used a considerably
bigger Web map, with 200 million pages, and obtained $\gamma_{in} = 2.1$. Pennock
\emph{et al.}~\cite{Pennock2002} divided the distribution of pages by category, such as
the set of company websites or the set of newspaper websites. Although the in-degree
distributions of these subnetworks displayed the heavy tail characteristic or power-laws,
the frequency of nodes with low degree $k_{in}$ significantly deviates from a pure
power-law. Other power-laws were found in the WWW for the distribution of
the sizes of connected components \cite{Broder2000} and for the distribution of the
number of pages in a website \cite{Huberman1999}. Moreover, self-similarity at different
scales was verified in the WWW using the \mbox{nd.edu} map \cite{Song2005}. In this case,
the network was divided into boxes of nodes of size $l_b$, where every pair of nodes in a
box was separated by a distance $l<l_b$. It was found that the number of boxes required
to group the network and the size of the box also follow a power-law. Furthermore, the
values given by PageRank, which is a webpage ranking method based on a random surfer
model used by the Google search engine~\cite{Brin1998}, follow a power-law distribution
with exponent $2.1$ \cite{Donato2004}. PageRank is not statistically correlated with the
in-degree, which means that webpages with high in-degree (\emph{i.e.} referenced by many other
pages) do not necessarily appear at the top of a Google search.  The small-world effect
in the WWW was observed by Adamic \cite{Adamic1999} at the site level using approximately
250,000 sites. Hence, this WWW map presents high average clustering coefficient and low
average distances between nodes. Bianconi \emph{et al.}~\cite{Bianconi2007} developed a
theory to assess if a directed network has more or less loops than a random network.
Using the \mbox{nd.edu} map, the authors observed that the WWW has more short loops than
its randomized version (including loops of length 3, which is considered in the
computation of the clustering coefficient), thus pointing to the small-world effect. The
authors also found that other directed networks, including food webs and the
\emph{C.~elegans}' neural network, have less short loops than their randomized
counterparts, which means that the small-world effect vanishes when these networks have
their directionality preserved. In another experiment, random versions of observed data
that preserve the original degree distribution were used to estimate the average geodesic
length $\left\langle d \right\rangle$ of the entire WWW \cite{Albert1999,Barabasi2000}.
The size of the WWW was thought at that time (late 1990s) to be $N = 8 \times 10^8$,
which gave $\left\langle d \right\rangle \simeq 19$, a relatively small value. The
authors predicted a logarithmic growth of $\left\langle d \right\rangle$, indicating that
the average distance between nodes in the Web grows slowly with $N$. However, Broder
\emph{et
  al.}~\cite{Broder2000} verified that $\left\langle d \right\rangle$
is very different in their Web map of 200 million pages, with
$\left\langle d \right\rangle \simeq 500$. Moreover, they found that
a path between randomly chosen pairs of nodes exists in only 24\% of
the times, thus confirming the absence of a small-world effect in
their data. Even in the biggest strong connected component (a
central group of pages) they obtained $\left\langle d \right\rangle
\simeq 28$.

Before Bianconi \emph{et al.}~\cite{Bianconi2007} studied cycles in
the Web, Caldarelli \emph{et al.}~\cite{Caldarelli2004} performed a
similar investigation in some WWW maps. The latter authors
introduced the grid coefficient, a measurement that quantifies the
level of cycles of length $n$ in a network. The authors also
characterized cycles of length 4 (referred to as quadrilaterals) in some
networks, including the WWW map of the \mbox{nd.edu} domain. This
Web map has far more quadrilaterals than its randomized counterpart,
which indicates that some order of regularity, represented by
grid-like structures, is present (a similar behavior
was found in the Yeast protein interaction network and in a
scientific collaboration network). Moreover, the authors noticed
that the grid coefficients follow a power-law distribution in these
networks and in an AS map of the Internet. This observation leads to
the claim that the WWW and the other networks have a hierarchical
arrangement of cycles.  Broder \emph{et al.}~\cite{Broder2000}
reported an intricate organization of a WWW map collected in 1999,
which was divided into four groups of roughly similar size: (i)~the
largest strong connected component (SCC), where every page can be
reached starting from any other page; (ii)~the IN group, which
contains pages that can lead to SCC (even indirectly), but cannot be
reached from SCC; (iii)~the OUT group, with pages that do not lead
to SCC, but can be reached from SCC; and (iv)~the TENDRILS, with
pages separated from SCC that can only reach IN and OUT. Donato
\emph{et al.}~\cite{Donato2004} investigated the size of these
components in a WWW map collected in 2001, and obtained
different values to those reported by Broder \emph{et al.}
Other modular structures in the WWW were reported in the
literature, regarding groups of pages about similar topic. In
\cite{Flake2002} communities of related topics were found based on
the maximum-flow/min-cut theorem and in \cite{Eckmann2002} groups of
connected pages with high curvature (a geometrical interpretation of
the clustering coefficient) were associated with specific topics.

Social networks based on the Web can be analyzed using networks of personal pages, such
as the blogs where hyperlinks are frequently created between pages of friends or
acquaintances. Some connectivity properties between 200,000 pages of a Chinese blogspace
revealed a scale-free degree distribution and the small-world effect~\cite{Fu2006}.
Indeed, Milo \emph{et al.}~\cite{Milo2004} showed that the WWW and social networks are
included in the same super family, based on the significance profile of motifs (this
profile is defined in Section~\ref{sec:charac_superf}). This observation may be useful to
help understanding the WWW using models of social organization.

Another network related to the Web, though not built using pages as
nodes, was characterized by Shen and Wu \cite{Shen2005}. This
network, called folksonomy, uses semantic tags created to describe
the content of a database of hyperlinks (bookmarks). A folksonomy
represents the connectivity between tags, where two tags are linked
if they are associated with the same content. The employed data of
9,804 tagged bookmarks also exhibit scale-free and small-world
properties.

The dynamics in the WWW is usually associated with user surfing,
\emph{i.e.} the sequence of pages a user (or a group of them) visits when
following hyperlinks in the WWW. Dezs\"o \emph{et
al.}~\cite{Dezso2006}, for example, divided the pages of a news
portal into stable pages, \emph{i.e.} the overall fixed structure of the
portal, and news pages, which are pages frequently created. The rate
of visits to stable pages is constant, while news pages receives a
large number of visits after a few hours and decays as a power-law.
Moreover, unlike Poisson processes, a power-law distribution also
describes consecutive visits to the site performed by a single user.
Huberman \emph{et al.}~\cite{Huberman1998} compared real user
activity data with a model of user surfing that gives the
probability distribution of the number of visits to a Web page. The
model, which reproduced well real data, considers that there is a
cost associated with a surfing activity, and that the user continues
surfing while this cost remains below a threshold. Another surfer
model, the aforementioned PageRank, is used by Google to sort its
search results \cite{Brin1998}. PageRank simulates the behavior of a
user randomly surfing the Web, where the user goes from page to page
following hyperlinks and sometimes performs random jumps with
probability $d$ to any other page in the Web. The frequency of
visits of the random surfer per page is the PageRank score. Menezes
and Barab\'{a}si \cite{Menezes2004} recorded the visits to a group
of sites during 30 days, and the mean flux per node $\left\langle f
\right\rangle$ and the flux dispersion per node $\sigma$ were
estimated. Similarly to what happens with the Internet (see
Section~\ref{sec:Internet-dyn}), a scaling behavior of the form
$\sigma \sim \left\langle f \right\rangle^\xi$ was found in the
WWW, with exponent $\xi \simeq 1$. Results indicate
that the fluctuations in the WWW are dominated by external driving
forces, such as the variation in the number of surfers, rather than
by internal choices made by surfers.

Kleinberg \emph{et al.}~\cite{Kleinberg1999} and Kumar \emph{et
al.}~\cite{Kumar1999} modeled WWW growth applying intuitive ideas of
user behavior, where a user tends to create links to pages related
to some topic of interest. While the network grows, some pages
appear about a topic and users interested in this topic start
creating links to those pages. This makes it easier for other users
to find those pages, thus also establishing other links to them.
Ultimately, it allows the formation of groups of pages about similar
topics. The model performs this procedure by using four stochastic
processes, one for node creation and one for edge creation, along
with two other processes for node/edge removal. Kumar \emph{et
al.}~\cite{Kumar2000} created a model for generating an evolving
WWW, where new nodes are continuously added, connecting to previous
nodes with either uniform probability or with preference to older
nodes. The latter case simulates a behavior where newly created
nodes are likely to be unknown by most of page creators. Moreover,
some nodes reproduce the neighborhood of a randomly picked node. All
these models agreed with observed degree and bipartite core
distributions.

Bornholdt and Ebel \cite{Bornholdt2001} employed the Simon model,
created in 1955, to explain Zipf's law. This model accurately predicts
the in-degree exponent found in the WWW
\cite{Barabasi1999,Broder2000}, which is $\gamma_{in} = 2.1$. The
authors defined the following procedure, where a class $[k]$ contains
the nodes with the same degree $k$ and $f(k)$ is the number of nodes
in the class $[k]$: (i)~with probability $\alpha$ create a node and
connect it to any other node, or (ii)~with probability $(1-\alpha)$
connect any node to a node of class $[k]$ with probability $P_{[k]}=
\left[k f(k)\right] / \left[\sum_{i}if(i)\right]$. Tadi\'c
\cite{Tadic2001} suggested two general rules to generate the WWW:
(i)~growth, where at each time a new node $j$ is added to the network,
and (ii)~rearrangements, where at each time a number of outlinks are
created from $j$ to an old node and a number of links between old
nodes are created or removed. This model agreed with empirical data
regarding distributions of in- and out-degree and the distribution of
the size of connected components.  Pennock \emph{et
  al.}~\cite{Pennock2002} changed the BA model of preferential
attachment by including a uniform attachment probability. Thus, the
probability of connecting a new node $i$ to an old node $j$ is a
combination of the usual preferential attachment with a uniform
attachment rule. This model encompasses the rich-get-richer mechanism,
along with the creation of hyperlinks that are more influenced by
personal interests of the page owner than by the popularity of a
page. Moreover, the model is especially able to capture degree
distributions of specific subgraphs in the Web, such as the set of
company websites. Menczer \cite{Menczer2004} expanded this model and
included a parameter of lexical similarity between webpages, where the
probability of connecting two nodes includes a power-law that depends
on the lexical distance. Thus, this approach considers the content of
the documents and their connectivity. Besides approximating the degree
distribution of a snapshot of the Web with 100,000 documents, this
model can also estimate the content similarity distribution of
connected webpages.

In conclusion, as many real world complex networks, the WWW exhibits the scale-free
property concerning in- and out-degree distributions. Interestingly, in-degree hubs are
uncorrelated with the well known random surfer model named PageRank, meaning that highly
cited pages are not necessarily highly visited. The presence of the small-world effect in
the WWW is controversial, with results showing a fragmented Web lacking paths between the
majority of node pairs. Modularity seems to play an important role in the Web, with the
presence of general groups like SCC or specific communities related to some topic.
Moreover, several models for WWW evolution have already been proposed, all of them
reproducing some experimentally observed features. One of the main challenges of Web
researchers is to learn to obtain a more complete, accurate set of WWW data, a problem
faced by researchers of many other fields. Moreover, since more complete data means even
bigger Web maps, some computational complexity issues may arise when carrying out
experiments. Efficient search for information is now an important concern, thus a better
understanding of the structure and dynamics of WWW is fundamental. Merging page
interconnectivity and meta data (\emph{i.e.} information about page content) is a promising
approach because specific parts of the Web can be analyzed separately, hence allowing
more detailed insights about WWW that could help the design of search engines and
crawlers, and also guide webmasters to better place their pages in the cyberspace.


\section{Citations}

Although dynamically described by diffusion processes, information
transfer can also be quantified from static structures such as
citation networks. Any information contained in a paper or technical
report contributes to the global knowledge of the reader. However,
only the most relevant manuscripts tend to be remembered when one has
limited time and space. Therefore, an information filtering process is
already performed at this stage. By using data from citation
databases, one can investigate the information flow through specific
groups of people. The citation networks so far constructed derive from
technical reports (especially scientific papers) represented as nodes
and citations between those reports as directed edges. Those networks
are constantly growing. As the in-degree represents the citations from
other reports, this measurement carries information about the
importance of the report, \emph{i.e.} the number of times that report was
found to be relevant to other works. However, once published, the
report cannot aggregate new references and the out-degree is fixed
over time. Because of this intrinsic network property, the out-degree
distribution of scientific papers from the ISI database in the period
from $1991$ to $1999$ is characterized by a maximum value with strong
fluctuations at the left-hand side of the distribution. Most
importantly, there is a universal behavior at the two rightmost
classes, corresponding to journals of limited and unlimited number of
pages per paper~\cite{Vazquez01}.  In contrast, patent-to-patent
citations present a power-law with exponent $\gamma =
0.62$~\cite{Chen04}. Van Raan found a power law with exponent $\gamma
\sim 3.7$ in the number of references per publication in
$2001$~\cite{vanRaan05}.  This interesting unidirectional growth
mechanism has curious properties: though the average number of
citations increased over time~\cite{Redner98,Vazquez01}, the average
number of citations for papers published in a given year has decreased
slowly with time~\cite{Redner98}. This suggests that the relevance of
a paper might decrease over time. Analysing data from arXiv and PRL,
Hajra \emph{et al.} concluded that most papers are cited within a
period of $10$ years after publication. It also indicates that trendy
research topics are popular during this same period and afterwards,
become less important~\cite{Hajra05:physicaA}. Redner constructed two
networks, one considering only the citation of papers published in
$1981$ and available at the ISI database, and another network covering
Physical Review D papers published and cited in a period of nearly
$22$~years \cite{Redner98}. By means of a Zipf plot (cumulative
distribution), he found a power-law citation distribution $P(k)\sim
k^{-\gamma}$ with $\gamma \approx 3$ for large $k$. Other power-law
distributions with exponent $\gamma = 2.7$ were verified in the SPIRES
database by Bilke and Peterson~\cite{Bilke01}, now with exponent
$\gamma \sim 3.1$ for papers published and cited in
$2001$~\cite{vanRaan05}. Such distributions, together with other
evidence, led Redner to suggest that minimally and highly-cited papers
obey different statistics where most part of papers are forgotten a
short time after publication. In fact, nearly $47\%$ of the papers in
the ISI database were not cited and $80\%$ were cited only 10 times or
less. Conversely, about $0.01\%$ were cited over $1,000$
times~\cite{Redner98}. On the other hand, Tsallis and
Albuquerque~\cite{Tsallis00} suggested that the entire distribution
found by Redner might emerge from the same phenomenon since it could
be well fitted by a single function derived from the non-extensive
formalism. Power-law citation distributions were also found on an
extensive US patents database granted between January $1963$ and
December $1999$. Chen and Hicks~\cite{Chen04} showed that the
patent-to-patent citation distribution had exponent $\gamma = 2.89$,
and $\gamma = 2.31$ when the data were restricted to a specific field
related to both basic and applied research.  Interestingly, while
analyzing only citations of papers in US patents, they concluded that
papers with explicit funding acknowledgements tended to be cited
relatively more often ($\gamma = 1.97$) than papers without explicit
acknowledgements ($\gamma = 2.16$), possibly because those articles in
the former category are more likely to have impact on inventions
described by patents. Using a random walk procedure, Bilke and
Albuquerque~\cite{Bilke01} found that the spectral dimension of the
ISI citation network is about $3.0$, having a tree-like structure with
the small-world property.

The level of similarity between two reports can be inferred not only by keywords, but
also by the number of common references. Since the references are supposed to be directly
related to the subject of the report, similar reports should cite similar previous works.
This property is obtained by the neighborhood of a specific node, but it could also be
obtained from a network such as that of van Raan who used databases from several citation
indexes in $2001$. He considered that two publications were connected if they had at
least one reference in common~\cite{vanRaan05}. Although the network was unweighted, the
same data could now be used to quantify the level of similarity between two reports. The
degree (named \emph{bibliographically coupled publication cluster size}) distribution is
also described by a power-law function and an exponential cut-off for values larger than
$1,000$. Analogously to other works, a memory effect was identified in the age of the
references, the distribution function changed from a power-law to an exponential if the
network was built only with older references. He suggested that reports with low degrees
are related to very specific themes which are typically recent, and therefore cite recent
articles. In contrast, old references are generally more general and connect more parent
publications, giving rise to a more uniform distribution.

The self-citation issue was addressed by Hellsten \emph{et
al.}~\cite{Hellsten07:scientometrics,Ausloos07} adopting the optimal percolation
method~\cite{Lambiotte05:pre,Lambiotte06:epjb}. They proposed a methodology over the
self-citation evolution patterns of a specific author and compared this pattern to the
co-authors and keywords in the articles. The method was used to detect emerging research
fields and to trace mobility of scientists through different fields and critical moments
in the academic career. For one author, changing co-authorship drives the changing
research interests and move to new research topics~\cite{Hellsten07:scientometrics} while
for another author, this inter-field movement is an effect of maintenance of the same
collaborators~\cite{Ausloos07}.

The databases reviewed in this section are considerably large and
reliable in terms of an interaction rule between their
components. However, some inaccuracies are present since little
standardization is adopted and individual entries are made by
different people at different times. Redner suggested that
inaccuracies as incorrect page numbering for citations, citations of
specific pages and input errors in transferring citation data supplied
by authors, have minor effects in the citation
distribution~\cite{Redner98}.


\section{Transportation}

Transportation networks are important for the development of a
country and may be seen as indicator of economic growth. The tourism
industry and transport of goods and people are particularly
dependent on transportation networks, which include airports,
railways, highways, subways, and other forms of public transport. Studying this
kind of network can help one to understand the movement of people
around the world and predict how diseases can spread, in addition to
design optimal networks for the flow of people. Ultimately, it may
give insights on how to improve the economy of a country.

\subsection{Airports}

In airport networks, cities with airports are considered vertices,
and flights between them are the arcs. This type of network
is naturally directed and weighted because of the direction and
number of passengers in flights. The arc weights are given by the
number of passengers of flights or number of flights itself in a day
or in a week. Guimer\`{a} \emph{et al.}~\cite{Guimera05:PNAS},
however, argued that the connections for the network of all
airports in the world are almost symmetrical, with minor asymmetries
arising from a small number of flights following a ``circular''
path.  Therefore, there is no need to consider the arc directions.

He \emph{et al.}~\cite{He04:IJMPB} studied the Chinese airport
network in 2002 without considering either the directions or the
number of passengers in a flight. They concluded that this network
is small-world without the scale-free property, as the vertex degree
distribution is exponential. They explained the topology of the
network with a model based on the population and gross domestic
product of the cities. Li and Cai~\cite{Li04:PRE} also analyzed the
Chinese airport network using a connectivity matrix of size $N
\times N \times 7$, where $N$ is the number of airports for the 7
days of the week. Besides the scale-free and small-world properties,
they also observed for the Chinese airport network that (i)~the
daily and weekly cumulative vertex degree distribution of undirected
and directed Chinese airport networks obey two regime power-laws
with different exponents (the so-called Pareto
law~\cite{Pareto1897}); (ii)~the cumulative distribution of flight
weights has power-law tails; (iii)~the diameter of a subcluster,
consisting of an airport and all of its neighbors, is inversely
proportional to its density of connectivity; and (iv)~the
transportation efficiency of the subclusters of the airport network
increases with the density of connectivity.

The worldwide airport network~\cite{Guimera05:PNAS,Guimera04:EPJB}
is a scale-free, small-world network.  Nevertheless, the most
connected cities are not necessarily the most central, in contrast
to other scale-free networks. This is a consequence of the community
structure of this network~\cite{Guimera05:PNAS}, on which a global
role of cities was developed. This measurement indicated that the
vertices connecting different communities are hubs in their own
community. Guimer\`{a} \emph{et al.}~\cite{Guimera05:PNAS} pointed out
that the community structure cannot be fully explained only by
geographical constraints, but geopolitical considerations have to be
taken into account. A model to account for these findings was
developed by Guimer\`{a}  and Amaral~\cite{Guimera04:EPJB}, which
explains why the most connected cities are not the most central. In
the model, not all cities of a country can establish connection with
cities of another country.

Bagler~\cite{Bagler04:condmat} studied the airport network of India
considering edge weights, and found that this network is
hierarchical with the small-world property, and characterized by a
truncated power-law degree distribution. The network has
disassortative mixing, in contrast to the worldwide network of
airports. This difference can be related to the local and global
scales of airport networks.

An analysis of the evolution of the Brazilian airport network~\cite{Rocha08} showed that
its structure changed within a 12-year period. The number of airports and routes
decreased while the betweeness centrality increased over time. The overall behavior
suggests that companies focus on concentrating the operation in the most profitable
routes, increasing the number of flights and removing the less profitable ones in a
dynamic way such that airports gain and lose importance through the years. One of the
consequences is the possible increase in the vulnerability of the network for both random
failures and targeted attacks~\cite{Rocha08}.

In addition to analyzing the networks, it is important to design
optimal networks for this kind of transport. Kim and
Motter~\cite{Kim2008iop, Kim2008njop} analyzed networks with
Regard to the resource allocation, \emph{i.e.} the seat-occupation in the
flights, and showed that airport networks have a very efficient
capacity distribution, which is a consequence of the high cost of
air transportation. A new route network to optimize the operational
cost for the French airports network, based on graph coloring for
the air traffic flow management, was proposed
in~\cite{Barnier04:AOR}.

Problems associated with airport networks are related to
difficulties in the air traffic flows, \emph{e.g.}, caused by heavy fogs or
snowstorms. In order to assess the performance of the network under
such circumstances, Chi and Cai~\cite{Chi04:IJMPB} analyzed the
errors caused by an attack to the structure of the US airport
network. Analogously to other scale-free networks, the US airport
network is tolerant to errors and random attacks, but extreme
vulnerable to a target attack to hubs. While topological properties,
including average vertex degree, clustering coefficient, diameter,
and efficiency are almost unaffected by the removal of a few
airports with few connections, the same properties are drastically
altered if a few hubs are removed.

\begin{table}
  \begin{center}
    \caption{Papers that considers different airport networks.}
    \begin{tabular}{c|c}
      \textbf{Network} & \textbf{Reference}\\
      \hline
      Chinese airport network & \cite{He04:IJMPB,Li04:PRE} \\
      French airport network & \cite{Barnier04:AOR} \\
      Indian airport network & \cite{Bagler04:condmat} \\
      Japan airport network & \cite{Hayashi05} \\
      US airport network & \cite{Chi04:IJMPB} \\
      Brazilian airport network & \cite{Rocha08} \\
      Worldwide airport network & \cite{Guimera05:PNAS,Guimera04:EPJB} \\
      \hline
   \end{tabular}
\end{center}
\end{table}

\subsection{Roads and urban streets}

These networks deserve special attention in our daily life since
they directly influence our travel times and transport costs. Over the
last years, we have witnessed an increasing number of vehicles on
the roads which resulted in slower traffics and more frequent jams.
Although possible in the past, building new roads or streets is not
feasible today because of spatial constraints imposed by the local
buildings. The best solution is, therefore, to make better planning
for the traffic or even to manage it in real-time so that time wasted on the roads is minimized.

In the context of complex network theory, many questions related to
the representation of the metric distances of the roads arise. These
include what kind of graph representation to use, which topological
features to study, as well as the correlations between structural
measurements and the dynamics of the traffic flows. The
representation of road networks can be: \emph{primal} or \emph{dual}
graph (\emph{e.g.}~\cite{Porta2006pa, kalapala2006pre}). In the former
case, while intersections are nodes, the roads are edges. This is a
natural way of representing these networks since it captures the
most important feature of geographical dimensions, \emph{i.e.} the
distance, and it was used in~\cite{Cardillo2005pre, Gastner06:EPJB}.
A more detailed study of networks of this kind can be found
in~\cite{Cardillo2005pre}, where the authors analyze twenty samples
of street patterns of several world cities.

There is a limitation, however, with the primal representation,
since it does not express the difficulty to walk on
them~\cite{rosvall2005prl}, which can be obtained by the \emph{dual}
representation, where roads are nodes and intersections are edges.
Such a representation was defined in the seminal work of
Hillier and Hanson~\cite{hillier1984sls}, known as ``axial
mapping''. Although it does not give the geographical distance
between two arbitrary points in the network, it expresses the
information (\emph{i.e.} the number of road changes) needed to travel
between those points. Therefore, the smaller the number of road changes
to reach a specific destination, the easier to find it.

Rosvall \emph{et al.}~\cite{rosvall2005prl} expressed the difficulty
of navigation on these networks in terms of the ``search
information''~\cite{sneppen2005has}. The higher this measurement,
the more difficult it is to find a destination. With this approach,
the authors showed that new cities, as \emph{e.g.}, Manhattan, are better
planned than old cities, as \emph{e.g.}, Stockholm, since the search
information for the former is less than for the second city. Another
conclusion of~\cite{rosvall2005prl} is that it is preferable to
replace a big number of streets with a few long, provided it reduces
the number of road changes necessary to connect any two points of
the network. The accessibility of places in towns and cities was
also investigated with self-avoiding random walk
dynamics~\cite{Travenccolo2008oau}, and Traven\c{c}olo
and Costa showed that the dynamics of transportation through towns
and cities is strongly affected by the topology of the connections
and routes.

Following the second definition for road networks, Kalapala \emph{et
al.}~\cite{kalapala2006pre} found that this kind of network,
besides the small-world property~\cite{Porta2006pa}, has topological
and geographic scale invariance. In other words, for sufficiently
large geographical areas, the degree distribution is scale-free,
with journeys having identical structures, regardless of their
length. Indeed, the driver starts on small roads, moves to
progressively larger, faster roads, until the fastest is reached,
where he/she covers most of the trip before descending back to
progressively smaller roads. The authors also proposed a simple
fractal model to reproduce the above features. Other models
(geographical scale-free networks) can be found in the
review~\cite{Hayashi05}.

In~\cite{Gastner06:EPJB}, Gastner and Newman compared geographical
networks such as an airline network, the US interstate highway
network, and the Internet at the autonomous system level. They
showed that networks, which are fundamentally two-dimensional such
as the highways in the primal graph representation, have a close
relation between their geographical and topological features, as
\emph{e.g.}, the geographical and geodesic distance. A model to reproduce
their findings was also proposed.

The dynamics on road networks has been studied for years, with many
models proposed~\cite{nagel1992cam, knospe2000trm, helbing2001tar,
schadschneider2002tfs, nagel2000lst}, ranging from macroscopic
models based on the kinetic gas theory or fluid dynamics to
microscopic approaches with equations for each car in the network.
In~\cite{Schadschneider2005pa}, the authors analyzed the German
highway network using cellular automata as a model for the traffic
flow, and showed that the simulations are faster than real time
making this model suitable for traffic forecasting.
It was also possible to find the bottlenecks of the highway network.

\subsection{Other transportation networks}

There are many other transportation networks, but their properties
are quite similar to the airport and road networks mentioned above.
Therefore, we shall not discuss networks such as the
railways~\cite{Sen2003pre, Seaton2004pa, Li2006ijmpc}, subway
networks~\cite{Latora-Marchiori02}, and public
transportation~\cite{sienkiewicz2005pts, Sienkiewicz2005pre,
  Ferber2005cond-mat}.

The analysis of transportation networks is important because it provides manners to
investigate the economy of countries and can be used to improve their infrastructure. The
majority of this kind of network are small, generally with less than 1,000 vertices, has
the small-world phenomena with the scale-free degree distribution, and has hierarchies.
The air transportation network is also modular. Because to this property, the most
important vertices in the airport networks are the hubs inside each community, and not
the global ones. In the case of the road networks, there are two ways to be represented,
depending on how to represent roads and intersections. If roads are considered edges and
the intersections are the vertices, the representation is known as \emph{primal} graph.
Otherwise, it is referred to as \emph{dual} graph. In the first case, there is a close
relation between the topological and the geographic features, while in the other case the
representation is more related to the difficulty of traveling inside the network. If the
primal graph representation is used, the corresponding networks does not have the
small-world property and scale-free degree distribution.

The main results obtained for the airport networks were that they
are very efficient in terms of resource allocation, due to the
expensive cost of air transport, and tolerant to random attacks,
while vulnerable to targeted ones, analogously to many scale-free
networks. In the case of road networks, the best planned cities are
those with low search information, \emph{i.e.} it is not difficult
to travel inside them.

The majority of the models developed for transportation networks are based on
geographical and economic constraints, with the main aim of reaching optimal, robust
networks. The development of better models is crucial for achieving such endeavors.


\section{Electric power transmission systems}

The electric power transmission system is one of the most complex
human-made networks. It comprises transmission lines and several
substations, that include generators -- the electricity power
source; transmission substations, which connect high-voltage
transmission lines; and load centers, which delivery the electricity
to consumers (\emph{e.g.}~\cite{albert2004svn}). Colloquially known as
power grid, the electric power transmission system has a complex
structure including redundant paths to route power from any
generator to any load center. The main reason for this redundancy is
to guarantee that every load center can be supplied by any
generator. In other words, if one generator fails, the load centers
will receive the necessary power from the other generators. In the
same way, if one transmission substation fails, the others have to
be able to handle the additional load and keep the whole network
working.  Nonetheless, even with the redundancy of lines, some
cascading failures and blackouts happen and several load centers
stop receiving power. One of the most serious was the Northeast
Blackout that affected 50 million people in the U.S.A and
Canada on August 14th, 2003, and resulted in a huge loss of money (at
about 30~billion US~dollars)~\cite{Bai06:CCSP}. Obviously, this kind
of network is crucial for the economy of a country and deserves
special attention in engineering and science.

The first studies in the power transmission systems relied on
creating simple dynamical models which simulate each component of
the network to understand the blackout dynamics of the whole
system~\cite{dobson2001imf, Carreras01:IEEE, Carreras02:Chaos,
  Carreras04:Chaos, carreras2004eso}. The networks were simple
structures such as rings, trees or mathematical grids, and the
blackouts were considered instantaneous events caused by cascading
failures of the transmission lines. When one transmission line
failed, all the power flow was redistributed to the other lines, but
new failures may happen due to overflow, leading to a cascading
effect. These dynamical processes were simulated
in~\cite{dobson2001imf, Carreras01:IEEE, Carreras02:Chaos,
  Carreras04:Chaos, carreras2004eso}, from which it was inferred that
the size of the blackouts follows a power-law tail, which means that
big blackouts are not so uncommon.

Although useful for predicting blackouts and finding the critical
components of the power transmission network, the approach above was
limited to simple network topologies which do not correspond to
those of the real power transmission networks. In fact, these
networks have the small-world property, high clustering coefficient
(\emph{e.g.}~\cite{Bai06:CCSP}), degree distribution in an exponential
form~\cite{albert2004svn, Amaral2000:pnas}, and the ``bow-tie''
configuration~\cite{chassin2005ena}. For a power network of 314,123
nodes, Chassin \emph{et al.}~\cite{chassin2005ena} considered any
vertex regardless of voltage (note that the size of the power
networks used in the other studies did not take into account nodes
of small voltages, see Table~\ref{tab:powergrid} for more power
network data), and showed that the network has a radial form. The
generators are in the center, with transmission substations in the
middle and the load centers at the border, in the so-called
``bow-tie'' configuration~\cite{Broder2000}. The degree distribution
was power law with a cutoff term.

\begin{table}
  \begin{center}
    \caption{Electric power grid network databases. Complete networks
      contain all substations regardless of voltage.}
    \label{tab:powergrid}
    \begin{tabular}{c|c|c}
      \textbf{Network} & \textbf{Size} & \textbf{Reference} \\
      \hline
      Western US power network & 4,941 & \cite{Watts98:Nature} \\
      US power network & 14,099 & \cite{albert2004svn} \\
      Complete US power network & 314,123 & \cite{chassin2005ena} \\
      Italian power network & 341 & \cite{crucitti2004tai} \\
      Chinese power network & 8,092 & \cite{sun2005cnt} \\
      \hline
    \end{tabular}
  \end{center}
\end{table}

In the analysis of the topology of these networks, it was shown that
the removal of highly connected nodes (without distinguishing their
types) can lead to blackouts of certain regions of the
networks~\cite{Motter02:PRE, Crucitti04:PRE, Kinney05:EPJB}. Each
substation was assumed to have a transmission capacity that depends
on the number of shortest paths passing through it. Albert \emph{et
al.}~\cite{albert2004svn}, however, showed that only the removal of
highly connected transmission substations can provoke blackouts,
since the other transmission substations can fail because of the
additional overload. This is not the case of generator failures,
because even the removal of the highly connected generator is unable
to cause blackouts~\cite{albert2004svn}. This is a consequence of
the redundancy of these networks where all generators can be routed
to all load centers. When one generator fails, the others provide
the additional power to supply the whole network.

A more realistic and complex model was developed by Anghel \emph{et
al.}~\cite{Anghel07:IEEE}. It describes an electric transmission
network under random perturbations (\emph{e.g.}\ line or substation
failures and overloads) and the operator's response to the
contingency events (a system to repair the damaged lines or
substations). In this model, each random event was stochastic and
could be triggered at any time. The model was able not only to
predict blackouts but also to find the optimal strategy for
minimizing the impact of random component failures.

Overall, the power transmission systems represent very large networks, with hundreds of
thousand vertices, and exhibit the small-world property, with high clustering coefficient
and the degree distribution in an exponential form and bow-tie configuration. Blackouts
are caused by removal of highly connected transmission substations as the other
substations cannot take the additional overload. The first models developed for
transmission power systems had simple topologies, aimed at simulating blackouts. Now, a
more realistic model exists which includes more sophisticated topologies and is based on
stochastic events.


\section{Biomolecular Networks}\label{Sec:biological}

In June 2000, the Human Genome Project and Celera Genomics decoded the
human genome, providing the so-called ``book of life'', after which
the genetic code of many organisms has been discovered. The Genome
project is only a starting point~\cite{Collins03:Nature}, as the
post-genomic era should be concerned with modeling biological
interactions instead of analyzing the genetic code by itself. As the
behavior of living systems can seldom be reduced to their molecular
components, in the post-genomic area of \emph{system
biology}~\cite{Kitano02:Science} one has to assemble the parts
unfolded by genome projects. For instance, Vogelstein, Lane and Levine
\cite{Vogelstein00:Nature} concluded that more significant results can
be obtained by analyzing connections of the p53 gene (a tumor
suppressor) than by studying this gene separately.

Processes in living organisms are basically divided and linked in
three levels of complexity~\cite{Oltvai02:Science}:
(i)~\emph{metabolic and signaling pathways}, which are determined by
(ii)~\emph{the network of interacting proteins}, whose production is
controlled by (iii)~\emph{the genetic Regulatory
network}. Understanding life processes therefore requires one to:
(a)~analyze how energy is obtained from cell biochemical reactions by
interactions among metabolites, products and enzymes; (b)~understand
how proteins take part in various processes, as in the formation of
protein complexes; (c)~understand how information is transmitted from
a transcription factor to the gene that regulates this transcription
\cite{Barabasi03:linked}.

Modeling these biological systems may be performed by considering
complex network theory, because of its universality and ability to
mimic systems of many interacting parts
\cite{Amaral04:EPJB}. Biological networks comprise metabolites,
enzymes, proteins, or genes as vertices, which are linked depending on
their interaction. Using complex network theory, fundamental
properties of biological networks have been discovered, including
power law connectivity distributions, small-world properties and
community structures \cite{Barabasi04:RNG}. Furthermore, the crucial
concepts to understand biological systems, namely emergence,
robustness and modularity, are also ingredients of complex networks
theory \cite{Amaral04:EPJB,Kitano02:Science}.

In the following, we shall discuss major developments in applying
complex network theory to biological systems.

\subsection{Protein-protein Interaction Networks}

In modeling protein interactions, one considers a network comprising
proteins (nodes) connected by physical interaction (undirected
edges). This requires reliable databases, which have become available
since the 1980s because of high throughput methods, and a set of
measurements to characterize the network structure and dynamics. The
connections are generally not weighted, but some databases provide
indexes of reliability associated with each link as the confidence
score~\cite{Krogan06:Nature}.

The study of protein interactions using complex networks theory may be
divided into four basic areas: (i)~characterization of the network
structure; (ii)~prediction of protein functions; (iii)~modeling of the
interactome and (iv)~modeling and characterization of protein-domain
interactions.

The structure of protein interaction networks has been studied to
determine the relative importance of vertices and their organization
in modular structures or subgraphs. Some of these structures are
conserved along the evolution of many organisms, which allows one to
infer their importance for cellular maintenance.  The structure of
protein interactions was believed to be completely random. However, in
2001 Jeong \emph{et al.}~\cite{Jeong01:Nature} discovered that these
networks were far from uniform, upon analyzing the \emph{S.
cerevisae} protein interaction networks. In fact, a few proteins are
able to physically attach to a huge number of other proteins, but most
proteins play a very specific role, interacting with just one or two
other proteins. The distribution of connections of Yeast follows a
power-law with an exponential cutoff whose exponent is about
$2.5$~\cite{Jeong01:Nature}. Other systems, such as the simple
bacterium \emph{H.  pylori}~\cite{Rain01:Nature} and the insect
\emph{D. melanogaster}~\cite{Giot03:Science}, also present a power-law
dependence for the connectivity in the protein-protein
interaction. Therefore, it seems that the scale-free nature of protein
interaction networks is a general feature of all organisms.

The protein interaction networks have also been shown to be
small-world, with a small average shortest path length, and to present
many loops of order three~\cite{Ravasz03:PRE}.  Their structure is
formed by modules of highly connected proteins
\cite{Palla05:Nature,Rives03:PNAS,Albert05:JCS}. The hubs possibly
play a critical function in the network maintenance, being strongly
related to lethality. They are believed to be the oldest proteins in
the network. Jeong \emph{et al.}~\cite{Jeong01:Nature} showed a
correlation between lethality and connectivity. While among the
proteins with five or fewer links, 93\% are non-lethal, among the
proteins with more than 15 links, 62\% are lethal. The highly uneven
structure of protein interaction networks imparts robustness against
random failures~\cite{Jeong01:Nature}. When random failures occur in
scale-free networks, there is only a small probability that the
removed node is a hub, and the network structure is not affected
severely~\cite{Albert:2000}. In contrast, when highly connected
proteins are removed, the network breaks up in several disconnected
components, which reinforces the relation between lethality and
connectivity. This explains why scale-free networks are very abundant
in nature, being associated with the evolution of cellular processes
and genes~\cite{Albert05:JCS}. In spite of the fact that the loss of
hub proteins can lead to death of the organism, some network-based
strategies to restore cellular functions caused by specific
mutations have been developed~\cite{Motter08:MSB}.

The importance of hubs was also addressed by Maslov and
Sneppen~\cite{Maslov02:Science} who showed that the functional modules
inside the cell are organized around individual hubs. In addition, the
proteins are uncorrelated according to their connectivity which
indicates that the neighborhood of highly connected proteins tends to
be sparser than the neighborhood of less connected proteins
\cite{Maslov02:Science}.  Schnell and Fortunato~\cite{Schnell2006}
investigated the relation between structural features of hubs and
their number of connections. They concluded that the disorder of a
protein (or of its neighbors) is independent of its number of
protein-protein interactions. The presence of hubs and the local
cohesiveness of networks are also related, as suggested by Ravasz
\emph{et al.}~\cite{Ravasz03:PRE}. The latter authors showed that the
relations between the scale-free architecture and the high degree of
clustering in protein networks are consequences of a hierarchical
organization, suggesting that small groups of nodes organize in a
hierarchical manner to make large groups.

The functional importance of proteins can also be addressed in terms
of network entropy~\cite{Manke05:GISWGI} and betweenness centrality
\cite{joy2005hbp}. Joy \emph{et
al.}~\cite{joy2005hbp} observed that proteins with high betweenness,
but low connectivity, are abundant in the Yeast interaction
network, with a positive correlation between the fraction of
lethal proteins and their betweenness centrality.

Some structural properties of protein interaction networks seem to be
conserved during evolution. Wuchty \emph{et al.}~\cite{Wuchty03:NG}
observed that some types of subgraphs referred to as network motifs
may be preserved during natural evolution. These motifs were found
more frequently in real networks than in their random versions,
generated by the rewiring method that maintains the network degree
distribution~\cite{Milo02:Science}. Wuchty \emph{et al.}\ analyzed the
conservation of $678$ proteins of Yeast with an ortologus in five
eukaryote species, namely \emph{Arabidopsis thalianam},
\emph{Caenorhabditis elegans}, \emph{Drosophila melanogaster},
\emph{Mus musculus} and \emph{Homo sapiens}. They observed that some
motifs are conserved from simple organisms to more complex
ones~\cite{Wuchty03:NG}. Each of the motifs conserved was suggested to
perform specific roles, \emph{e.g.}\ in forming protein complexes, where
smaller parts are represented by fully connected $n$-node
motifs. Because some biological functions emerge from modules of many
connected proteins~\cite{Hartwell99:Nature,Oltvai02:Science}, instead
of single proteins, such structures are conserved throughout the
natural selection. In fact, as the cell is organized in modules, the
structures should be conserved along evolution, and not only the
individual cellular components
\cite{Hartwell99:Nature,poyatos2004bri}.  The conservation of network
motifs can be used to predict protein interactions. Using machine
learning algorithms, Albert and Albert~\cite{Albert04:Bioinformatics}
showed that conserved properties of protein networks can be used to
identify and validate protein interactions.

The Baker's Yeast (\emph{Saccharomyces cerevisiae}) has about 6,300
genes, which encode about the same number of
proteins~\cite{Williams96:Science}. Therefore, the determination of
interaction among such proteins requires checking about 6,300 times
6,300 pairs, which is close to forty million potential
interactions. In 1989, Stanley Fields proposed a revolutionary
technique to detect protein-protein interactions using the GAL4
transcriptional activator of Yeast \emph{Saccharomyces
cerevisiae}~\cite{Fields89:Nature}. This ``Yeast two-hybrid''(Y2H)
method~\cite{Bartel97,Ito01:PNAS} is based on the fact that a protein
with DNA-binding domain may activate transcription when it simply
binds to another protein containing an activation domain.  Basically,
Y2H detects interaction using two hybrid proteins referred to as the
``bait'' and the ``prey''.  The ``bait'' contains a query protein X
fused to a DNA binding domain and the ``prey'' is a fusion of a second
protein Y to a transcriptional activation domain. If the proteins X
and Y interact, then their DNA-binding domain and activation domain
will combine to form a functional transcriptional activator, which
will proceed to transcribe the reporter gene that is paired with its
promoter~\cite{Legrain00:FEBS,Mukherjee01:CS}. Further information
about this method and its extensions may be found
in~\cite{Mukherjee01:CS} and~\cite{Lesk2001}.

Y2H allowed the global analysis of protein-protein interactions
and the birth of \emph{interactome} (one of the next steps after the
genome)~\cite{Eisenberg00:Nature,Xenarios01}. The success of this
method lies on the identification of interactions without antibodies
or the need to purify proteins~\cite{Fields89:Nature}.  However, the
main drawback is that Y2H generates many false positives, \emph{i.e.}
interactions identified in the experiment but that never take place in
the cell~\cite{Mrowka01,Saito02:NAR}. This limitation has motivated
the development of other methods, including biochemical
techniques such as affinity~\cite{Krogan06:Nature} and molecular
size-based chromatography, affinity blotting, immuno-precipitation and
cross-linking~\cite{Bartel93:Biotech, Fields94:TG, Eisenberg00:Nature,
  Uetz00:Nature}. Also, computational methods are applied to identify
protein-protein interactions~\cite{vonmering02:Nature}, such as those
based on the genome sequence~\cite{Marcotte99:Science}. Sprinzak
\emph{et al.}\ proposed a method to asses the quality of protein
interaction databases using the cellular localization and
cellular-role properties to provide an estimative of true positives in
databases \cite{Sprinzak03:JMB}.  They suggested that the reliability
of the high-throughput Y2H is about 50\% and that the Yeast
interactome should be composed of $10,000$--$16,600$
interactions. Recently, Krogan \emph{et al.}~\cite{Krogan06:Nature}
made available a database of protein interactions based on tandem
affinity purification which identified $2,708$ proteins and $7,123$
interactions of the \emph{Saccharomyces cerevisiae}. This database
offers greater coverage and accuracy than the previous high throughput
studies related to Yeast protein-protein interaction. In
Table~\ref{Table:dataprot} the main protein-protein interaction
databases and their web addresses are presented.

\begin{table}[ht]
  \begin{center}
    \caption{Public databases of protein-protein interactions.}
      \begin{tabular}{c|c}
        \textbf{Database} & \textbf{URL}\\
        \hline
        DIP & \texttt{http://dip.doe-mbi.ucla.edu} \\
        IntAct & \texttt{http://www.ebi.ac.uk/intact} \\
        HPID & \texttt{http://wilab.inha.ac.kr/hpid} \\
        MIPS & \texttt{http://mips.gsf.de/services/ppi} \\
        Biogrid & \texttt{http://www.thebiogrid.org} \\
        BIND & \texttt{http://bind.ca} \\
        CYGD & \texttt{http://mips.gsf.de/proj/yeast/CYGD/interaction} \\
        iHOP & \texttt{http://www.ihop-net.org/UniPub/iHOP} \\
        JCB & \texttt{http://www.imb-jena.de/jcb/ppi} \\
        MINT & \texttt{http://mint.bio.uniroma2.it/mint} \\
        PathCalling & \texttt{http://curatools.curagen.com/pathcalling\_portal} \\
        String & \texttt{http://string.embl.de} \\
        InterDom & \texttt{http://interdom.lit.org.sg} \\
        \hline
      \end{tabular}\label{Table:dataprot}
  \end{center}
\end{table}

Assigning functions to unknown proteins is one of the most important
problems in the post-genomic era. This may be performed by genome
analysis methods that exploit domain fusion~\cite{Marcotte99:Science}
and phylogenic profiles~\cite{Pellegrini99:PNAS}. In the first method
one observes that pairs of interacting proteins have homologs in
higher species fused in a single protein chain. The second, on the
other hand, considers that proteins participating in structural
processes or metabolic pathways are functionally linked, evolving in a
correlated fashion. Other methods are based on interaction partners,
because proteins with the same functions tend to share
connections~\cite{Hishigaki01:Yeast}. This property arises from the
duplication and mutation mechanisms of protein evolution. When a
protein is duplicated, the daughter protein has the same features as
the mother protein. However, as the daughters suffer mutations, they
tend to differ in structure, but preserve similar functions and
connections. Another common approach for predicting protein function
is the majority rule assignment, which takes into account the
empirical finding that 70--80\% of protein interaction partners share
at least one function~\cite{Schwikowski00:NB}. Hishigaki \emph{et
  al.}~\cite{Hishigaki01:Yeast} proposed a methodology based on
analysis of $n$-neighborhood proteins, \emph{i.e.} for a given protein $i$,
the $n$-neighborhood is composed by proteins distant $n$ edges from
$i$. With this methodology, one can predict with 72.2\% accuracy the
subcellular localization, with 63.6\% accuracy the cellular role and
with 52.7\% accuracy the biochemical function of Yeast
proteins. However, such methods do not consider the links between
unknown proteins. Vazquez \emph{et al.}~\cite{Vazquez03:NB} extended
this method by minimizing the number of physical interactions among
different functional categories and considering the connections
between unknown proteins, thus achieving a more precise
identification of protein function.

Despite the many approaches to determine protein function using the
complex network of interactions, there are many limitations owing to
the large number of false positives and negatives in the interactions.
Thus, some connections between proteins may not occur in vivo and the
function associated with such interactions may not be real. Another
problem is related to multi-functional proteins. In this case, it may
be hard to determine the function of the proteins when their neighbors
have many functions.

Simple protein interaction models have been proposed using network
measurements~\cite{takemoto2005enm}. Eisenberg and
Levanon~\cite{Eisenberg03:PRL} proposed that the BA model is
relevant for protein interaction modeling. They applied a
cross-genome comparison and observed a correlation between proteins
age and their connectivity.  Then, they suggested that the protein
evolution is governed by the preferential attachment rule.
Pr\v{z}ulj \emph{et al.}~\cite{Prvzulj04:Bioinformatics}, on the
other hand, suggested that the protein evolution is governed by a
geometrical model and the scale-free degree distribution may be
caused by the high percentage of false negatives. The geometric
model is obtained by distributing $N$ proteins randomly in a metric
space and connecting two proteins according to the distance between
them. These results were later confirmed by Costa \emph{et
al.}~\cite{Costa07:AP}.

Even though these models reproduce some protein interaction
properties, other biological processes must be considered in
modeling. The modeling may be performed with addition and elimination of interactions between proteins, and gene duplication
increasing the number of proteins and interactions
\cite{Wagner01:MBE}. V\'{a}zquez \emph{et
al.}~\cite{Vazquez03:complexus} proposed a model in which each node in
the network represents a protein that is expressed by a gene, and the
network evolves following two steps: (i)~\emph{duplication}: a
randomly selected node $i$ is duplicated as $i'$ which is connected to
$i$ according to a probability $p$ and has all connections of $i$,
(ii)~\emph{divergence}: a node $j$ connected to $i$ and $i'$ loses the
connection $(i,j)$ or $(i',j)$ according to a probability $q$.
Therefore, this mode has two parameters $p$ and $q$, representing the
creation of a connection by node duplication and the loss of
interactions, respectively.  Sol\'{e} \emph{et al.}\ proposed a
similar model with the same concepts~\cite{Sole02:ACS}, which is given
in detail in~\cite{Pastor-Satorras02:JTB}. Sol\'{e} and Fern\'{a}ndez
showed that the models by V\'{a}zquez \emph{et
al.}~\cite{Vazquez03:complexus} and Sol\'{e} \emph{et
al.}~\cite{Sole02:ACS} reproduce the modular structure and the degree
correlation observed in protein interaction
networks~\cite{Sole03:condmat}.  In case of degree correlations, hubs tend to connect
with poorly connected proteins. Also, the modular scale-free structure
in protein networks emerge naturally from the duplication-divergence
rule~\cite{Sole03:condmat} or optimization
principles~\cite{pan2007mne}.

The rate of duplication and deleting of interactions was estimated by
Wagner~\cite{Wagner01:MBE} who concluded that every 300 million of
years, around half of all existing protein interactions are replaced by new
interactions. Wagner~\cite{Wagner03:PBS} also estimated the rate of
link dynamics and gene duplication using empirical data, showing that
the rate of link dynamics is at least one order of magnitude higher
than gene duplication. With this observation, Berg \emph{et
al.}~\cite{Berg04:BMC} proposed a model of protein interaction with link attachment and link detachment together with gene
duplication. In this case, the link dynamics is guided by a
preferential attachment rule supported by empirical data. Such a model
reproduces the scale-free degree distribution and the correlation
between interaction proteins, as observed
experimentally~\cite{Berg04:BMC}.

As domains can recombine to form multi-domain proteins, the domain
recombination may be the main mechanism to modify protein function and
increase the proteome complexity~\cite{vogel2005rbd}. Protein
interactions can be validated by domain-domain
interactions~\cite{Ng03:Bioinformatics} and proteins with similar
activities are likely to contain similar
domains~\cite{Betel:2004}. Therefore, the characterization of
domain-domain interactions has many applications and is crucial to
understand the evolution of protein interactions.

Domain-domain interaction networks are built from
(i)~protein complexes, (ii)~Rosetta Stone sequences, and by using
(iii)~protein interaction
networks~\cite{Wuchty01:MBE,Wuchty02:Prot,Ng03:Bioinformatics}.  With
the first methodology, domain information is inferred from the
intermolecular relationships in protein complexes. The second
considers domain fusion in different organisms, \emph{i.e.} domains that
appear separately in one organism and together in another one are
potentially interacting. Finally, the last approach considers the
protein interaction to determine domain interaction. In this case, all
domains belonging to two interacting proteins are also interacting. Ng
\emph{et al.}~\cite{Ng03:Bioinformatics} suggested a measure to
determine the potential of interaction between domains considering the
protein interaction network.

Wuchty~\cite{Wuchty01:MBE} studied domain interaction networks with
methodologies (i) and (ii). In both cases, he showed that protein
domain interactions follow a power law and exhibit the small-world
property with a high average clustering coefficient. In a subsequent
paper, Wuchty~\cite{Wuchty02:Prot} also investigated which factors
force domains to accumulate links to other domains. As the network of
domain interactions consists of hubs responsible for network
integrity, it is natural to investigate the importance of the most
connected domains. Costa \emph{et al.}~\cite{CostaRodrigues06:APL}
analyzed the structure of domain interactions and the relation between
protein connectivity and essentiality. In this case, the essentiality
of a domain is hypothesized in two ways: (i)~\emph{domain lethality in
a weak sense}: a domain is lethal if it appears in a lethal protein
and (ii)~\emph{domain lethality in a strong sense}: a domain is lethal
if it only appears in a single-domain lethal protein. Costa \emph{et
al.}\ showed that correlations between domains degree and lethality in
both the weak and strong senses are significantly higher than the
correlation obtained for proteins, which shows the importance of
domains in defining protein interaction and protein
lethality~\cite{CostaRodrigues06:APL}.

Complex networks theory has also been used to analyze protein
3D structures resulting from linear chains of amino acids, which is
a classical example of biological self-organization. Bagler and
Sinha~\cite{bagler2007amp} studied the role of topological
parameters in the kinetics of protein folding in two length
scales -- Protein Contact Networks (PCNs) and their corresponding
Long-range Interaction Networks (LINs), which are constructed by
ignoring the short-range interactions. They found that both PCNs
and LINs exhibit assortative mixing, which have been absent in most
biological networks, as discussed before. Besides, while the
assortativity coefficient provides a positive correlation with the
rate of protein folding at both short- and long-contact scale, the
clustering coefficients of the LINs exhibit a negative correlation.

\subsection{Metabolic Networks}

Metabolism is the complete set of chemical reactions that occur in living cells, allowing
cells to grow and reproduce, maintain their structures, and respond to their
environments. The essential energy and substrates for the function of a cell are obtained
by breaking large molecules into small ones. Hence, this process is the basis of life.
Metabolic networks are related to the chemical reactions organized into metabolic
pathways, in which one chemical compound is transformed into another by the action of a
sequence of enzymes. Basically, the reaction networks can have three possible
representations: (i) directed and weighted graphs, whose vertices can be of three types:
metabolites, reactions and enzymes, while there are two types of edges representing mass
flow (from reactants to reactions) and catalytic reactions (from enzymes to reactions),
(ii) connecting all reactants that participate in a same reaction, (iii) connecting two
reactions if they share a reactant. Major catalogues available on-line are the Kyoto
Encyclopedia of Genes and Genomes (KEGG, \texttt{http://www.genome.jp/kegg}) and the
EcoCyc~(\texttt{http://ecocyc.org/}).

Jeong \emph{et al.}~\cite{Jeong2000lso} investigated metabolic
networks of 43 organisms from all three domains of life, and found
that metabolic organization is not random, but follows the power law
degree distribution ---the probability that a given substrate
participates in $k$ reactions follows a power law, $P(k)\sim
k^{-\gamma}$, with $\gamma\simeq 2.2$ in all 43 organisms. Besides,
metabolic networks are small-world ($\ell \approx 3.2$), where two
metabolites can be connected by a small path. For instance, Fell and
Wagner~\cite{Wagner2001swi} showed that the center of \emph{E. coli}
metabolism map is glutamate and pyruvate, with a mean path length to
other metabolites of 2.46 and 2.59, respectively. An important
property discovered by Jeong \emph{et al.} is that the diameter of
metabolic networks is almost the same for all the 43 organisms. This
property is quite different from that for other types of network,
where the diameters increase logarithmically with the addition of new
vertices. A possible explanation is that as complexity of organisms
grows, individual substrates get more connections to maintain a
relative constant network diameter. Since metabolic networks are
scale-free, few hubs concentrate a high number of
connections~\cite{Wagner2001swi}. Such networks are thus tolerant to
random failures, but vulnerable to attacks in which there is a
sequential removal of nodes starting with the most connected one.  This procedure increases the network diameter and quickly breaks the
network into disconnected components. This expectation has been
confirmed experimentally in \emph{Escherichia coli}, where mutagenesis
studies performed \emph{in-silico} and \emph{in-vivo} showed that the
metabolic network is highly tolerant upon removal of a considerable
number of enzymes~\cite{edwards2000ecm}. Another feature of hubs is
their preservation along evolution. Only 4\% of all substrates
encountered in the 43 organisms are present in all species, which can
be considered generic as they are utilized by many
species~\cite{Jeong2000lso}. Such substrates are the hubs in metabolic
networks.

The structures of a metabolic network are organized in a modular,
hierarchical fashion~\cite{schuster2000gdm}. The modules are discrete
entities composed by several metabolic substrates densely connected by
biochemical reactions. Ravasz~\emph{et al.}~\cite{ravasz2002hom}
showed that the average clustering coefficient of the metabolic
networks of 43 organisms is independent of network size, being higher
than for scale-free networks generated by the Barab\'{a}si-Albert
model of the same size. In addition, the clustering
coefficient follows a scaling law with the number of links, $c(k)\sim
k^{-1}$, which indicates a hierarchical organization. Hence, metabolic
networks are characterized by a scale-free degree distribution,
average clustering coefficient independent of network size, and
hierarchical, modular organization. Ravasz~\emph{et al.} suggested
a model of metabolic organization that reproduces these properties. It
is built starting with a small cluster of four densely connected
nodes, followed by generation of three copies of this structure with
the three external vertices being connected to the central vertices of
the old clusters. This process is repeated until a network with the
same size as the real network is obtained. The networks thus generated
display the properties of metabolic networks discussed above.

Guimer\`{a} and Amaral~\cite{Guimera05:Nature} proposed a methodology to
find functional modules in complex networks and classify vertices
according to their patterns in intra- and inter-module
connections. They found that 80\% of substrates are only connected to
substrates within their modules. Also, substrates with
different roles are affected by different evolutionary constraints and
pressures. In contrast to the result by Jeong \emph{et
al.}~\cite{Jeong2000lso}, Guimer\`{a} and Amaral showed that
metabolites participating in a few reactions but connecting different
modules are more conserved during evolution than the most connected
substrates (hubs).

\subsection{Genetic Networks}

Living cells are governed by gene expression programs involving
regulated transcription of thousands of genes. Cell signaling and
differentiation can be investigated by pattern of gene expression,
which can be represented by complex networks~\cite{diambra2005cna}.
Transcription is controlled at many levels and the gene
regulation network fits into a network of networks that represent not
only interactions among transcription factors but also the factors
that modulate these interactions
biochemically~\cite{Alon2006isb}. High-throughput methods and
computational approaches have allowed important discoveries in genetic
networks. In order to analyze the topology and dynamics of genetic
networks, approaches were developed to evaluate identification and
expression level of interacting genes, how interactions change with
time and the phenotypic impact when key genes are disrupted. The
techniques aimed at elucidating transcriptional regulatory networks
are mainly based on genome-wide expression profiling and the
combination of chromatin immunoprecipitation (ChIP), which was discussed
in~\cite{Blais2005ctr,Siggia2005cmt}.

Transcriptional regulatory networks control gene expression in
cells and are composed of regulators and targets (nodes) and
regulatory interactions (edges). Edges are directed from a gene that
encodes a transcription factor protein (TF) to a gene
transcriptionally regulated by that transcriptional factor, referred
to as target gene (TG)~\cite{Shenorr2002nmt}. Such complex systems can
be analyzed as a multilayered system divided into four basic
levels~\cite{Babu2004sae}: (i)~the first level encompasses the
collection of transcriptional factors, its target genes with DNA
recognition site and regulatory interaction between them; (ii)~the
second level involves regulatory motifs, which are
patterns of interconnections that appear more frequently in real
networks than in randomized networks~\cite{Milo02:Science}; (iii)~the
third level considers the modular organization, where modules are
group of nodes that regulate distinct cellular processes; and (iv)~the
last level consists of the regulatory network composed by the whole
set of modules.

\begin{figure}
  \centerline{\includegraphics[width=0.5\linewidth]{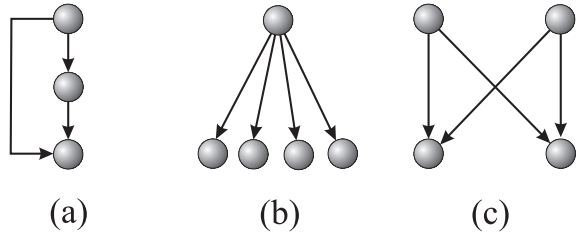}}
  \caption{Three types of motif found in transcriptional regulatory
    networks of \emph{Saccharomyces cerevisiae} and \emph{Escherichia
      coli}: (a) feed-forward loop, (b) simple input module and (c)
    multiple input.}
  \label{Fig:typemotifs}
\end{figure}

Because network motifs appear at frequencies much higher than in
random networks, they are expected to have special functions in
information processing performed by the
network~\cite{balcan2007icy}. The complexity of networks can be
reduced by considering their motifs. Figure~\ref{Fig:typemotifs}
presents the three main types of motifs found in regulatory networks
in the Yeast of \emph{Saccharomyces cerevisiae} and the bacteria
\emph{Escherichia coli}. In transcriptional regulatory networks, the
feed-forward motifs are defined by a transcription factor that
regulates a second one such that both, jointly, regulate a final
target gene. On the other hand, the simple input module and the
multiple input regulate their targets by a simple or multiple
transcriptions factors, respectively. All these targets are
controlled by the same sign (all positive or all negative) and have
no additional transcriptional regulation \cite{Shenorr2002nmt}.
Other important motifs are found in the transcriptional regulatory
networks of \emph{Saccharomyces cerevisiae}, for which a detailed
discussion is presented in~\cite{Lee2002trn}.

Modules are organized by interconnections of network motifs. In general, the modules
refer to a group of physically or functionally linked genes. Distinct cellular processes
are regulated by discrete, separable modules~\cite{Guelzim2002tac}. Ma \emph{et
al.}~\cite{Ma2004hsa} identified 39 modules in transcriptional regulatory networks of
\emph{E. coli}, which showed well-defined functions. Also, Dobrin \emph{et
al.}~\cite{dobrin2004atm} showed that in \emph{E. coli} multiple input and feed-forward
loops motifs do not exist in isolation, but share transcription factors or target genes.
A review about the modular topology in transcriptional regulatory networks is presented
in~\cite{Babu2004sae}.

The structure of transcriptional regulatory networks exhibits two main
properties: (i)~the fraction of genes with a given incoming
connectivity decreases exponentially, and (ii)~the outgoing
connectivity distribution follows a power
law~\cite{Guelzim2002tac}. The exponential character of the incoming
connectivity indicates that most target genes are regulated by a
similar number of factors. On the other hand, the scale-free
distribution of the outgoing connections points to a few transcription
factors participating in regulation of a large number of target
genes. In addition, such genes tend to be lethal if
removed~\cite{yu2004gae}. The structure of transcriptional regulatory
networks changes with time or environmental conditions, because not
all nodes are active at a given time. Dynamical processes can be
analyzed in genetic networks, as in Li \emph{et al.}~\cite{li2004ycc}
who observed that the cell-cycle of the regulatory network in Yeast
was extremely stable and robust. Klemn and
Bornholdt~\cite{klemm2005tbn} investigated the reliability of gene
regulatory networks and found a distinction between reliable and
unreliable dynamical attractors, showing that biological signaling
networks are shaped by selective advantage of the ability of robust
signaling processing. Other dynamical processes investigated in
genetic networks are discussed in~\cite{Bornholdt2005lmm}.

The evolution of transcriptional regulatory networks occurs via
three processes: (i)~duplication of the transcription factor, which
results in both copies regulating the same gene, (ii)~duplication of
the target gene with its regulatory region, where the target gene
will be regulated by the same transcription factor, and
(iii)~duplication of both the factor and the
target~\cite{Babu2004sae}. Such a mechanism of duplication and
inheritance determines a large fraction of the interactions in
regulatory networks. Furthermore, a large number of interactions are
gained or lost after gene duplication. This process can result in
network motifs and specific connectivity distribution in the
network. Teichman and Babu showed that the evolution of regulatory
networks is mainly defined by gene
duplication~\cite{teichmann2004grn} and that 45\% of regulatory
interactions in \emph{E. coli} and \emph{S. cerevisiae} arose from
duplication with inheritance of interactions. Balcan \emph{et
al.}~\cite{balcan2007icy} proposed an information theoretical approach
for modeling transcriptional regulatory networks based on statistics
of the occurrence of binding sequences of given specificity in
random promoter regions and a ``sequence-matching'' rule. They
showed that the degree distributions, clustering coefficient, degree
correlations, rich-club coefficient and the k-core structure
obtained from the model agree with the measurements
in the real network of \emph{S. cerevisiae}.

Though the structure of biological networks has been extensively studied, the
understanding of temporal activity of a cell is still embryonic. Databases with time
evolution of gene expression activity are being
developed~\cite{holter2000fpu,alter2000svd}. When data on cell network processing and
intercommunication become available, new insights about the mechanisms of life will be
possible.


\section{Medicine}

Complex networks have been useful in medicine, as networks
representing biomolecular systems can be applied to understand disease
principles and spreading~\cite{Lee08:PNAS}.
Barab\'asi~\cite{Barabasi2007nmo} suggested the treatment of diseases
by taking into account different levels of complex biological networks.
The basic level considers a complex network connecting cellular
components (\emph{e.g.}\ metabolic, protein-protein interaction and genetic
networks), the intermediate level represents the networks of diseases (two
diseases are connected if they have a common genetic or functional
origin), and the top level represents the society, composed by human
interactions, such as friendship, geographical localization, and so on.  In
case of obesity, the analysis should consider biomolecular networks,
because some genes or metabolic disfunctions may be of fundamental
importance. At the network disease level, obesity is related to other
diseases, such as diabetes, asthma and insulin resistance. Finally, at
the top level, the social interactions contribute to spreading of
eating habits and lack of physical activities.

In addition to disease origin, complex networks can be useful in epidemiology. Indeed,
the spreading of infections caused by virus or bacteria has been studied by dynamical
analysis in networks. Studies on spreading virus are discussed in
Section~\ref{Sec:spreading}.

Complex networks theory is potentially suitable to be applied in
anatomy. For instance, the bone structure in mammals are composed by a
complex network of channels (Havers and Volkmann channels), which are required to nourish the bone
marrow cells. The vertices represent intersection of two or
more channels and these interconnecting channels are expressed as
edges~\cite{dafontouracosta2006ccn}. Costa \emph{et
al}.~\cite{dafontouracosta2005hfs} characterized a network
from a cortical bone structure in terms of complex networks
measurements. Three important findings were reported: (i) that the
channel branching density resembles a power law implies the existence
of distribution hubs; (ii) the conditional node degree density
indicates a clear tendency of connection between nodes with degrees 2
and 4; and (iii) the application of the
hierarchical clustering coefficient allows identification of
typical scales of channel redistribution. The biological implications of such findings were also discussed.


\section{Ecology}

The characterization, modeling and simulation of ecological systems are major challenges
of natural sciences today. In ecological communities, each species interacts in different
ways, forming networks of interacting species.  Important ecological relations are
competition, parasitism, mutualism and predatory-prey relationships. Food webs are
composed by $S$ trophic species connected directly by $L$ trophic
links~\cite{Drossel03:HGN}. The links are represented by arrows from $i$ to $j$ if
species $i$ is eaten by species $j$. Such rows indicate the flow of resources. Trophic
species is a collective designation for all species having a common set of predator and
prey. Thus, species such as sponges and clams may belong to the same trophic species. The
use of terms such as ``detritus'' and ``dead organic material'' indicates how hard it is
to decide on what can be included or omitted in the food web chain. Before the 1990s,
food webs were very small, with sizes varying from 5 to 48~\cite{cohen1990cfw}. The
analysis of such networks suggested that the ratio $L/S$ was independent of $S$ and
therefore this relation was scale-invariant. However, further works showed that this
belief is not correct~\cite{murtaugh1997vtf}. Food webs are small-world networks, obeying
the two degree of separation rule~\cite{williams2002tds}. This property implies that the
loss of biodiversity and species invasions may affect ecosystems strongly. Garlaschelli
\emph{et al.} showed that differently from biomolecular networks, food webs are not
scale-free and do not present high average clustering
coefficient~\cite{Ggarlaschelli2003usr,Garlaschelli2004ufw}. The modeling of food webs
can be divided into three basic levels: static models, dynamical models, and species
assembly and evolutionary models, which are discussed in the review by Drossel and
McKane~\cite{Drossel03:HGN}.  Ecological relationships were also explored by considering
dynamical processes, such as metapopulation dynamics~\cite{hanski1998md, vuorinen2004nmd}
and epidemic spreading~\cite{guimaraesjr2007vkw}. In addition, food webs have been
studied in terms of robustness~\cite{Camacho2002rpf} using concepts of scaling and
universality. This particular dynamics is important to quantify species extinction in
face of habitat modification and global warming~\cite{Wilmers2007uer}. Using the
May-Wigner stability criterion~\cite{Sinha2005cvs},Sinha and Sinha~\cite{Sinha2005eum}
showed that increasing complexity in terms of size and connectivity results in great
instability, leading to extinction of more species. Therefore, previously unconnected
ecosystems that start to interact by anthropogenic or natural means are highly vulnerable
to large losses in species.

In addition to predatory-prey relationships, species can interact
according to mutualism --- plants and animals establish beneficial
interactions, where both individuals derive a fitness
benefit~\cite{boucher1988bme}. A classical example of mutualism
involves unicellular algae and corals, where the coral provides the
algae with shelter and inorganic nutrients, while the pigmented algae
provide photosynthesis.  Mutualism is fundamental in ecology and
evolutionary biology, driving the evolution of much of the biological
diversity~\cite{rezende2007nrc}. The structure of networks
representing mutualistic relationship are different from those
obtained for antagonistic interactions (predator-prey,
herbivore-plant). While antagonistic networks are highly
compartmentalized~\cite{guimaraes2006asa}, mutualistic networks are
often nested, characterized by species with many interactions forming
a kernel, and species with few interactions, which interact just with
the highly connected species. This kind of structure was observed in
mutualistic relationship between animals and plants (\emph{e.g.}\ flower and
bees)~\cite{bascompte2003nap} and cleaner and client species (cleaning
networks)~\cite{guimaraes2007nsm}, as the cleaning networks are
more asymmetric than plant-animal
networks~\cite{guimaraes2007nsm}. The nested structure of mutualistic
networks is related to phylogenetic relationships
\cite{rezende2007nrc}. Indeed, phylogenetic related species tend to
have similar biological roles in mutualistic networks~\cite{rezende2007nrc}.


\section{Neuroscience}

Complex networks theory is a useful framework for the study of large
scale brain networks. The network models, topological measurements,
and the dynamical analysis (\emph{e.g.}\
avalanches~\cite{Fontouracosta2008aaa,beggs2003nan}) can be considered
for brain studies. Representing brain connections with networks
is useful to study brain diseases, which are related to network
attacks and failures, and brain functions, where it is possible to
associated a particular brain architecture with specific brain
functions.

Brain networks can be investigated at many scales, ranging
from individual neurons to large brain
areas~\cite{sporns2002nac,dafontouracosta2005mcn}. In the former case,
neurons are connected by synapses. In the latter, regions are
linked by pathways or functions. Brain networks are generally directed
and unweighted, though the description of neuronal systems may be
enhanced by incorporating weighted networks.

An important question in neuroscience regards the relationship between
brain function and the structure of neural connectivity. The brain is
composed of a complex interconnection system whose organization is
aimed at optimizing resource allocation and minimizing
constraints~\cite{koch1999can}. A limitation of studies in brain
connectivity is the proper choice and integration of the spatial and
temporal resolutions, because a given dataset may reflect individual
neuron activities, neural assemblies dynamics and other macroscopic
brain activities~\cite{hamalainen1993mti,logothetis2001nib}. The main
types of connectivity are: (i)~\emph{anatomical connectivity}, related
to the connections between two brain sites; (ii)~\emph{functional
connectivity}, which is defined as the temporal correlation between
spatially distant neurophysiological events; and (iii)~\emph{effective
connectivity}, which is a more abstract approach defined as the
influence that a neural system may have over another in a direct or
indirect way.

Anatomical connectivity data can be obtained at different scales, from neurons to brain
regions. This analysis can be considered static when the time of analysis varies from
seconds to minutes, but can be dynamic for large time scales, such as hours and days,
because of learning or development activities. At a more microscopic spatial scale, the
connectivity of the brain is related to the geometry and spatial position of the
individual neuronal cells (\emph{e.g.}\ \cite{dafontouracosta2005tmc, costa2003mhn,
dafontouracosta2005mcn}). For instance, networks involving more intricate neurons are
likely to imply enhanced connectivity, therefore affecting the corresponding dynamics
(\emph{e.g.}\ \cite{CostaBarbosa04}). An interesting open question is to predict connectivity
and dynamics of neuronal networks from the geometric features of specific types of
neurons~\cite{CostaBarbosa04, dafontouracosta2005mcn}, for the form of neurons is highly
correlated to their functions~\cite{CostaBarbosa04}. One of the first studies using graph
theory to represent brain connections was performed by Felleman and Van
Essen~\cite{felleman1991dhp} in 1991. They mapped the hierarchical organization of the
cerebral cortex of primates, reporting 305 connections among 32 visual and
visual-association areas. Watts and Strogatz in 1998 analyzed neuroanatomical networks
for the nervous systems of \emph{C. elegans}~\cite{Watts98:Nature}. This network, formed
by $N=282$ neurons and $M=3,\!948$ synaptic connections, revealed a small-world
characteristic structure, but not a scale-free distribution of links. The topology of
such network is suitably represented by the Watts-Strogatz small-world model. Similar
results were obtained by Hilgetag~\cite{hilgetag2000acd}, who studied compilations of
corticocortical connections data from macaque -- the whole cortex, the visual and the
somatosensory -- and cat. Other studies suggested that the large scale organizations of
the brain cortex of rat~\cite{burns2000aco}, cat~\cite{scannell1995acc} and
monkey~\cite{felleman1991dhp} are neither regularly nor randomly connected. Most cortical
networks have a multi-cluster structure, presenting the small-world property (\emph{e.g.}\ in
\emph{C. elegans} $\langle c \rangle = 0.28$ and $\ell = 2.65$, in macaque visual cortex
$\langle c \rangle = 0.59$ and $\ell = 1.69$, in macaque cortex $\langle c \rangle =
0.49$ and $\ell = 2.18$, and in cat cortex $\langle c\rangle = 0.60$ and $\ell = 1.79$).
Therefore, the neuronal and cortical connectivity obtained are small-world networks. This
particular topology might be chosen by selection pressure to minimize wiring costs. The
small-world structure allows the network to have a modular organization and connections
between such modules by adding several long-distance connections~\cite{karbowski2001owp,
hilgetag2000usa}. Costa and Sporns~\cite{dafcosta2005hfl} characterized cortical networks
with hierarchical measurements and identified principles for structural organization of
networks. In addition, Costa \emph{et al.}~\cite{dafcosta2007pcp} proposed a
computational reconstruction approach to the problem of network organization in cortical
networks and showed that the organization of cortical networks is not entirely determined
by spatial constraints. Cortical areas have also been analyzed dynamically. Costa and
Sporns~\cite{dacosta2007dcs} applied Metropolis dynamics on four configurations of the
cat thalamocortical systems, \emph{i.e.} (i) only cortical regions and connections; (ii) the
entire thalamocortical system; (iii) cortical regions and connections with the thalamic
connections rewired so as to maintain the statistics of node degree and node degree
correlations; and (iv) cortical regions and connections with attenuated weights of the
connections between cortical and thalamic nodes. They identified substructures determined
by correlations between the activity of adjacent regions when only cortical regions and
connections were taken into account. In addition, two large groups of nearly homogenous
opposite activation were observed in cases (ii) and (iii). The effects from uniform
random walks on the dynamic interactions between cortical areas in the cat brain
thalamocortical connections were investigated, from which such connections were found to
be organized to guarantee strong correlation between the out-degree and occupancy
rate~\cite{dafontouracosta2006ctc}.

The small-world topology has a direct influence in the dynamical complexity of the
network. As discussed by Barahona and Pecora~\cite{barahona2002ssw} and Hong and
Choi~\cite{hong2002ssw}, information propagates faster on many small-world networks of
undirect uniformly coupled identical oscillators. Thus, the topology of neuroanatomical
networks provides a better propagation of activities than regular or equivalent random
networks.  Also, Lago-Fern\'{a}ndez \emph{et al}.~\cite{lagofernandez2000fra} showed that
nonidentical Hodgkin-Huxley neurons coupled by excitatory synapses present coherent
oscillations in regular networks, fast system response in random networks, and both
advantages in small-world networks. On the other hand, Percha \emph{et
al.}~\cite{Percha2005tlg} showed that small-world neural networks suffer a transition
from local to global phase synchrony depending on the rewiring parameter for network
construction ($p\approx 0.3-0.4$~\cite{Percha2005tlg}). Indeed, the authors suggest that
neural systems may form networks whose structures lie in the critical regime between
local and global synchrony. In this way, the appearance of connections in injured regions
may lead to the onset of epileptic seizures. Sinha \emph{et al.}~\cite{Sinha2007ess}
showed that the different ratios of long-range connections in small-world networks can
result in strikingly different patterns. Indeed, recent studies suggest the functional
role of neuronal-glia ratio in neuro-dynamical patterns, \emph{e.g.},
epilepsy~\cite{Nadkarni2003sod}.  Chatterjee and Sinha~\cite{Chatterjee2007umw} analyzed
the hierarchical structure of the \emph{C. elegans} nervous system through k-core
decomposition and showed evidence that the assortativity increases as one goes to the
innermost core of the network. The authors suggested that such assortative nature of the
inner core can help in increasing communication efficiency while turning the network more
robust at the same time. Therefore, the topological organization of neuroanatomical
networks is directly related to network function. A good review about dynamical modeling
of brain functions is~\cite{rabinovich2006dpn}.

Other approaches of brain functional networks are based on the concept of functional or
effective connectivity. The network is obtained by recordings of brain physiological
functions instead of brain anatomy. Aertsen \emph{et al.}~\cite{aertsen1989dnf}
introduced graph theory in the study of brain functions. Three basic approaches are used
to obtain the connection structures: (i) electroencephalographic (EEG), (ii)
magnetoencephalographic (MEG), and (iii) functional magnetic resonance imaging (fMRI).
The data from the two first methods are suitable for graph analysis because of the high
spatial resolution. Stam~\cite{stam2004fcp} presented one of the first works considering
MEG and networks.  By performing MEG recordings of five healthy human subjects, the graph
was formed by $N=126$ vertices and $1,\!890$ edges. For frequency bands smaller than $8$
Hz and larger than $30$ Hz, the synchronization patterns displayed the features of
small-world networks. Bassett~\emph{et al.}~\cite{bassett2006car} detected small-world
properties in brain functional networks obtained from 22 subjects using wavelet
decomposition of MEG time series. Dodel \emph{et al.}~\cite{dodel2002fcc} applied graph
theoretical analysis to identify functional clusters of activated brain areas during a
task.

Fallani \emph{et al.}~\cite{Fallani08:Neuroinformatics} investigated the cortical network
dynamics during foot movements and showed that circular motor areas act as network hubs,
presenting a large number of outgoing links. In another work, Fallani \emph{et al.}
investigated the cortical network structure of spinal cord injured patients (SCI),
comparing with health subjects~\cite{Fallani07:HBM}. They analyzed the structure of
cortical connectivity obtained by the Direct Transfer Function (DFT) applied to the
cortical signals estimated from high resolution EEG recordings achieved during the
attempt to move a paralyzed limb by a group of five SCI patients. The consideration of
the DFT allows to determine the directional influences between each pair of channels in a
multivariate data set, allowing to construct a directed network~\cite{Fallani07:HBM}. The
analysis of network structure showed that spinal cord injuries do not affect the global
efficiency. Nevertheless, the authors observed significant differences in the cortical
functional connectivity between SCI and normal patients, which can be related to the
internal organization of the network and its fault tolerance (these networks present
hubs). In addition, SCI patients revealed a higher local efficiency than healthy
subjects, which can be associated to more robustness of the cortical networks of these
patients.

Egu\'{\i}luz \emph{et al.}~\cite{eguiluz2005sfb} showed that the functional network
obtained by fMRI for seven subjects follows a power law. The unweighted networks are
obtained by correlation matrices, which are derived from functional correlations between
brain sites (called ``voxels''). Two voxels are connected if their temporal correlation
exceeds a pre-defined threshold $r_c$. The main drawback of this approach is related to
the choice of the right $r_c$. For $r_c = 0.6$, $0.7$ and $0.8$, scale-free topologies
were seen in the networks, regardless of the type of task. High clustering coefficient
($\sim 0.15$), relative small-world ($\ell=11.4$ for $N = 31,503$, $\ell = 12.9$ for
$N=17,174$, and $\ell = 6.0$ for $N=4,891$), well defined coefficient of the power law
($\sim 2.2$), and hierarchical organization ($cc(k) \sim k^{-\alpha}$) were observed.
Therefore, the functional networks were scale-free and small-world. Similar studies were
performed by Salvador \emph{et al.}~\cite{salvador2005ugf} with five subjects, but using
the frequency rather than the time domain. Achard \emph{et al.}~\cite{achard2006rlf}
concluded that the brain regions are so resilient to random failures as to target attacks
(removal of the largest hubs), thus indicating that the brain networks are not scale-free
as suggested by Egu\'{\i}luz \emph{et al.}~\cite{eguiluz2005sfb}.

MEG and EGG techniques were also used to study the
relationship between brain networks topology and brain pathologies. The
matrices obtained by pairs of synchronization likelihood values were
converted in unweighted networks by taking a
threshold. Stam~\cite{stam2004fcp} chose the threshold to keep the
network with average degree equal to 15. Stam \emph{et
al}.~\cite{stam2007swn} compared a group of 15 Alzheimer patients to a
healthy control group of 13 patients and showed that the networks
obtained for the first group have less pronounced small-world features
than for the healthy patients. Bartolomei \emph{et
al.}~\cite{bartolomei2006dfc} compared the networks of spatial
patterns of functional connectivity of the brain measured at rest by
MEG obtained for 17 patients with brain tumors to 15 healthy
patients. They concluded that brain tumors alter the functional
connectivity and the network topology of the brain. While pathological
networks are closer to random networks, healthy networks are closer to
small-world networks. Therefore, randomization in network structure
can be potentially associated with brain damage.

In summary, the complex network theory is useful for neuroscience,
opening up new opportunities and creating new challenges. For
instance, one may try to understand how brain topology changes during
animal growth or evolution. Also, the implications of the genetic and
environmental factors for brain formation can be addressed by network
representation. Studies related to brain structure and pathologies may
help understand brain diseases.


\section{Linguistics}

The creation of language is one of the greatest accomplishments of
the human beings. Understanding the evolution and organization of a
specific language is useful because it sheds light into cognitive
processes, as the way people think strongly affects the organization
of a language. Conversely, the language influences how humans think.
It is also important to compare the properties of various languages,
to study their evolution. Moreover, the use of linguistic data is
crucial for automated systems such as Web search engines and machine
translators. Various linguistic structures can be treated as
networks, including texts and thesauri~\cite{Sole2005}. The Natural
Language Processing (NLP) \cite{Jurafsky2000} community has
traditionally used such network representations to develop
applications for automatic language understanding and generation.
For instance, Graph Theory concepts have been applied to sentiment
analysis \cite{Pang2004} and tools from Spectral Graph Theory have
been used in word sense disambiguation and text summarization
\cite{Mihalcea2006}. More recently, linguistic networks were
included in the context of complex networks research. Novel
techniques frequently based on statistical physics are now used in
language studies \cite{Dorogovtsev2001,Cancho2001}, providing new
approaches for NLP applications and linguistic theories. Indeed,
there has been an increasing effort to join statistical physics and
language research, as demonstrated by the workshop inside STATPHYS~23, the largest conference for statistical
physics~\cite{Loreto2007}.

A linguistic network can be formed by, for example, a group of
interconnected words or syllables. The many ways these elements can
be linked in a network lead to a division of two main groups of
linguistic networks: semantic and superficial. The former group
comprises networks such as the ones constructed from dictionaries or
lexicons, which usually contain information about semantic
relationships between words, such as synonyms or antonyms. The
latter group mainly uses the inner structure of words or the
position of words in texts to build networks. For example, in a
word-adjacency network, words are connected if they appear as
neighbors in a text. These linguistic networks are frequently
analyzed according to well-known criteria, such as the ones
characteristic of small-world and scale-free networks. Consequently,
the tools most often used to study linguistic networks are the
measurements referred to as degree, clustering coefficient and
length of the shortest paths. Examples of these investigations are
presented and discussed in the next subsections.

\subsection{Semantic Networks}

Semantic networks usually encode relationships between a subset of
words of a language, as in a thesaurus where each entry (frequently
a word) is followed by a list of words that express concepts similar
to the entry. An example of English thesaurus used to build semantic
networks is the Roget's Thesaurus, originally published in 1852. The
Wordnet database is another common resource for semantic networks.
This lexicon groups words into sets of synonyms and stores other
semantic relationships between words, such as antonymy, hypernymy
and hyponymy. A semantic network can also be obtained from
word-association experiments, where someone sequentially and freely
provides words that he/she thinks are semantically related.

Semantic networks frequently exhibit small-world, scale-free properties. As already
mentioned, a small-world network presents high average clustering coefficient and low
average shortest path, while a scale-free network is characterized by a power-law degree
distribution. English \cite{Steyvers2001,Motter2002,Holanda2004} and
Turkish~\cite{Strori2007} synonym networks are known as small-world and scale-free
networks, and a Polish synonym network was verified to be scale-free~\cite{Makaruk2008}.
A directed network was built for an English thesaurus in \cite{Holanda2004}, and the
$k^{\mathrm{in}}$ and $k^{\mathrm{out}}$ distributions were described by two power-law
regions. It was even suggested that associative memory is originated in a process of
efficiency maximization in the retrieval of information (also related to the fact that
the neural network is a small-world network)~\cite{Motter2002}. Cycles of tourist walks
were used to give complementary information to the clustering coefficient and the
shortest path (a tourist walker goes to the nearest site that has not been visited in the
last $s$ steps) \cite{Kinouchi2001}. The English Wordnet was scale-free, small-world
\cite{Steyvers2001,Sigman2002}, with polysemic links (which indicate the different
meanings of a word) being important for the small-world effect because they tend to
shorten distances, and hubs typically represent abstract concepts. Word-association
networks are also scale-free, small-world \cite{Steyvers2001,Costa2004,Ferreira2006}.
When the flow of sequentially induced associations is considered, other properties such
as context biasing and edges asymmetry are present \cite{Costa2004}. Evoked words from
two economically distinct populations were used to build word-association networks that
were scale-free an independent of the population's economic level~\cite{Ferreira2006}.

The process of lexical development for an individual was modeled in
\cite{Steyvers2001}, using a preferential attachment strategy that
produces both scale-free and small-world structures. The network
obtained with this model encodes associations, which can be thought
of as semantic, between words or concepts. Three main rules guide
this model: (i)~a process of differentiation, in which a new
word/concept is defined as the variation of the meaning of an
existing word/concept through the acquisition of part of the pattern
of connectivity of an existing node; (ii)~the probability to
differentiate a node is proportional to its degree; and (iii)~each
node has a ``utility'' value which gives the probability that it
will be connected to new nodes. As shown in \cite{Steyvers2001},
this model produces networks similar to those obtained from
real-world data of word-associations.

\subsection{Superficial Networks}\label{sec:charac_superf}

The networks discussed in this subsection do not employ semantic
information; instead, they use morphological properties of words or
their position in sentences, or even syntactic structures. One of
these networks is based on a list of consonants pertaining to
specific natural languages. The set of syllables of a language may
be used to build another type of superficial network, where the
nodes are the syllables and the edges indicate that two syllables
co-occur in the same word.  Another network of interest encodes on
its structure the proximity of words in texts. In such a network,
referred to as a word-adjacency or co-occurrence network, words that appear
in a text near to each other are connected, as word proximity
frequently indicates a syntactic relationship~\cite{Cancho2001}.
Syntactic properties can be encoded in the network
explicitly if a grammar formalism is employed, such as dependency
grammars.

As for semantic networks, the concepts of degree distribution,
clustering coefficient and length of the shortest paths are
used to characterize superficial networks. A
bipartite consonant network displayed a two regime power-law
distribution for the occurrence of consonants over languages
\cite{Choudhury2006}. This network presents two sets of nodes, one
for the 317 languages and one for the 541 consonants, where a
consonant node is connected to a language node if the former is
present in the latter. A Portuguese syllable network, built from the
entire work of the eminent Brazilian writer Machado de Assis, also
presents a power-law degree distribution, along with the small-world
effect~\cite{Soares2005}. Although some superficial networks were
studied regarding community structure (see~\cite{Mukherjee2006}
that employs a non-bipartite consonant network), the most
investigated network properties are indeed the small-world effect
and the scale-free degree distribution. Some word-adjacency networks
obtained from the British National Corpus (BNC) were shown to be
small-world and scale-free \cite{Cancho2001}, where the authors also
associated language disorders to hubs disconnection. Word-adjacency
networks built from technical and literary texts in English and
Portuguese were shown to be scale-free and small-world
\cite{Caldeira2006}. Moreover, the George Orwell's novel ``1984''
was used to build a word-adjacency network \cite{Masucci2006}, which
exhibited a composite power-law behavior when the node degree was
correlated to the average neighbor's degree and to the average
clustering coefficient. Word-adjacency networks from different
languages (English, French, Spanish and Japanese) were found to
belong to a super family of networks based on the significance
profile of specific subgraphs (motifs)~\cite{Milo2004}. The
significance profile compares the number of a given motif in a real
network with its number in a random network with the same size and
degree distribution. In another work, syntactic networks were
obtained for Czech, German and Romanian texts using a dependency
grammar formalism, where binary relations were defined between
lexical nodes~\cite{Cancho2004}. The authors showed that these
networks present the small-world effect, a scale-free degree
distribution and disassortative mixing. Using this syntactic
dependence network for the Romanian language \cite{Cancho2005}, it
was also shown that spectral methods developed to detect community
structure \cite{Capocci2004} can be useful to group words of the
same class.

Word-adjacency networks have been applied to some natural language processing tasks. For
instance, choosing the synonym most expected in a given context can be done using a word
co-occurrence network \cite{Edmonds1997}. An automatic assessment of text quality was
implemented with network measurements, in which correlation was established between
measurements (such as degree) and text quality scores assigned by humans to essays
written by high-school students in Portuguese \cite{Antiqueira2006}. A more specific task
of quality assessment was studied in \cite{Pardo2006,Pardo2006a}, where network
measurements were associated to the evaluation of the quality of Portuguese summaries. A
similar approach was adopted in \cite{Amancio2008} to assess the quality of machine
translations, where the original and translated texts, or more specifically, their word
adjacency networks, were compared. The degree of structural change from source to
translation was measured and correlated with translation quality. Another NLP task
studied with word-adjacency networks was the authorship characterization, where texts in
English created by famous writers were related to network
measurements~\cite{Antiqueira2006a}. In measurements taken from English, Bengali and
Hindi, lexical networks were correlated with some issues related to the construction of
spell checkers~\cite{Choudhury2007}. These networks were weighted, where each edge was
associated with the orthographic distance between words (this distance consider the
number of character substitutions, deletions or insertions necessary to transform one
word into another). The authors found that the weighted clustering coefficient is
correlated with the difficulty to correct non-word errors (\emph{i.e.} spelling errors resulting
in nonexistent words). Real word errors, which are misspelled words still valid in the
given language, were found to be higher in words with higher weighted degree. In another
applied research, a framework based on syntactical networks for the study of language
acquisition was proposed~\cite{CorominasMurtra2007}. The author defined a set of criteria
to build syntactical networks from child utterances collected at successive time
intervals. From these networks, measurements such as degree, clustering coefficient,
length of shortest paths and assortativeness were proposed to assess language development
in children.

A growth model for word-adjacency networks was reported in \cite{Caldeira2006}, which
adds to a network full sentences instead of single words. Every word of a new sentence is
represented by a node that is connected to the other nodes of the sentence (\emph{i.e.} the
sentence is represented by a complete subgraph -- a clique). The ratio between new and
old words in a sentence decreases over time according to a power-law, with the old words
being the hubs. This model is in good agreement with real data, and presents both
scale-free and small-world properties. A preferential attachment approach for building
consonant networks was reported in \cite{Choudhury2006}. The degree distribution fits
well with the network obtained from a real consonant inventory. The models of language
development created by Dorogovtsev and Mendes~\cite{Dorogovtsev2001} and
Marko\v{s}ov\'{a}~\cite{Markosova2008} also employ the degree-based preferential
attachment rule to make connections between newly created words and words already
inserted in the vocabulary. In these cases, the vocabulary is a web of words that
interact in sentences, thus it is a positional (or co-occurrence) word web. Dorogovtsev
and Mendes also defined that, as new nodes are created, new edges are created between old
words, again using the preferential attachment rule. Marko\v{s}ov\'{a}, instead,
preferentially rewired some old edges, thus both models incorporate changes in vocabulary
use. Their models fit the experimental data with a two- regime power-law distribution of
degrees.

The papers reviewed in this section cover many of the possibilities to represent
linguistic structures as networks, ranging from word-adjacency to synonym networks. Most
exhibit small-world and scale-free features, despite being rather different from each
other, like syllable networks and Wordnet. Many studies on linguistic networks focus on
experimental global features of the structures, while some works are devoted to model the
evolution of these networks, with results agreeing with experimental data. Another
important issue is the application of linguistic networks to NLP, for helping develop
spell checkers and automatically assess text quality. As far as we know, networks using
more sophisticated resources such as argument structure have not been studied yet,
perhaps mainly because of the difficulty in obtaining large amounts of related data.
Modeling the development of language with specific individuals has also its drawbacks,
since we do not know how language vocabulary or syntactical rules are stored in the
brain. As for applied research, there is a great deal of possibilities unexplored, such
as improving parsers and machine translation systems. The great challenge in this case is
to choose the right type of network and analysis method to retrieve the desired
information from interlinked structures.


\section{Earthquakes}

The study of earthquake occurrences and its spatial distribution may be done through
complex networks. Abe and Suzuki \cite{ABE_EARTHQUAKE_EL_2004,ABE_EARTHQUAKE_NPG_2006}
proposed a method in which the Earth surface was divided into cells that were the nodes
of the network. Two successive earthquakes (i.e without any other between them) establish
a new link in the network between the cells where these shocks occurred. Upon analyzing
seismic data from Japan and California, the authors showed that both earthquake networks
have a scale-free topology with different scale coefficients. This difference may
characterize the distinct tectonic plates analyzed. The hubs were found to be related to
the place where a mainshock occurred (i.e earthquakes with large magnitude). In another
paper \cite{ABE_EARTHQUALE_PHYSA_2004}, the earthquake network was shown to have a very
small average path length and high clustering coefficients, in comparison to the random
network version. These results suggest the presence of the small-world effect and a
hierarchical organization, as indicated by the power-law behavior of the clustering
coefficient with respect to the node degree \cite{ABE_EARTHQUAKE_CONDMAT_2006}.  The
seismology network is therefore different from the Barab\'asi-Albert networks, which also
has a scale-free degree distribution but no hierarchical structure. Structured networks
contain features that are linked to the physical properties of the earthquake dynamics.
The mixing property was investigated using the concept of nearest-neighbor average
connectivity, in which Abe \emph{et al}.\ showed that highly connected nodes are linked
to each other, for both seismology networks, namely from California and Japan. Abe and
Suzuki \cite{ABE_EARTHQUAKE_PHYSICS_2006} discovered that in the growth of a seismology
network after the main shock the clustering coefficient remains constant during the time
$\Delta t$, which depends on how the network was built, typically of the order of hours.
After $\Delta t$, the clustering coefficient has a steep drop and then decays slowly
according to a power law before it becomes steady again.

The seismology networks were also studied by Baiesi and Paczuski
\cite{BAIESI_EARTHQUAKE_PRE_2004}, who used correlation between events
to classify them systematically as foreshocks, main shocks and
aftershocks.  The network was built with each earthquake (node)
receiving one incoming link from its most correlated predecessor,
while outgoing links referred to aftershocks. It displayed the
scale-free behavior with exponent $\gamma=2$. The metrics used
consider the magnitude and the spatial and temporal distances between
events. Two spatially distant earthquakes can be related if the temporal
distance is short, which is consistent with the self-organized nature
of earthquake dynamics.

Networks built from the seismology data from different places in Earth have about $10^2$
and $10^3$ nodes, with scale-free topology. The results from network analysis may help us
understand the collective behavior and perhaps predict these natural events.


\section{Physics}

Many phenomena in physics may be modeled as networks.  For instance, in 2001, Bianconi and Barab\'asi
\cite{BIANCONI_BOSE_PRL_2001} discovered that a Bose-Einstein
condensate was related to complex networks. The authors mapped the
network in a Bose gas, where the nodes corresponded to energy levels
and the links were particles. The condensation state was reached when
a single node had a macroscopic fraction of the links. Burioni
\emph{et al.}\ \cite{BURIONI_BOSE_JPB_2001} reported a complementary
work with topological heterogeneities in the network.

\subsection{Energy landscapes}

In 2005 Doye and Massen \cite{DOYE_LANDSCAPE_JCP_05} used networks to study the
configurational space of the Lennard-Jones potential. To build the network, called
reaction graph, the authors took each minimum of the potential function as a node, with
the surrounding of each node defining its basin of attractions. The nodes were connected
when there was a direct state transition between their corresponding minima, not
considering auto-loops and multiple edges. The results indicated that the network
topology is scale-free and small-world. The authors found an interesting negative
correlation between the degree and the potential energy of the minima, indicating that
the low energy minima act as hubs, the signature of scale-free networks. Interestingly,
the scale-free nature of this network is not due to preferential attachment, but a direct
consequence of the potential landscape topography, because the basins of attraction
related to the low-energy minimum are larger in general and more likely to have
surrounding transition states. The authors also suggested that the small-world effects
can be useful to explain the Levinthal paradox (which states that the time of a protein
to arrive at its native conformation is larger than the age of the universe) because it
reduces dramatically the number of steps needed to find the native conformation. For
example, in a typical problem with $10^{21}$ minima, algorithms based on global
optimization converge, in average, after only 150 minima. Reaction graphs were already
used in previous papers, such as~\cite{TAKETSUGU_LANDSCAPE_MP_02}, to investigate the
isomers of water dimer, and \cite{RAO_LANDSCAPE_JMB_04} to study interconnection of the
protein conformal space. Gfeller \emph{et al.}\ \cite{GFELLER_LANDSCAPE_PNAS_07} also
studied energy landscapes through networks (weigthted in this case). Their findings
include some analytical results for low-dimensional models and they discuss how the
network approach is useful to study the isomerization problem. The landscape of
spin-glasses mapped in networks by Seyed-allaei \emph{et
al.}~\cite{SEYED_LANDSCAPE_CONDMAT_07} had a Weibull distribution for the degree, instead
of the power-law found previously for the Lennard-Jones potential
\cite{DOYE_LANDSCAPE_JCP_05}. The authors concluded that the topology of the landscape
networks is not universal because they depend on the physical properties of the model
system.

\subsection{Astrophysics}

Magnetic effects were also explored through complex network
concepts. In \cite{LU_SOLAR_AJL_91} the authors suggested that the
solar corona can be thought of as a critical self-organized system.
Hughes and Paczuski
\cite{HUGHES_CORONAL_ASTRO_03,HUGHES_CORONAL2_PHYSA_03} studied the
spatial structure of a network generated from this critical system
after the transient phase. To build this network, an approach similar
to that of earthquakes studies was used, with the solar surface being
a discrete mesh with each mesh cell being a node. Two nodes were
connected when a magnetic arc (solar flare) emerged between the
cells. The network obtained has scale-free topology with power law
distributions of degree and strength.

\subsection{Ising Model}

With the increase of popularity of complex networks, systems that were
only studied in regular networks started to be described in networks
with arbitrary topologies. This is the case of the Ising, Potts and
$XY$ models. Gitterman \cite{GITTERMAN_ISING_JPA_2000} showed that a
phase transition occurs when there is a minimum number of long range
connections per node. Barrat and Weigt
\cite{BARRAT_ISING_EPJB_2000} found that at high temperatures the
system behaves as in the conventional 1D case. At low, non-zero
temperatures, a ferromagnetic phase transition was found in the
small-world Ising model. Therefore, the phase transition in the
small-world Ising model is of a mean-field nature. In 2001, Hong
\emph{et al.}\ \cite{HONG_ISING_PRE_2002} used Monte Carlo simulation
to confirm the results of Gitterman~\cite{GITTERMAN_ISING_JPA_2000},
Barrat and Weigt~\cite{BARRAT_ISING_EPJB_2000}, and found critical
exponents $\alpha=0$, $\beta=1/2$, $\gamma=1$ and $\nu=2$ ($\alpha$ is
the critical exponent for specific heat, $\beta$ is the critical
exponent for magnetization. $\gamma$ and $\nu$ are the critical
exponent for the susceptibility). S. Roy and
S. M. Bhattacharjee \cite{ROY_ISING_PLA_2006} also studied the Ising
model under small-world topology.

The two and three dimensional cases of the small-world Ising model were
studied by Herrero \cite{HERRERO_ISING_PRE_2002}. He noted that, in 2D
and 3D, the small-world geometry changes the universality class of the
phase transition. In 1D, the critical temperature is given by
$T_c\propto|\log p|^{-1}$ where $p$ is the probability of rewiring an
edge. Herrero found that in 2D and 3D, the dependence between $T_c$
and $p$ is given by the power law $T_c-T_c^{0}\propto p^{1/{\nu d}}$,
where $T_c0$ is the critical temperature in the regular lattice and
$d$ is the spatial dimension. The bidimensional Ising model was
also studied by Cai and Li~\cite{CAI_ISING_IJMPB_2004}.

The Ising spin system under a scale-free topology was investigated by
Aleksiejuk \emph{et al.} in 2002 \cite{ALEKSIEJUK_ISINGBA_PHYSA_02}
and Bianconi \cite{BIANCONI_ISING_CONDMAT_02}. A
ferromagnetic-paramagnetic phase transition was observed, with the
critical temperature depending on the network size as $T_c\propto|\log
N|$. The analytical solution for this model in a network with
arbitrary degree distribution was found in
\cite{DOROGOVTSEV_ISING_PRE_02,LEONE_ISING_EPJB_02}. In 2004, a spin
system under a directed version of the Barab\'asi-Albert network was
investigated by Sumour and Shabat~\cite{SUMOUR_ISING_CONDMAT_04}, who
used Monte Carlo simulation to show the existence of spontaneous
magnetization of the system.

The p-Potts model was investigated by Dorogovtsev \emph{et al.} in 2004
\cite{DOROGOVTSEV_POTTS_EPJB_04} for networks with degree
distribution $P(k)=k^{-{\gamma}}$. The authors showed that when the
second moment of this distribution diverges, the phase transition is
continuous and of infinite order instead of the first-order phase
transition.


\section{Chemistry}

While applying complex networks concepts in chemistry, Jiang \emph{et al.}\
\cite{JIANG_AMMONIA_PHYSICS_2005} showed how an ammonia plant can be mapped in a complex
network with small-world properties. The ammonia plant network is weakly self-similar and
possesses a modular structure, with each community representing a modular section in
chemical plants. Another interesting result is that the ammonia complex network exhibits
excellent allometric scaling, guaranteeing a better fluid flow. Amaral and
Barth\`{e}lemy~\cite{AMARAL_POLYMER_EL_2001} showed that the small-world effect emerges
on the phase space of polymer conformations. For this, each conformation was represented
by a single node in a complex network and two nodes were connected if the Monte Carlo
distance between them was unitary. Another work involving polymer analysis through
complex networks was developed by Kabakçioglu and Stella
\cite{KABAKCIOGLU_POLYMER_PRE_2005}. Chemical reactions were studied as a complex network
topology.  For example, Lazaros and Argyrakis \cite{GALLOS_REACTION_PRL_04} investigated
the classic model $A+A\rightarrow 0$ and $A+B\rightarrow 0$ under scale-free topology.
Their results indicate that the speed of the reaction is rather faster than in lattice
models. The authors also observed that the diffusion behavior on very sparse scale-free
networks is the same as for regular lattices.

Another network built from chemical reactions can be found in
\cite{SOLE_ASTRONETS_EL_2004}, in which the UMIST Database for
Astrochemistry was used to generate networks. The nodes were reactants
and products of each relevant reaction found in the interstellar
medium and in planetary atmospheres (including Earth, Mars, Titan,
Venus). Among the results, two basic topologies were revealed and
associated with the presence or absence of life. According to the
authors, the Earth chemical network is the only one to present a
scale-free topology, although all the others also displayed
small-world properties. The community structure was investigated in
the Earth chemical reaction network, with two main communities being
identified. They are related to reactants and products of the
reactions with OH and Cl, and one of the products is H$_2$O. The
authors suggest that the community structure in this network results
from action of the most reactive free radicals of Earth atmosphere,
namely Cl and OH.

The range of applications of complex networks in chemistry is particularly high. In this
section we showed applications in unitary processes, polymers, chemical reactions and
astrochemistry. We have seen how the networks theory can help one identify the main
features in industrial plants, such as small-world topology, modular structure and
allometric scaling. Also, the use of network concepts may allow the determination of the
lowest energy state, in spite of the millions of possible polymer conformations, because
the phase space can be mapped quickly owing to the small-world topology.


\section{Mathematics}

Complex networks have been used to represent sets of natural and prime
numbers connected by some specific rules. Corso
\cite{CORSO_NUMBERS_PRE_2004} used a theorem of number
theory, which states that each natural number has a unique
decomposition in prime factors, to build a network. Two natural
numbers (nodes) were connected if they shared at least one prime
factor. The networks were nonsparse and small-world ($d\approx 2$,
$N\approx 10^{5}$). The degree distribution was invariant with network
size (number of natural numbers considered to build the network) and
presented a plateaux related to families of prime numbers. Chandra and
Dasgupta \cite{CHANDRA_NUMBERS_PHYSA_2005} used a similar method to
connect two prime numbers according to a free parameter
($\alpha$). An even number $n$ can be written as a
sum of two prime numbers p and q: $n=p+q$. The probability of
establishing a connection between p and q is $|p-q|^{\alpha}$. The
authors observed a phase transition in the network topology by varying
the free parameter. For $\alpha > \alpha_0$, the network exhibits a
scale-free topology and high clustering coefficient, characterizing a
small-world effect. Outside this regime, the power law of the degree
distribution was not observed, but the hubs still existed. The size of
the network studied is $5\times 10^{3}$ nodes.

\section{Climate networks}

The extension of network theory to climate sciences was performed by Tsonis \emph{et al.}
\cite{Tsonis06:BAMS, Tsonis03:MAP, Tsonis05:GRL}, where the networks is constructed
considering the climate system, represented by a grid of oscillators -- each of them is a
dynamical system varying in some complex way. The nodes in the network represent cells of
a grid of 5$^o$ latitude $\times$ 5$^o$ longitude that covers the globe. The connections
between the nodes were defined through a threshold of correlation coefficient of a time
series of each node. The connections are obtained by considering the database from global
National Center for Environmental Prediction/ National Center for Atmospheric Research
(NCEP/NCAR). Generally, data from 500 hPa value, which indicates the height of the 500
hPa pressure level and provides a good representation of the general circulation of the
atmosphere, are used for network construction. If the correlation coefficient for 500-hPa
time series of two nodes exceeds 0.5, these nodes are connected. The resulting network
revealed to be scale-free with small-world behavior~\cite{Tsonis06:BAMS}.

Climate networks can be considered to identify global change and for investigation and
representation of climate dynamics~\cite{Tsonis06:BAMS}. For instance, Tsonis and
Swanson~\cite{Tsonis08:PRL} constructed the El Ni\~{n}o and El Ni\~{n}a networks by
taking into account the global temperature. They show that El Ni\~{n}o network network is
less communicative and less stable than El Ni\~{n}a network, because the former present
fewer links and lower clustering coefficient and characteristic path length than the
latter. Due to these applications, complex networks provide a new and useful tool in
climate research.


\section{Security and Surveillance}

Latora and Marchiori \cite{LATORA_TERRORISM_CSF_2004} studied the
consequence and prevention of terrorist attacks in a given network,
and suggested a method to detect critical nodes (\emph{i.e.} the most
important nodes for efficient network functioning). The network
efficiency is related to the shortest path between all nodes of the
network. The importance of a particular node is given in terms of the
change in network efficiency when this node is removed. The deletion
of an important node leads to a large decrease in network efficiency.
The authors illustrated this concept in the communication network
Infonet, which is the US and European Internet backbones. After
deactivating all nodes, one by one, the most important nodes detected
were New Jersey and New York. The results indicated that the
destruction of these nodes can reduce network efficiency by more than
$50\%$. Another important result indicated that the most connected
nodes are not necessarily the most important. For example,
deactivation of Chicago, with degree 15, only reduces network
efficiency by $28\%$, while New Jersey and New York, both with degree 9,
reduces network efficiency by $57\%$ and $53\%$, respectively.

In the same paper, the authors gave a second example in which they
built a network where the nodes were the terrorists related directly
or indirectly with hijackers of September 2001 attacks. The links are
the knowledge interplay among the hijackers. The data were obtained
from information taken from major newspapers. The most critical node
had the largest number of direct connections with other
terrorists. The second most critical node had degree of only half of
the maximum. This shows that even nodes with low connectivity can play
a crucial role for the network efficiency. In 2002,
Krebs~\cite{KREBS_TERRORIST_CONN_2002} also analyzed a network of
hijackers linked to events of September 11th. According to the author,
using network theory to prevent criminal activity is difficult, but it
is an important tool for prosecution purposes. Maeno and Ohsawa
\cite{MAENO_COVERT_COND_2008} proposed a method to solve the node
discovery problem in complex networks, which may be applied to
identify an unobserved agent behind the perpetrators of terrorist
attacks.


\section{Epidemic spreading}\label{Sec:spreading}

Epidemics of computer viruses have been studied with the aid of graphs
and random graphs for at least three decades. In 1991, Kephart and
White \cite{KEPHART_VIRUS_IEEE_91} extended epidemiological models to
investigate the spreading of computer viruses using a directed random
graph. They showed that proliferation can be controlled if the
infection rate does not exceed a critical epidemic
threshold. Kleczkowski and Grenfell \cite{KLECZKOWSKI_VIRUS_PHYSA_99}
applied a cellular automata model to the spreading of diseases in
small-world networks, where the dynamics of the clusters was
investigated as a function of the network order parameter (the
fraction of the long range links). In the model, $N_a$ agents were
placed in the nodes of a two-dimensional square lattice. The nodes were
classified as infected, infectious and immune. At each time step
($\Delta t=1$ $\mathrm{week}$) each agent interacts with its $z$ nearest
neighbors. Moreover, the authors included a mixing pattern in the
model, \emph{i.e.} they allowed two agents to swap their positions at any
time, in order to reflect the rules of the social structure. The
results showed that upon increasing the mixing the disease spreads
faster. The problem of epidemic spreading in small-world networks was
also explored in \cite{MOORE_DISEASE_PRE_00}, where the authors found
the exact values for the epidemic thresholds as a function of the infection
and transmission probabilities. Also studying this system, Kuperman and
Abramson \cite{KUPERMAN_EPIDEMIC_PRL_01} found an interesting
oscillatory behavior of the size of the infected subpopulation. The
authors showed that the number of infected agents changed from an
irregular, low-amplitude state, to a spontaneous, high amplitude state,
when the order parameter changed from 0 to 1. Recently, Small \emph{et al.}\
\cite{SMALL_VIRUS_IEICE_04} showed that only with the introduction of
a small-world topology in the epidemiological model the random
fluctuation of the real data could be explained.

Pastor-Satorras and Vespignani \cite{SATORRAS_EPIDEMIC_PRE_01} showed
that, in contrast to the pure small-world lattices, the uncorrelated
scale-free networks with exponent $0\leq \gamma\leq 1$ do not have a
critical threshold, thus indicating that in these networks diseases
spread regardless of the infection rate of the agents. For the scale
exponent in the range $1<\gamma\leq 2$ the critical threshold appeared
but no critical fluctuations were observed. Only for $\gamma > 2$ has
the traditional behavior been observed. For correlated scale-free
networks with $P(k'|k) \propto k$ there is a non zero critical
threshold for the spreading dynamics
\cite{BOGUNA_EPIDEMIC_PRE_02,VOLCHENKOV_EPIDEMIC_PRE_02}. This result
was modified in subsequent papers
\cite{BOGUNA_EPIDEMIC_PRL_2003,MORENO_EPIDEMIC_EPJB_02}, where the
authors used analytical arguments to show that in absence of higher
order correlations, the epidemic threshold is null for scale-free
networks with $2 < \gamma \leq 3$. The analytical solution of the SIR
model (susceptible-infectious-recovered), including bipartite nodes
and non-uniform transmission rate, was found by
Newman~\cite{NEWMAN_EPIDEMIC_PRE_02}. The effects imposed by the
finite size of the networks to epidemic threshold was studied by
Pastor-Satorras and Vespiagnani in \cite{SATORRAS_EPIDEMIC_PRE_02}.

The lack of epidemic threshold in social networks with scale-free
topology is worrying because for any transmission rate the disease can
propagate to all nodes of the network \cite{SATORRAS_EPIDEMIC_PRE_01,
WANG_REVIEW_CSMIEEE_03, LILJEROS_SEXUAL_NATURE_03,
ZHOU_REVIEW_PNS_06}. This is the case of the sexual contact network of
Sweden, as shown by Liljeros \emph{et al.}\
\cite{LILJEROS_SEXUAL_NATURE_03}.  The solution to eradicate
spreading of virus in these classes of network was proposed in
\cite{DEZSO_EPIDEMIC_PRE_02, SATORRAS_IMMUNIZATION_PRE_02,
SATORRAS_IMMUNIZATION_BOOKS_03}, which is based on immunizing the hubs
of the network (\textit{target immunization}). Through computer
simulation they showed that this approach is able to eradicate the
disease, being more effective than a random immunization. Though
efficient, this method requires a priori information about the whole
network structure, such as the connectivity of all nodes.  Instead of
this global approach, Cohen \emph{et al}.\
\cite{COHEN_IMMUNIZATION_PRL_03} suggested a local method referred to
as \textit{acquaintance immunization}, where a subset of nodes was
randomly selected and, depending on the neighborhood of each node,
immunization was performed. In
\cite{MORENO_IMMUNIZATION_EPJB_06}, G\'{o}mez-Garde\~{n}es \emph{et al.}\
studied various immunization strategies for Internet maps at the
Autonomous Systems (AS) and Routers networks. They suggested a new
immunization method that is neither local nor global, in which each
vertex looks at its neighborhood (at maximum distance $d$) and immunizes
the highest connected neighbor. The results confirmed this to be the
most efficient method.

Spreading on growing networks was investigated by Hayashi \emph{et
al}.\ \cite{HAYASHI_EPIDEMIC_PRE_04} who studied the oscillatory pattern
for the number of infected computers on scale-free networks as new
users joined. The authors simulated the spreading dynamics on networks
with exponent of degree distribution extracted from real data of
received and sent emails. Random immunization was not able to
eradicate the virus. The number of infected computers oscillated with
a period depending on the fraction of vertices that receive
immunization.  It also occurs in the scale-free networks, but in this
case, it is possible to find a set of parameters to stop the
oscillatory pattern and eliminate the virus. For example, an
immunization of $20\%$ of the hubs in a growing scale-free network
prevents virus infection.

Real data of \textit{Mycoplasma pneumoniae} infection was used in
\cite{MEYERS_EPIDEMY_EID_01} to build a mathematical model and investigate the spread and
control of this disease in closed communities. Other approaches,
including percolation theory, have also been used to study epidemic spreading with complex networks
\cite{SANDER_PERCO_MB_02,SANDER_EPIDEMY_PHYSA_03}. Small \emph{et al.}
\cite{SMALL_SPREADING_IEICE_2004} compared the data from the SARS cases in Hong Kong in
2002 with various models. While the classical SIR model is able to
describe epidemic spreading at a large scale, the small-world network model is
better to describe the local variability.  The Hong Kong SARS data
were also used in \cite{SMALL_SPREADING_PHYA_2005} where a new 4-state
model was used to simulate the disease transmission under a
small-world topology.  The model included as agents those who were
infected but not yet infectious, and as the main result the author
found that outbreaks could be prevented if the patients with symptoms
were isolated as soon as possible. Another model applied to SARS spreading of virus in Hong Kong can be found in
\cite{SMALL_SPREADING_PHYD_2006}. In 2007, Small \emph{et al.}
\cite{SMALL_SPREAD_PRL_2007} studied the distribution of Avian
Influenza virus among wild and domestic birds, and obtained a network
with scale-free topology with no epidemic threshold. Therefore, local
methods could not be used to eradicate the disease, thus pointing to
attacks to the hubs as a possible strategy.

Li and Wang \cite{LI_EPIDEMIC_IJSS_07} studied the SIR dynamics on
small-world networks with a delay-time recovery. The results indicate
that actions to recover the network should be executed as soon as
possible to avoid spread of the infection. The recovery of geographic
networks depended on the action radius (local region immunization),
where a critical radius existed, above which the epidemic spreading
vanished ~\cite{GUO_EPIDEMIC_PHYSA_07}. A particular type of
scale-free network with self-similar structure, high clustering
coefficient, "large-world" and dissortative behavior was used by Zhang
\emph{et al.} \cite{ZHANG_EPIDEMIC_CONDMAT_08} to solve the SIR
model. They used a renormalization approach, and found an epidemic
threshold, with the degree distribution being insufficient to
characterize the epidemic dynamics.


\section{Collaboration Network of the Papers Cited in this Review}

\begin{figure}[t]
  \centerline{\includegraphics[width=\linewidth]{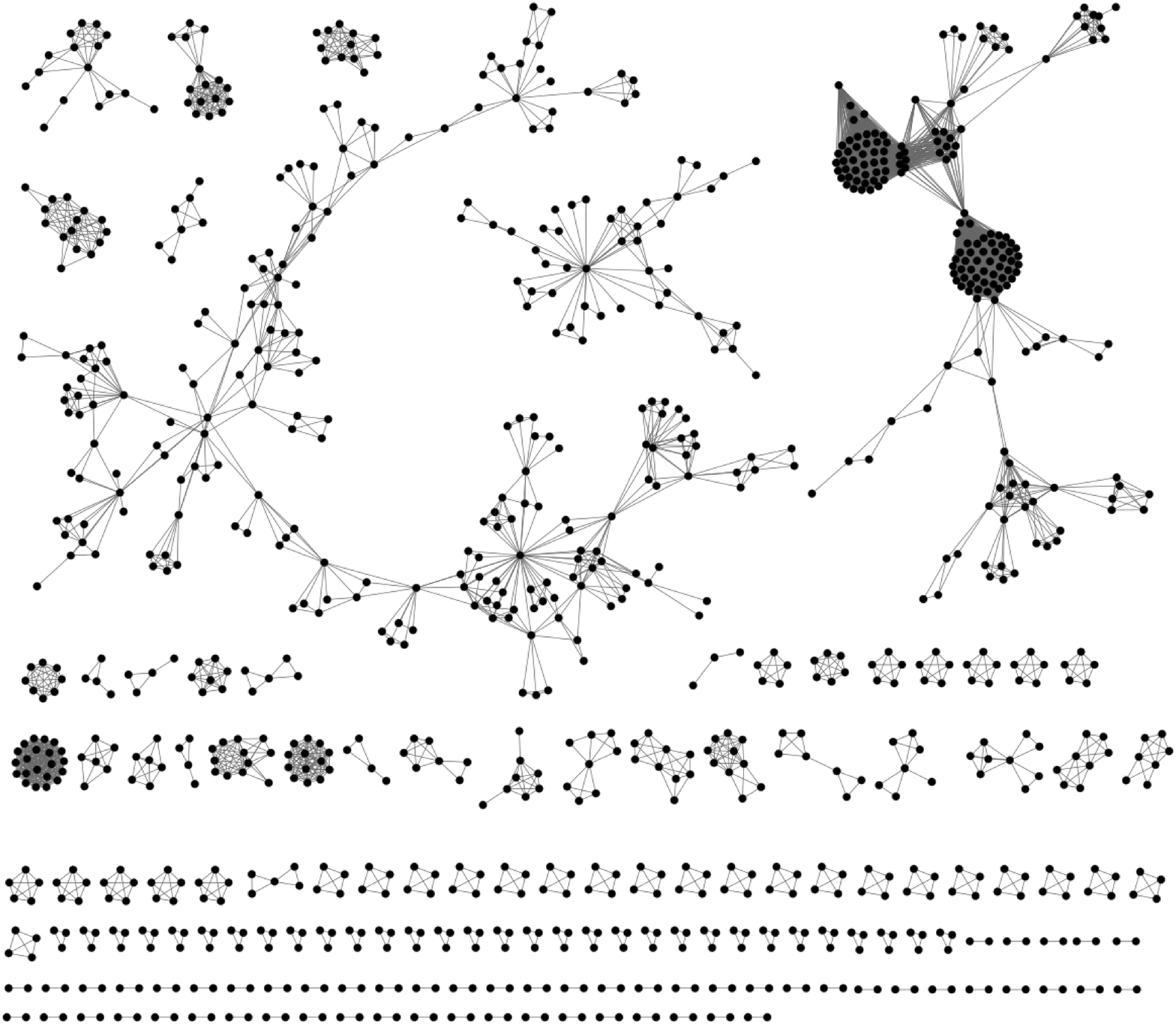}}
  \caption{Collaboration network considering the papers cited in this
    review. Graph visualization  obtained using Cytoscape~\cite{shannon2003cse}.}
  \label{fig:graph}
\end{figure}

In order to have a more complete perspective of the works addressing
applications of complex networks, we constructed a network of
collaborations taking into account the references listed in the
present survey.  More specifically, each author in each reference
entry was mapped into a node, while co-authorship was considered to
implement the connections (\emph{i.e.} an edge was used to
interconnect each pair of co-authors).  A program was especially
implemented to filter and organize the Bibtex database.  The network
contains 1,028 nodes and 4,707 edges, and is illustrated in
Figure~\ref{fig:graph} with the main areas covered being identified by
red circles. Several features can be seen in this figure.  First, the
overall network is highly fragmented, except for two larger connected
components.  This high fragmentation indicates that application of
complex networks is still performed by distinct communities of
researchers, in a little integrated manner. As a consequence, the
concepts and methods used by each of the communities are not likely
known or applied by the other communities.  It is felt that important
and interesting advances could be achieved through a more integrated
research and more intense collaboration between the several
communities in Figure~\ref{fig:graph}.  Table~\ref{tab:meas} shows the
main measurements of the topology of the collaboration network.  The
way in which the network was built naturally implies in a high
clustering coefficient.  The assortative mixing coefficient ($r$) is
particularly high, implying that highly connected researchers tend to
be co-authors with few connected collaborators..

\begin{figure}[t]
  \centerline{\includegraphics[width=\linewidth]{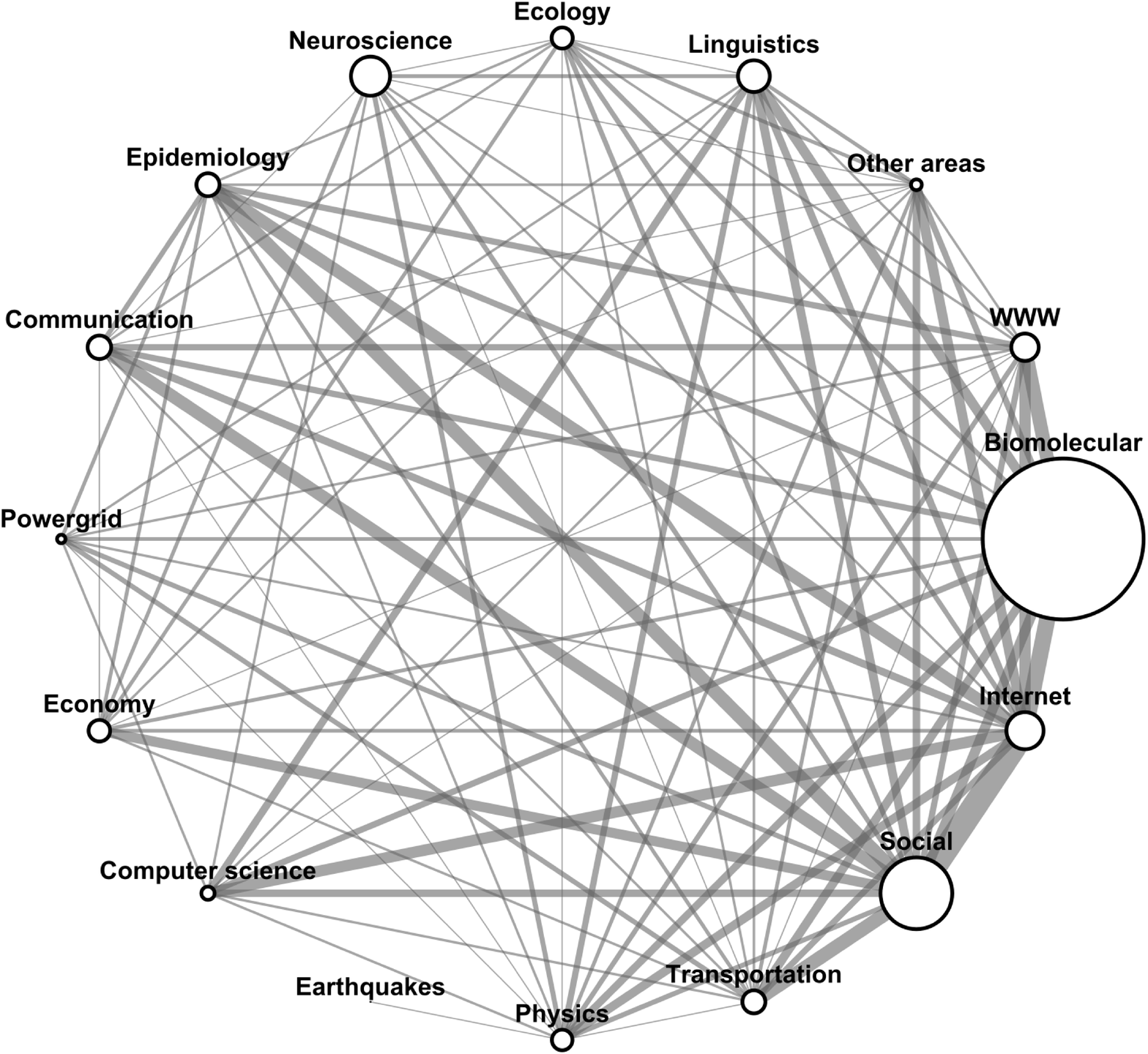}}
  \caption{Network representing the connections between the research areas covered by complex networks theory. The size of the vertices represent the number of authors and the strength of links, the number of researches shared by areas.}
  \label{fig:areas}
\end{figure}

Figure~\ref{fig:areas} presents the connections between the research
areas covered by complex networks.  The vertex sizes represent the
number of researchers by area, and the strength of the links stand for
the number of researchers shared by pair of areas. The numbers of
authors appearing in two areas are shown in Table~\ref{tab:areas}.
The percentage of authors by area is shown in
Table~\ref{tab:areas_percent}.  The Biomolecular area concentrates
many authors mainly due to the importance of complex networks theory
to model biological interactions~\cite{Oltvai02:Science}. It is
interesting to note that the strongest connections occur between the
four pioneers areas in complex networks, \emph{i.e.}  Social,
Internet, Biomolecular and WWW (these areas also include a larger
number of authors). Other strong connections occur between affine
areas, such as Social sciences and Epidemiology, Social sciences and
Communication, Social sciences and Transportation, Social sciences and
WWW, Internet and Computer Science, Internet and WWW, Internet and
Transportation. Distinct areas, such as Power Grid and Economy or
Linguistics and Epidemiology, have only a few or even no links. The
Social sciences, Internet and Biomolecular areas act as hubs in the
network, sharing many authors with other areas. The large percentage
of researchers of these areas (see Table~\ref{tab:areas_percent})
suggests that the studies related to society, Internet and cellular
process has been particularly relevant in complex networks
investigations.

\begin{table}
  \caption{Topological properties of the collaboration network.}
  \label{tab:meas}
  \begin{center}
    \begin{tabular}{l | r}
      \textbf{Measurement} & \textbf{Value} \\
      \hline
      Number of nodes & 1028 \\
      Number of edges & 4707 \\
      Average degree & 9.16 \\
      Maximum degree & 70 \\
      Average clustering coefficient & 0.78 \\
      Assortative mixing coefficient & 0.92 \\
      Size of the giant component & 209 \\
      Average shortest path length$^{*}$ & 4.89 \\
      \hline
    \end{tabular}
    \\
  \small{* Inside the giant component.}
  \end{center}
\end{table}

\begin{table}
  \caption{Percentage of the number of authors by area.}
  \label{tab:areas_percent}
  \begin{center}
    \begin{tabular} {l | r}
      \hline
      Area & Size \\
      \hline
      Biomolecular & 32.88\% \\
      Social & 15.08\% \\
      Neuroscience & 8.56\% \\
      Internet & 8.17\% \\
      Linguistics & 7.10\% \\
      WWW & 6.23\% \\
      Communication & 5.45\% \\
      Epidemiology & 5.35\% \\
      Transportation & 5.35\% \\
      Economy & 5.06\% \\
      Physics & 4.86\% \\
      Ecology & 4.47\% \\
      Computer science & 3.40\% \\
      Other areas & 2.92\% \\
      Powergrid & 2.43\% \\
      Earthquakes & 0.39\% \\
      \hline
    \end{tabular}
  \end{center}
\end{table}

\begin{table}
  \caption{Number of common authors to two areas.}
  \label{tab:areas}
  \begin{center}
    \begin{tabular} {l | l | c}
      \hline
      Area 1 & Area 2 & Number of authors in common \\
      \hline
      Biomolecular	&	Internet	&	13	\\
      Biomolecular	&	WWW	&	11	\\
      Biomolecular	&	Social	&	8	\\
      Biomolecular	&	Linguistics	&	8	\\
      Biomolecular	&	Transportation	&	7	\\
      Biomolecular	&	Physics	&	6	\\
      Biomolecular	&	Computer science	&	5	\\
      Biomolecular	&	Communication	&	5	\\
      Biomolecular	&	Epidemiology	&	5	\\
      Biomolecular	&	Ecology	&	4	\\
      Communication	&	Social	&	10	\\
      Communication	&	Internet	&	6	\\
      Communication	&	WWW	&	5	\\
      Communication	&	Epidemiology	&	4	\\
      Computer science	&	Internet	&	9	\\
      Computer science	&	Social	&	6	\\
      Computer science	&	Linguistics	&	6	\\
      Ecology	&	Social	&	4	\\
      Ecology	&	Internet	&	4	\\
      Economy	&	Social	&	8	\\
      Epidemiology	&	Social	&	11	\\
      Epidemiology	&	Internet	&	10	\\
      Epidemiology	&	WWW	&	5	\\
      Internet	&	Social	&	20	\\
      Internet	&	WWW	&	9	\\
      Internet	&	Physics	&	7	\\
      Internet	&	Linguistics	&	6	\\
      Internet	&	Transportation	&	5	\\
      Linguistics	&	Social	&	7	\\
      Linguistics	&	Physics	&	5	\\
      Neuroscience	&	Social	&	4	\\
      Neuroscience	&	Physics	&	4	\\
      Physics	&	WWW	&	4	\\
      Physics	&	Social	&	4	\\
      powergrid	&	Social	&	4	\\
      powergrid	&	Transportation	&	4	\\
      Social	&	Transportation	&	11	\\
      Social	&	WWW	&	5	\\
      \hline
    \end{tabular}
  \end{center}
\end{table}


\section{Conclusions and perspectives}

Frequently, the success of new areas of physics are judged not only
from their theoretical contributions, but also from their
potential for applications to real-world data and problems.  Despite
its relatively young age, the area of complex network research has
already established itself, especially through its close relationship
with formal theoretical fields such as statistical mechanics and graph
theory, as a general and powerful theoretical framework for
representing and modeling complex systems. It has been capable of
taking into account not only the connectivity structure of those
systems, but also intricate dynamics (see, for instance,
the surveys by Newman~\cite{Newman03:survey} and Boccaletti \emph{et
al.}~\cite{Boccaletti06:PR}).

Judging by the large number of areas and articles reviewed in the
present survey, complex networks have performed equally well (if not
better) with respect to their application potential. Indeed, the
generality and flexibility of complex networks extends to virtually
every real-world problems, from neuroscience to earthquakes,
encompassing at least the 22 areas considered in the present work.  In
addition, most of the studies report detailed,
comprehensive investigations where not only complex systems are
represented as networks, but their topological properties are also
quantified with respect to a number of measurements.  Also, frequently
real-world networks are compared with some of the existing theoretical
models such as scale-free and small-world.  While the application
potential of complex networks is clearly substantiated, it is
important to conclude the present survey by conducting a global
analysis of the reported applications.

Table~\ref{tab:summ} lists the 27 application areas considered here
and the corresponding \emph{number of reviewed papers}, the \emph{size
of the networks}, the \emph{number of applied measurements}, as well
as the \emph{number of theoretical models} adopted for comparative
purposes.  Despite the bias implied by the choice of articles to be
reviewed, this table still provides a representative snapshot of the
state of the art in complex networks applications. The first remarkable feature in this table is the fact that protein
applications is the area where complex networks have been more
intensively applied (50 applications), while the applications in
Internet (one of the important initial motivations for complex
networks research) is limited to 42 articles.  Other areas that
received particular interest from complex network research include
World Wide Web, transportation,  neuroscience, and
linguistics, all with at least 30 reviewed articles.  The
areas with the smallest number of applications (less than 10 revised
articles) include actors, acquaintances, trust,
sexual contacts, sports, financial market, metabolic,
chemistry, mathematics, and earthquake.  It is an interesting,
difficult issue to identify which of these areas will remain little
investigated, while others may become the focus of increasing
attention.

Interestingly, the measurements to characterize quantitatively the
connectivity of the networks vary considerably from one area to the
other, reaching as many as 20 different measurements in the case of
organizational management. Most applications, however, involve only 4
or 5 distinct measurements such as node degree and clustering
coefficient. Surprising trends can also be identified in
Table~\ref{tab:summ} regarding the number of theoretical models
considered or proposed as part of the investigation of the real-world
structures.  Reinforcing the major motivation and importance the
Internet has had in complex networks, a total of 12 models appeared in
the application articles.  This seems to corroborate the fact that a
definitive model of the Internet has not been reached yet. Several
papers with applications have considered 8 or more models,
including linguistics, WWW, transportation and computer science.

The results contained in the revised works, as well as the several
measurements and models adopted, provide unquestionable evidence about
the importance and dynamics of the complex network field. Because of
their intrinsic potential for representing, characterizing and
modeling real-world complex systems, complex networks are poised to
play a key role in an ever-increasing number of areas. The
developments which may characterize the future application of complex
networks include the use of additional measurements, required for a
more comprehensive characterization of investigated structures, as
well as additional theoretical models that may be created.

\begin{table}[ht]
  \caption{Number of papers by research area and network sizes, total number of measurements and number of models covered by the
    works considered in the review.} \label{tab:summ}
    \begin{center}
    \begin{tabular}{c|c|c|c|c}
      \textbf{Research field}  & \textbf{Papers} & \textbf{Network size} & \textbf{Measurements} & \textbf{Models} \\
      \hline
      Actors & 4  & 225,000 & 3 & 1 \\

      Citation & 11 & 1,000--$7,10^6$ & 18 & 0 \\

      Acquaintance & 7 & 1,500--375,000 & 12 & 0 \\

      Trust & 2 & 60,000 & 5 & 0 \\

      Sexual contacts &9 & 1,500--3,000 & 2 & 1 \\

      Sports & 3 & 100--13,500 & 4 & 0 \\

      Music &11 & 200--51,000 & 12 & 0 \\

      Collaboration & 11 & 300--$1.5\,10^6$ & 18 & 0 \\

      Communication & 22 & 15,000--$5\,10^7$ & 5 & 5 \\

      Economy & 17 & 180 & 3 & 0 \\

      Financial market &8  & 200--30,000 & 10 & 1 \\

      Computer science & 18 & 20--50,000 & 9 & 8 \\

      World Wide Web & 30 & 1,000--$271\,10^6$ & 7 & 9 \\

      Internet & 42 & 30--$1\,10^6$ & 12 & 12 \\

      Transportation & 30 & 100--7,000 & 8 & 8 \\

      Transcription & 20 & 500-2,000 & 5 & 2 \\

      Power grid & 15 & 500--15,000 & 4 & 2 \\

      Protein interaction & 50 & 2,000-20,000 & 5 & 3 \\

      Metabolic & 9 & 500-1,000 & 6 & 0 \\

      Ecology & 14 & 50--300 & 5 & 2 \\

      Neuroscience & 31 & 100-1,000 & 5 & 0 \\

      Linguistics & 35 & 100--500,000 & 13 & 8 \\

      Physics & 27 & 210,000 & 2 & 2 \\

      Chemistry & 3 & 400--3,200 & 5 & 2 \\

      Mathematics & 2 & 100--5,000 & 5 & 1 \\

      Earthquake & 6 & 1,200--60,000 & 6 & 2 \\

      Epidemic spreading & 19 & 1,000--$1\,10^8$ & 2 & 5 \\
      \hline

    \end{tabular}
  \end{center}
\end{table}

\section*{Acknowledgments}

L. da F. Costa is grateful to FAPESP (proc. 05/00587-5), CNPq
(proc. 301303/06-1) for financial support.  F. A. Rodrigues
acknowledges FAPESP financial support (proc. 07/50633-9). P. R.
Villas Boas is grateful to CNPq sponsorship (141390/2004-2). L. E.
C. Rocha is grateful to CNPq and to the Swedish Research Council
(grant 621-2002-4135) for financial support. L. Antiqueira
acknowledges FAPESP sponsorship (proc. 06/61743-7). M. P. Viana
is grateful to FAPESP sponsorship (proc. 07/50882-9 ). We are also
grateful to S. Abe  , M. Ausloos, R. Baggio, Y. Bar-Yam, G.
Bianconi, V. Blondel, D. Braha, R. Criado, F. D. Fallani, S. Dos
Reis, B. Kahng, J. Kert\'{e}sz, R. Lambiotte, B. Li, F. Liljeros, D.
Lusseau, M. M. Babu, Y. Maeno, G. Michailidis, A. Motter, C. Oosawa,
A. Penn, S. Roy, A. S\'{a}nchez, S. Sinh\'{a}, M. Small, F.
Sorrentino, D. Stauffer, F. Tamarit, M. Tugrul, D. Volchenkov, Y.
Xiao, and T. Zhou for comments and suggestions.

\bibliographystyle{unsrt}
\bibliography{survey}

\end{document}